\documentclass{emulateapj}
\shorttitle{Multi-Wavelength LFs at $1.9\la z\la 3.4$}
\shortauthors{Reddy et~al.}
\slugcomment{Received 2007 May 21; Accepted 2007 June 12}

\begin{document}

\newcommand{\umg}{U_{\rm n}-G}
\newcommand{\rmk}{{\cal R}-\ks}
\newcommand{\ebmv}{E(B-V)}
\newcommand{\sfr}{{\rm M}_{\odot} ~ {\rm yr}^{-1}}
\newcommand{\ks}{K_{\rm s}}
\newcommand{\ugr}{U_{\rm n}G{\cal R}}
\newcommand{\bzk}{BzK}
\newcommand{\rs}{{\cal R}}
\newcommand{\gmr}{G-\rs}
\newcommand{\lya}{Lyman~$\alpha$}
\newcommand{\lyb}{Lyman~$\beta$}
\newcommand{\za}{$z_{\rm abs}$}
\newcommand{\ze}{$z_{\rm em}$}
\newcommand{\cmtwo}{cm$^{-2}$}
\newcommand{\nhi}{$N$(H$^0$)}
\newcommand{\degpoint}{\mbox{$^\circ\mskip-7.0mu.\,$}}
\newcommand{\kms}{\,km~s$^{-1}$}      % note leading thinspace
\newcommand{\minpoint}{\mbox{$'\mskip-4.7mu.\mskip0.8mu$}}
\newcommand{\peryr}{\mbox{$\>\rm yr^{-1}$}}
\newcommand{\secpoint}{\mbox{$''\mskip-7.6mu.\,$}}
\newcommand{\sqdeg}{\mbox{${\rm deg}^2$}}
\newcommand{\squig}{\sim\!\!}
\newcommand{\subsun}{\mbox{$_{\twelvesy\odot}$}}
\newcommand{\et}{{\rm et al.}~}

%\twocolumn[\
\def\ltsima{$\; \buildrel < \over \sim \;$}
\def\simlt{\lower.5ex\hbox{\ltsima}}
\def\gtsima{$\; \buildrel > \over \sim \;$}
\def\simgt{\lower.5ex\hbox{\gtsima}}
\def\arcs{$''~$}
\def\arcm{$'~$}
\def\erf{\mathop{\rm erf}}
\def\erfc{\mathop{\rm erfc}}
%
%

%\slugcomment{DRAFT: \today}

\title{MULTI-WAVELENGTH CONSTRAINTS ON THE COSMIC STAR FORMATION HISTORY FROM SPECTROSCOPY:
THE REST-FRAME UV, H$\alpha$, and INFRARED LUMINOSITY FUNCTIONS AT REDSHIFTS $1.9\la z\la 3.4$\altaffilmark{1}}
\author{\sc Naveen A. Reddy\altaffilmark{2,3}, Charles C. Steidel\altaffilmark{2}, 
Max Pettini\altaffilmark{4}, Kurt L. Adelberger\altaffilmark{5}, Alice E. Shapley\altaffilmark{6},
Dawn K. Erb\altaffilmark{7}, and Mark Dickinson\altaffilmark{3}}

\altaffiltext{1}{Based, in part, on data obtained at the W.M. Keck
Observatory, which is operated as a scientific partnership among the
California Institute of Technology, the University of California, and
NASA, and was made possible by the generous financial support of the
W.M. Keck Foundation.  Also based in part on observations made with the
{\it Spitzer Space Telescope}, which is operated by the Jet Propulsion
Laboratory, California Institute of Technology, under a contract with NASA.}
\altaffiltext{2}{California Institute of Technology, MS 105--24, Pasadena, CA 91125}
\altaffiltext{3}{National Optical Astronomy Observatory, 950 N Cherry Ave, Tucson, AZ 85719}
\altaffiltext{4}{Institute of Astronomy, Madingley Road, Cambridge CB3 OHA, UK}
\altaffiltext{5}{McKinsey \& Company, 1420 Fifth Avenue, Suite 3100, Seattle, WA 98101}
\altaffiltext{6}{Department of Astrophysical Sciences, Peyton Hall-Ivy Lane, Princeton, NJ 08544}
\altaffiltext{7}{Harvard-Smithsonian Center for Astrophysics, 60 Garden Street, Cambridge, MA 02138}

\begin{abstract}

We use a sample of rest-frame UV selected and spectroscopically
observed galaxies at redshifts $1.9\le z<3.4$, combined with
ground-based spectroscopic H$\alpha$ and {\it Spitzer} MIPS
$24$~$\mu$m data, to derive the most robust measurements of the
rest-frame UV, H$\alpha$, and infrared (IR) luminosity functions (LFs)
at these redshifts.  Our sample is by far the largest of its kind,
with over $2000$ spectroscopic redshifts in the range $1.9\le z<3.4$
and $\sim 15000$ photometric candidates in 29 independent fields
covering a total area of almost a square degree.  Our method for
computing the LFs takes into account a number of systematic effects,
including photometric scatter, Ly$\alpha$ line perturbations to the
observed optical colors of galaxies, and contaminants.  Taking into
account the latter, we find no evidence for an excess of UV-bright
galaxies over what was inferred in early $z\sim 3$ LBG studies.  The
UV LF appears to undergo little evolution between $z\sim 4$ and $z\sim
2$.  Corrected for extinction, the UV luminosity density (LD) at
$z\sim 2$ is at least as large as the value at $z\sim 3$ and a factor
of $\sim 9$ larger than the value at $z\sim 6$, primarily reflecting
an increase in the number density of bright galaxies between $z\sim 6$
and $z\sim 2$.  Our analysis yields the first constraints anchored by
extensive spectroscopy on the infrared and bolometric LFs for faint
and moderately luminous ($L_{\rm bol}\la 10^{12}$~L$_{\odot}$)
galaxies.  Adding the IR to the emergent UV luminosity, incorporating
independent measurements of the LD from ULIRGs, and assuming realistic
dust attenuation values for UV-faint galaxies, indicates that galaxies
with $L_{\rm bol}<10^{12}$~L$_{\odot}$ account for $\approx 80\%$ of
the bolometric LD and SFRD at $z\sim 2-3$.  This suggests that
previous estimates of the faint-end of the $L_{\rm bol}$ LF may have
underestimated the steepness of the faint-end slope at $L_{\rm
bol}<10^{12}$~L$_{\odot}$.  Our multi-wavelength constraints on the
global SFRD indicate that approximately one-third of the present-day
stellar mass density was formed in sub-ultraluminous galaxies between
redshifts $z=1.9-3.4$.

\end{abstract}

%]

\section{INTRODUCTION}

Constraining the star formation history and stellar mass evolution of
galaxies is a central component of understanding galaxy formation.
Observations of the stellar mass and star formation rate density, the
QSO density, and galaxy morphology at both low ($z\la 1$) and high
($z\ga 3$) redshifts indicate that most of the activity responsible
for shaping the bulk properties of galaxies to their present form
occurred in the epochs between $1\la z \la 3$ (e.g.,
\citealt{dickinson03, rudnick03, madau96, lilly96, lilly95, steidel99,
shaver96, fan01, dimatteo03, conselice04, papovich03, shapley01,
giavalisco96}).  While this period in the Universe was perhaps the
most active in terms of galaxy evolution and accretion activity, it
was not until recently that advances in detector sensitivity and
efficiency, the increased resolution and light-gathering capability
afforded by larger 8-10 meter class telescopes, and a number of new
powerful imagers and spectrographs on space-based missions such as
$HST$, $Spitzer$, and $Chandra$, allowed for the study of large
numbers of galaxies at $z\sim 2$.  These developments have prompted a
spate of multi-wavelength surveys of high redshift galaxies from the
far-IR/submm to IR, near-IR, optical, and UV, enabling us to examine
the SEDs of star forming galaxies over much of the 7 decades of
frequency over which stars emit their light either directly or
indirectly through dust processing (e.g., \citealt{steidel03,
steidel04, daddi04, daddi04b, franx03, vandokkum03, vandokkum04,
abraham04, chapman05, smail03}).

The first surveys that efficiently amassed large samples of high
redshift galaxies used the observed $\ugr$ colors of galaxies to
identify those with a deficit of Lyman continuum flux (e.g.,
\citealt{steidel95}) in the $U_{\rm n}$ band (i.e., U ``drop-outs'')
for galaxies at $z\sim 3$.  Those initial results have been adapted to
select galaxies at higher redshifts ($z>4$; e.g.,
\citealt{bouwens05, bouwens04, dickinson04, bunker04, yan03}) and
moderate redshifts ($1.4\la z\la 3$; \citealt{adelberger04,
steidel04}).  Combining these high redshift results with those from
{\it GALEX} (e.g., \citealt{wyder05}), we now have an unprecedented
view of the rest-frame UV properties of galaxies from the epoch of
reionization to the present, perhaps the only wavelength for which
star forming galaxies have been studied across more than $\sim 93\%$
of the age of the Universe.  The accessibility of rest-frame UV
wavelengths over almost the entire age of the Universe makes
rest-frame UV luminosity functions (LFs) useful tools in assessing the
cosmic star formation history in a consistent manner.

The foray of observations into the epoch around $z\sim 2$ has occurred
relatively recently, and with it have come various determinations of
the UV LFs at these epochs \citep{gabasch04, lefevre05}.
Unfortunately, such studies are often limited either because (1) they
are purely magnitude limited (resulting in inefficient selection of
galaxies at the redshifts of interest and even fewer galaxies with
secure spectroscopic redshifts and poorly determined contamination
fraction), (2) they generally rely on photometric redshifts that are
highly uncertain at $z\sim 2$, and/or (3) they are estimated over a
relatively small number of fields such that cosmic variance may be an
issue.  While purely magnitude limited surveys allow one to easily
quantify the selection function, as we show below, Monte Carlo
simulations combined with accurate spectroscopy can be used to
quantify even the relatively complicated redshift selection functions
and biases of color-selected samples of high redshift galaxies.  This
``simulation'' approach allows one to assess a number of systematics
(e.g., photometric imprecision, perturbation of colors due to line
strengths, etc.)  and their potential effect on the derived LF; these
systematic effects have been left untreated in previous calculations
of the LFs at $z\sim 2-3$ \citep{gabasch04, lefevre05}, but are
nonetheless found to be important in accurately computing the LF
(e.g., \citealt{adel00, bouwens04, bouwens05, bouwens06}).

For the past several years, the main focus of our group has been to
assemble a large sample of galaxies at the peak epoch of galaxy
formation and black hole growth, corresponding to redshifts $1.5\la
z\la 2.6$, in multiple independent fields.  The selection criteria aim
to identify actively star-forming galaxies at $z\sim 2$ with the same
range in intrinsic UV color and extinction as Lyman break galaxies
(LBGs) at $z\sim 3$ \citep{steidel03}.  The color selection criteria
are described in \citet{adelberger04, steidel04}.  Initial results
from the survey, including analyses of the star formation rates,
stellar populations, stellar and dynamical masses, gas-phase
metallicities, morphologies, outflow properties, and clustering are
presented in several papers (e.g., \citealt{shapley05, adelberger05a,
adelberger05b, erb06a, erb06c, erb06b, reddy06a, reddy05a, reddy04,
steidel05}).  With a careful accounting of extinction, photometric
imprecision, and systematic effects caused by observational
limitations, rest-frame UV selected samples can be used to estimate
the cosmological star formation history (e.g., \citealt{steidel99,
adel00, bouwens04, bouwens05, bouwens06, bunker04, bunker06}).

There are primarily two methods by which one can attempt to construct
``complete'' luminosity functions that make a reasonable account of
all star formation at a given epoch.  The first method is to observe
galaxies over as wide a range in wavelengths as possible in order to
establish a census of all galaxies which dominate the star formation
rate density.  For example, the union of rest-frame UV
\citep{adelberger04, steidel04}, rest-frame optical \citep{franx03,
daddi04}, and submillimeter-selected samples (e.g., \citealt{smail97,
hughes98, barger98, blain02, chapman05}) should account for optically-bright
galaxies with little to moderate dust extinction as well as the
population of optically-faint and heavily reddened galaxies.  One can
then examine the intersection between these various samples and,
taking into account overlap, compute the total star formation rate
density \citep{reddy05a}.  Unfortunately, this technique poses several
challenging problems, not the least of which are the practicality of
obtaining multi-wavelength data in a large number of uncorrelated
fields, disparate data quality and photometric depth between optical
and near-IR images, and the inefficiency of spectroscopically
identifying galaxies in near-IR selected samples to properly quantify
the selection function.

The second approach, and the one which we adopt in this paper, is to
estimate sample completeness by way of simulations.  This method
involves simulating many realizations of the intrinsic distribution of
galaxy properties at high redshift, subjecting these realizations to
the same photometric methods and selection criteria as applied to real
data, and adjusting the simulated realizations until convergence
between the expected and observed distribution of galaxy properties is
achieved.  The method thus corrects for a large fraction of galaxies
that might be ``missing'' from the sample, just as long as some of
them are spectroscopically observed.  The obvious disadvantage of this
method is that some (e.g., optically-faint) galaxies will never be
scattered into our selection window and hence we cannot account for
such galaxies in our analysis.  However, multi-wavelength data in
several of our fields enable us to quantify the magnitude of, and
correct for, the incompleteness resulting from objects that never
scatter into our sample.  Simulations such as the kind presented in
this paper become even more important at higher redshift ($z\ga 4$)
where no corresponding multi-wavelength data exist to assess the
fraction of galaxies that are not recovered by color selection.  In our
case, applying the Monte Carlo method to joint photometric and
spectroscopic samples of high redshift galaxies allows one to assess
the systematic effects of photometric scattering and the intrinsic
variation in colors due to line emission and absorption with
unprecedented accuracy.  In this paper, we take advantage of a large
sample of spectroscopically-confirmed star-forming galaxies to arrive
at the first completeness-corrected spectroscopic estimate of the UV
LF and star formation rate density (SFRD) at $z\sim 2$, computed
across the many independent fields of our survey.  We extend our
results by using spectroscopy of Lyman Break galaxies in many new
independent fields to recompute the UV LF and SFRD at $z\sim 3$.

While considerable progress in quantifying the cosmic star formation
history can be achieved by UV observations alone, the most robust
determination can only come from an analysis at multiple wavelengths,
where systematic effects (e.g., extinction) can be corrected for.  In
addition, assessing the star formation history consistently at several
different wavelengths allows for a useful cross-check between results
and may reveal any underlying trends between the star-forming
properties of galaxies and redshift.  In this paper, we combine
extensive multi-wavelength data in our fields with our
spectroscopically-derived completeness corrections to measure the
rest-frame UV, H$\alpha$, and infrared luminosity functions at
redshifts $z\sim 2-3$.  The primary goal of this paper is to then
use these luminosity functions to evaluate the cosmic star formation
history in a consistent manner across $4$~decades of wavelength.

The outline of this paper is as follows.  In \S~\ref{sec:selection},
we describe the fields of our survey and the color criteria used to
selected candidate galaxies at $z\sim 2$.  We then proceed with a
description of the spectroscopic followup and quantify the fraction of
contaminants, including low redshift ($z<1$) star forming galaxies and
low and high redshift AGN and QSOs, within the sample.  We conclude
\S~\ref{sec:selection} by demonstrating that the redshift distribution
for the spectroscopic sample is not significantly biased when compared
with the redshift distribution of all photometric candidates at $z\sim
2$.  In \S~\ref{sec:ic}, we detail the Monte Carlo method used to
assess both photometric bias and error, the effect of Ly$\alpha$ line
perturbations on the observed rest-UV colors of galaxies, and the
procedure used to correct our sample for completeness.  Our results
pertaining to the intrinsic Ly$\alpha$ equivalent width and reddening
distributions of $1.9\le z<3.4$ galaxies are discussed in
\S~\ref{sec:results1}.  Results on the rest-frame UV, IR, and
H$\alpha$ LFs are presented respectively in \S~\ref{sec:uvlf},
\ref{sec:irlf}, and \ref{sec:half}.  Lastly, we discuss the
implications of our results for the luminosity and global star
formation rate densities in \S~\ref{sec:discussion}.  A flat
$\Lambda$CDM cosmology is assumed with
$H_{0}=70$~km~s$^{-1}$~Mpc$^{-1}$, $\Omega_{\Lambda}=0.7$, and
$\Omega_{\rm m}=0.3$.

\section{DATA: SAMPLE SELECTION AND SPECTROSCOPY}
\label{sec:selection}

\subsection{Fields}
\label{sec:fields}

Our $z\sim 2$ survey is being conducted primarily in fields chosen for
having $V\le 17.5$~mag QSOs with redshifts $2.5\la z\la 2.8$, ideally
placed to study the correlation between $z\sim 2$ galaxies and HI and
high-metallicity (e.g., CIV) absorbing systems in the IGM (see
\citealt{adelberger05b}).  We have extended our survey to include the
GOODS-North field \citep{dickinson03,giavalisco04}, encompassing the
original HDF-North field \citep{williams96,williams00}, and the
Westphal field (currently encompassed by the large Extended Groth
Strip survey) to take advantage of the multi-wavelength data amassed
for these fields.  Imaging was conducted under similar conditions as
the $z\sim 3$ fields of \citet{steidel03}.  The $14$ fields of the
$z\sim 2$ survey are summarized in Table~\ref{tab:fields} (instruments
used and dates of observation are shown in Table~1 of
\citealt{steidel04}).

\begin{deluxetable*}{lccccccc}
\tabletypesize{\footnotesize}
\tablewidth{0pc}
\tablecaption{Survey Fields}
\tablehead{
\colhead{} &
\colhead{$\alpha$\tablenotemark{a}} &
\colhead{$\delta$\tablenotemark{b}} &
\colhead{Field Size} &
\colhead{} &
\colhead{} & 
\colhead{} & 
\colhead{} \\
\colhead{Field Name} &
\colhead{(J2000.0)} &
\colhead{(J2000.0)} &
\colhead{(arcmin$^{2}$)} &
\colhead{$N_{\rm BX}$\tablenotemark{c}} &
\colhead{$N_{\rm BX}(1.9\le z<2.7)$\tablenotemark{d}} &
\colhead{$N_{\rm LBG}$\tablenotemark{e}} &
\colhead{$N_{\rm LBG}(2.7\ge z<3.4)$\tablenotemark{f}}}
\startdata
Q0000	&	00 03 25 & -26 03 37 &	18.9	&	... &	...	&	28 &	12 \\
CDFa	&	00 53 23 & 12 33 46 &	78.4	&	... &	...	&	100 &	30 \\
CDFb	&	00 53 42 & 12 25 11 &	82.4	&	... &	...	&	121 &	21 \\
Q0100	&	01 03 11 & 13 16 18 &	42.9	&	345 &	65	&	100 &	18 \\
Q0142	&	01 45 17 & -09 45 09 &	40.1	&	287 &	72	&	100 &	20 \\
Q0201	&	02 03 47 & 11 34 22 &	75.7	&	... &	...	&	87 &	13 \\
Q0256	&	02 59 05 & 00 11 07 &	72.2	&	... &	...	&	120 &	42 \\
Q0302	&	03 04 23 & -00 14 32 &	244.9	&	... &	...	&	191 &	29 \\
Q0449	& 	04 52 14 & -16 40 12 &	32.1	&	188 &	40	&	88 &	13 \\
B20902	&	09 05 31 & 34 08 02 &	41.8	&	... &	...	&	78 &	34 \\
Q0933	&	09 33 36 & 28 45 35 &	82.9	&	... &	...	&	211 &	47 \\
Q1009	&	10 11 54 & 29 41 34 &	38.3	&	306 &	33	&	137 &	25 \\
Q1217	&	12 19 31 & 49 40 50 &	35.3	&	240 &	26	&	65 &	11 \\
GOODS-N	&	12 36 51 & 62 13 14 &	155.3	&	909 &	138	&	210 &	62 \\
Q1307	&	13 07 45 & 29 12 51 &	258.7	&	1763 &	40	&	564 &	8 \\
Westphal &	14 17 43 & 52 28 49 &	226.9	&	612 &	39	&	334 &	177 \\
Q1422	&	14 24 37 & 22 53 50 &	113.0	&	... &	...	&	453\tablenotemark{g} &	92\tablenotemark{h} \\
3C324	&	15 49 50 & 21 28 48 &	44.1	&	... &	...	&	51 &	10 \\
Q1549	&	15 51 52 & 19 11 03 &	37.3	&	243 &	49	&	119 &	46 \\
Q1623	&	16 25 45 & 26 47 23 &	290.0	&	1348 &	209	&	580 &	24 \\
Q1700	&	17 01 01 & 64 11 58 &	235.3	&	1472 &	92	&	438 &	38 \\
Q2206 	&	22 08 53 & -19 44 10 &	40.5	&	213 &	49	&	52 &	22 \\
SSA22a	&	22 17 34 & 00 15 04 &	77.7	&	... &	...	&	146 &	47 \\
SSA22b	&	22 17 34 & 00 06 22 &	77.6	&	... &	...	&	89 &	28 \\
Q2233	&	22 36 09 & 13 56 22 &	85.6	&	... &	...	&	94 &	36  \\
DSF2237b &	22 39 34 & 11 51 39 &	81.7	&	... &	...	&	176 &	45 \\
DSF2237a &	22 40 08 & 11 52 41 &	83.4	&	... &	...	&	121 &	30 \\
Q2343	&	23 46 05 & 12 49 12 &	212.8	&	1018 &	128	&	428 &	25 \\
Q2346	&	23 48 23 & 00 27 15 &	280.3	&	1069 &	37	&	171 &	1 \\
\\
{\bf TOTAL}	&	{\bf ...} & {\bf ...} &	{\bf 3186.1}	& {\bf 10013} &	{\bf 1017} & {\bf 5452} & {\bf 1006 }\\
\enddata
\tablenotetext{a}{Right ascension in hours, minutes, and seconds.}
\tablenotetext{b}{Declination in degrees, arcminutes, and arcseconds.}
\tablenotetext{c}{Number of BX candidates to $\rs=25.5$.}
\tablenotetext{d}{Number of spectroscopically confirmed BX candidates 
with redshifts $1.9\le z<2.7$, excluding those whose spectra indicate
an AGN/QSO.  Note that the total numbers of galaxies, excluding AGN and QSOs,
 with spectroscopic
redshifts $1.9\le z<2.7$ in the survey (total in all fields is 1288) are larger than the numbers given
here since a significant fraction of LBG and BM candidates lie at these redshifts.}
\tablenotetext{e}{Number of LBG candidates to $\rs=25.5$.}
\tablenotetext{f}{Number of spectroscopically confirmed LBG candidates
with redshifts $2.7\le z<3.4$, excluding those whose spectra indicate
an AGN/QSO.  Note that the total numbers of galaxies, excluding AGN and QSOs,
with spectroscopic redshifts $2.7\le z<3.4$ in the survey (total in all fields is $1058$) 
are larger than the numbers given 
here since a small fraction of BX candidates lie at these redshifts.}
\tablenotetext{g}{Number includes 180 galaxies with $25.5<\rs \le 26.0$.}
\tablenotetext{h}{Number includes 10 galaxies with $25.5<\rs \le 26.0$.}
\label{tab:fields}
\end{deluxetable*}

We have expanded significantly the number of spectroscopically
confirmed LBGs beyond the original sample from $17$ fields presented
in \citet{steidel03}, by including those LBGs selected in the newer
fields of the $z\sim 2$ survey.  We have used these new spectroscopic
redshifts, in addition to those previously published in
\citet{steidel03}, to re-evaluate the UV LF and SFRD at $z\sim 3$.
Fields where we have carried out LBG selection are also listed in
Table~\ref{tab:fields}.

One of the unique advantages of our analysis is that we use a large
number of uncorrelated fields (14 and 29 for the $z\sim 2$ and $z\sim
3$ surveys, respectively), combined with a large sample of
spectroscopic redshifts between $1.9\le z<3.4$, in order to compute
the LF, negating the need for uncertain normalization corrections to
account for clustering and cosmic variance.  For example, we find
evidence for significant large scale structure within several fields
of the $z\sim 2$ survey (e.g., \citealt{steidel05}), generally
characterized by over-densities in redshift space above what would be
expected given our redshift selection function.  By averaging results
over many fields well distributed throughout the sky, we can estimate
the LF insensitive to variations in large scale structure, and
furthermore estimate the magnitude of the effect of cosmic variance on
the results.  The total area of all of the independent fields of the
$z\sim 3$ survey is $\sim 3200$~square~arcmin, or close to a full
square degree.  The $z\sim 2$ survey area is $\sim
1900$~square~arcmin.  Despite the smaller area covered by the $z\sim
2$ survey, there are roughly twice as many BX candidates as LBGs given
the larger surface density of the former.

\subsection{Photometry}

Photometry was performed using a modified version of FOCAS
\citep{valdes82}. Object detection was done at $\rs$ band, and $\gmr$
and $\umg$ colors were computed by applying the $\rs$-band isophotal
apertures to the $G$ and $U_n$ images (see
\citealt{steidel03,steidel04} for further details).  The optical
images have typical depth of $\rs\sim 27.5$ as measured through a
$\sim 3\arcsec$ diameter aperture ($3$~$\sigma$).  Field-to-field
variations in photometry are dominated by systematics due to the
different instruments, filter sets, and slightly varying observing
conditions when the fields were imaged.  These field-to-field
systematics are negligible compared to measurement errors.  We have
incorporated some of these effects (e.g., seeing, airmass of the
observation, CCD response, and filter shape) in computing the expected
colors of galaxies with known intrinsic properties.  Modeling all the
field-to-field variations in photometry rapidly becomes a very complex
problem, infeasible to resolve within a reasonable time frame.  The
remaining biases (e.g., errors in the zeropoints used) are discussed
in \S~\ref{sec:photerrsec}.

\subsection{Color Selection}
\label{sec:colorselection}

Even with {\it a priori} knowledge of the intrinsic properties of all
$z\sim 2$ galaxies, constructing a practical set of selection criteria
to select all galaxies in any desired redshift range and reject all
others is an intractable problem.  One extreme is to select all
objects down to a given magnitude limit, such as in flux-limited
surveys of high redshift galaxies, but unfortunately such studies
suffer from significant amounts of foreground contamination.
Color-selected samples have the advantage of allowing one to
specifically target a desired redshift range while minimizing the
number of interlopers.  Perhaps the most successful of the various
color criteria that have been designed to select high redshift
galaxies is rest-frame UV color selection, initially used to target
galaxies at $z\sim 3$ \citep{steidel95}, and extended to higher
redshifts (e.g., \citealt{bouwens05, bouwens04, bunker04, dickinson04,
yan03}).  The success of this technique is partly due to its
simplicity in that only a few broadband filters are required to
assemble such samples.  Further, at lower redshifts ($1.4\la z\la
3.5$), galaxies can be spectroscopically observed and precise
redshifts can be obtained in a short amount of observing time on
$8-10$m class telescopes.  Color-selected high redshift galaxy surveys
will, as a consequence, have rather complex selection functions.  The
approach described in \S~\ref{sec:ic} allows one to quantify such
selection functions with relative ease.

The criteria used to select galaxies with redshifts $1.9\la z \la 2.7$
based on their rest-frame UV colors were designed to recover objects
with intrinsic properties similar to those of $z\sim 3$ Lyman Break
Galaxies.  The colors at $z\sim 2$ were estimated from spectral
synthesis analysis of $70$ LBGs with broadband $U_{\rm n}G{\cal
R}JK_{\rm s}$ photometry and spectroscopic redshifts
(\citealt{adelberger04}; \citealt{steidel04}).  Initial spectroscopy
of $z\sim 2$ candidates led to a refinement of the criteria used to
select galaxies at redshifts $1.9\la z \la 2.7$ to their present form:
\begin{eqnarray}
G-{\cal R} &\geq& -0.2\nonumber\\
U_n-G      &\geq& G-{\cal R}+0.2\nonumber\\
G-{\cal R} &\leq &0.2(U_n-G)+0.4\nonumber\\
U_n-G      &\leq& G-{\cal R}+1.0,
\label{eq:bx}
\end{eqnarray}
termed as ``BX'' selection (\citealt{adelberger04};
\citealt{steidel04}), with fluxes in units of AB magnitudes
\citep{oke83}.  Candidates were selected to $\rs=25.5$ to ensure a
sample of galaxies amenable to spectroscopic followup.  This limit
corresponds to an absolute magnitude at observed $\rs$-band that is
$0.6$~mag fainter at $z=2.2$ (the mean redshift of BX candidates with
$z>1$) than at $z\sim 3$.  Additionally, we exclude all sources with
${\cal R}<19$ that are saturated in our images, all of which are
stars.  The above criteria yielded $10013$ candidates in the $14$
fields, with an average surface density of $\sim 5$~arcmin$^{-2}$,
uncorrected for contamination from objects with redshifts $z<1.4$ (see
\S~\ref{sec:interlopers}).  The number of candidates in each field are
summarized in Table~\ref{tab:fields}.  Note that we did not select BX
galaxies in $15$ fields of the $z\sim 3$ survey \citep{steidel03}
since spectroscopy of these fields was carried out before the $z\sim
2$ survey began.

The color criteria used to select LBGs at redshifts $2.7\la z\la 3.4$
are published in \citet{steidel03} and are summarized here for convenience:
\begin{eqnarray}
G-{\cal R} & \leq & 1.2\nonumber\\
U_n-G & \geq & G-{\cal R} + 1.0.
\label{eq:lbgsel}
\end{eqnarray}
These criteria form the superset of the individual sets of criteria
for ``C'', ``D'', ``M'', and ``MD'' candidate types given in
Table~4 of \citet{steidel03}.  Hereafter, we will refer to all these
different candidate types as LBGs.  Candidates were selected to
$\rs=25.5$, except in the field Q1422 where the photometric depth
allowed selection of candidates to $\rs=26.0$.  The number of $z\sim 3$
candidates in each field are also summarized in Table~\ref{tab:fields}.
Given the constraints of the color criteria and the $\rs=25.5$
spectroscopic limit, the combined BX and LBG samples constitute $\sim
25\%$ of the total $\rs$ and $\ks$-band counts to $\rs=25.5$ and
$\ks({\rm AB})=24.4$, respectively.

\subsection{Spectroscopic Followup}
\label{sec:specfollow}

\begin{figure}[tbh]
\plotone{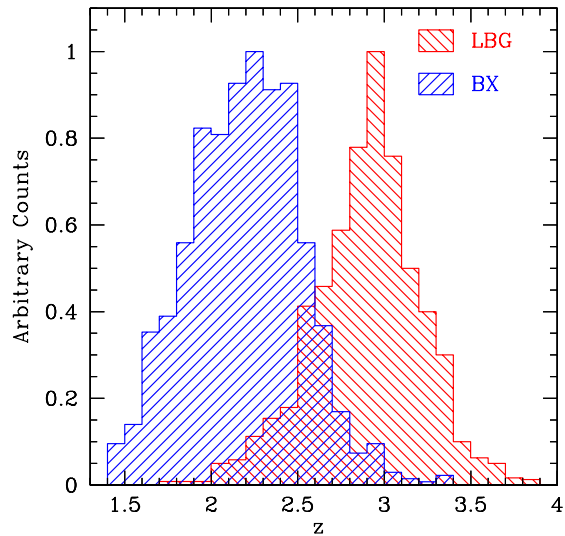}
\caption{Arbitrarily normalized
  spectroscopic redshift distributions of galaxies with $z>1.4$ in the
  BX and LBG samples.  The total number of galaxies represented
  here is 2569.
\label{fig:zhistall}}
\end{figure}

The spectroscopic followup of candidates is discussed extensively in
\citet{steidel03} and \citet{steidel04}.  Of the 10013 BX candidates,
we have targeted $24\%$ (2382 out of 10013) with spectroscopy,
yielding 1711 redshift identifications, or a $72\%$ success rate
averaged over all fields.  As discussed in \S~\ref{sec:speccomp}, the
spectroscopic success rate is primarily determined by the observing
conditions, and subsequent spectroscopy of spectroscopic failures
indicates they have similar redshift distribution as successes.
Similarly, of the 5452 LBG candidates, 1903 were targeted with 1492
successful redshifts.  Figure~\ref{fig:zhistall} shows arbitrarily
normalized redshift distributions for the BX and LBG samples.  The
mean spectroscopic redshifts for the BX and LBG samples, when
restricted to those objects with $z>1$, are $\langle z\rangle
=2.20\pm0.32$ and $\langle z\rangle = 2.96\pm0.26$, respectively.
Preliminary versions of these histograms, along with sample spectra of
BX galaxies and LBGs, are presented in \citet{steidel03,steidel04}.
Table~\ref{tab:specfrac} lists the spectroscopic fractions relevant
for the BX and LBG samples.

\begin{deluxetable*}{lcccccccccc}
\tabletypesize{\footnotesize}
\tablewidth{0pc}
\tablecaption{Spectroscopic and AGN/QSO Fractions of the BX and LBG Samples}
\tablehead{
\colhead{} &
\colhead{} &
\colhead{} &
\colhead{BX} &
\colhead{} &
\colhead{} &
\colhead{} &
\colhead{} &
\colhead{LBG} &
\colhead{} &
\colhead{} \\
\colhead{${\cal R}$} &
\colhead{$N_{\rm phot}$\tablenotemark{a}} &
\colhead{$N_{\rm spec}$\tablenotemark{b}} &
\colhead{$f_{\rm spec}$\tablenotemark{c}} &
\colhead{$f_{\rm AGN}$\tablenotemark{d}} &
\colhead{$f_{\rm AGN}(z \ge 1.4)$\tablenotemark{e}} &
\colhead{$N_{\rm phot}$\tablenotemark{f}} &
\colhead{$N_{\rm spec}$\tablenotemark{g}} &
\colhead{$f_{\rm spec}$\tablenotemark{h}} &
\colhead{$f_{\rm AGN}$\tablenotemark{i}} &
\colhead{$f_{\rm AGN}(z \ge 1.4)$\tablenotemark{j}}}
\startdata
$19.0-22.0$ & 620 & 74 & 0.12 & 0.12 & 0.78 & 142 & 30 & 0.21 & 0.33 & 1.00 \\
$22.0-22.5$ & 162 & 31 & 0.19 & 0.00 & 0.00 & 34 & 7 & 0.21 & 0.17 & 1.00 \\
$22.5-23.0$ & 252 & 77 & 0.31 & 0.05 & 0.20 & 71 & 25 & 0.35 & 0.23 & 0.67 \\
$23.0-23.5$ & 466 & 178 & 0.38 & 0.01 & 0.02 & 137 & 62 & 0.45 & 0.07 & 0.10 \\
$23.5-24.0$ & 1053 & 330 & 0.31 & 0.03 & 0.03 & 392 & 177 & 0.45 & 0.02 & 0.03 \\
$24.0-24.5$ & 1894 & 511 & 0.27 & 0.01 & 0.01 & 881 & 398 & 0.45 & 0.02 & 0.03 \\
$24.5-25.0$ & 2741 & 341 & 0.12 & $<0.01$ & $<0.01$ & 1617 & 442 & 0.27 & 0.01 & 0.01 \\
$25.0-25.5$ & 2819 & 169 & 0.06 & 0.02 & 0.02 & 1994 & 336 & 0.17 & $<0.01$ & $<0.01$ \\
$25.5-26.0$ & ... & ... & ... & ... & ... & 180\tablenotemark{k} & 15\tablenotemark{k} &  0.08\tablenotemark{k} & 0.00\tablenotemark{k} & 0.00\tablenotemark{k} \\
\\
{\bf Total} & {\bf 10007} & {\bf 1711} & {\bf 0.17} & {\bf 0.02} & {\bf 0.02} & {\bf 5448} & {\bf 1492} & {\bf 0.27} & {\bf 0.03} & {\bf 0.03} \\
\enddata
\tablenotetext{a}{Number of BX candidates.}
\tablenotetext{b}{Number of BX candidates with spectroscopic redshifts.}
\tablenotetext{c}{Fraction of BX candidates with spectroscopic redshifts.}
\tablenotetext{d}{Fraction of AGN/QSOs in BX sample with spectroscopic redshifts.}
\tablenotetext{e}{Fraction of AGN/QSO in BX sample with $z_{\rm spec}\ge 1.4$.}
\tablenotetext{f}{Number of LBG candidates.}
\tablenotetext{g}{Number of LBG candidates with spectroscopic redshifts.}
\tablenotetext{h}{Fraction of LBG candidates with spectroscopic redshifts.}
\tablenotetext{i}{Fraction of AGN/QSOs in LBG sample with spectroscopic redshifts.}
\tablenotetext{j}{Fraction of AGN/QSO in LBG sample with $z_{\rm spec}\ge 1.4$.}
\tablenotetext{k}{Numbers are for Q1422 field.}
\label{tab:specfrac}
\end{deluxetable*}

\subsection{Interloper Contribution and AGN}
\label{sec:interlopers}

The region of color space defined by BX selection (e.g.,
Figure~\ref{fig:ewdist}) is also expected to include galaxies outside
of the targeted redshift range, including star forming galaxies at
$z\la 0.2$ and stars (see Figure~10 of \citealt{adelberger04}).
Spectroscopy shows that there is indeed a subset of BX candidates
that are interlopers --- candidates with redshifts $z<1.4$ --- with a
much higher contamination rate among candidates with ${\cal R}<23.5$,
as indicated in Table~\ref{tab:interlopertab}.  One can impose a rough
magnitude cutoff to only consider those candidates with ${\cal R}\ge
23.5$, but this would preclude the analysis of the bright-end of the
BX and LBG luminosity distributions, as well as more detailed studies
of the UV spectra of optically-bright objects.  Other options to
reduce the contamination fraction include using the ${\cal R}-K$ color
where the associated bands no longer bracket strong spectral breaks
for low redshift sources.  For example, the $\bzk$ criteria of
\citet{daddi04} can be used to reduce the foreground contamination
fraction in color-selected samples.

\begin{deluxetable*}{lcccccc}
\tabletypesize{\footnotesize}
\tablewidth{0pc}
\tablecaption{Interloper ($z<1.4$) Statistics of the BX and LBG Samples }
\tablehead{
\colhead{} &
\colhead{} &
\colhead{BX} &
\colhead{} &
\colhead{} &
\colhead{LBG} &
\colhead{} \\
\colhead{${\cal R}$} &
\colhead{$N_{\rm z\ge0}$\tablenotemark{a}} &
\colhead{$N_{\rm 0\le z<1.4}$\tablenotemark{b}} &
\colhead{$f_{\rm 0\le z < 1.4}$\tablenotemark{c}} & 
\colhead{$N_{\rm z\ge0}$\tablenotemark{a}} & 
\colhead{$N_{\rm 0\le z<1.4}$\tablenotemark{b}} &
\colhead{$f_{\rm 0\le z < 1.4}$\tablenotemark{c}}}
\startdata
$19.0-22.0$ & 74 & 65 & 0.88 & 30 & 20 & 0.67 \\
$22.0-22.5$ & 31 & 28 & 0.90 & 7 & 5 & 0.71 \\
$22.5-23.0$ & 77 & 56 & 0.73 & 25 & 17 & 0.68 \\
$23.0-23.5$ & 178 & 65 & 0.37 & 62 & 19 & 0.31 \\
$23.5-24.0$ & 330 & 58 & 0.18 & 177 & 19 & 0.11 \\
$24.0-24.5$ & 511 & 37 & 0.07 & 398 & 13 & 0.03 \\
$24.5-25.0$ & 341 & 19 & 0.06 & 442 & 1 & $<0.01$ \\
$25.0-25.5$ & 169 & 4 & 0.02 & 336 & 2 & $<0.01$ \\
$25.5-26.0$ & ... & ... & ... & 15 & 0 & $<0.01$ \\
\\
{\bf TOTAL} & {\bf 1711} & {\bf 332} & {\bf 0.19} & {\bf 1492} & {\bf 96} & {\bf 0.06} \\
\enddata
\tablenotetext{a}{Number of sources with spectroscopic redshifts.}
\tablenotetext{b}{Number of sources with $z < 1.4$.}
\tablenotetext{c}{Fraction with $z<1.4$.}
\label{tab:interlopertab}
\end{deluxetable*}

The interloper fractions are apt to decrease as the survey progresses
and we become more adept at excluding them from masks based on other
multi-wavelength data, such as their ${\cal R}-K$ colors.  However,
until now, we have not used any of the techniques discussed above to
actively discriminate against placing possible interlopers on
slitmasks; doing so would complicate our ability to apply the observed
contamination fractions to determine the interloper rate among all
BX/LBG sources.  Therefore, the fractions in columns (4) and (7) of
Table~\ref{tab:interlopertab} are assumed to represent the overall
fraction of interlopers as a function of ${\cal R}$ for the
photometric samples.  For the BX sample, most of the contamination at
bright magnitudes arises from foreground galaxies.  For the LBG
sample, most of the contamination arises from stars.  Applying a
bright magnitude limit of $\rs=23.5$ will reduce the overall
contamination fractions of the BX and LBG samples to $9\%$ and $3\%$,
respectively.

The BX sample also includes a small number of broad-lined QSOs and
broad and narrow line ($\sigma < 2000$~km~s$^{-1}$) AGN whose rest-UV
colors are similar to those of high redshift star forming galaxies,
but which show prominent (and in some cases broad) emission lines such
as Ly$\alpha$, CIV, and NV.  The detection rate of such sources is
$\sim 2.8\%$ (similar to the rate found among UV-selected $z\sim 3$
galaxies; \citealt{steidel02}), but is a strong function of apparent
magnitude where all but two of the objects with $\rs<22.0$ and $z>1$
are QSOs.  The fractions of spectroscopically confirmed BXs and LBGs
that show high ionization UV lines indicative of an AGN or QSO are
listed in Table~\ref{tab:specfrac}.  As discussed in \citet{reddy06b},
we have found the presence of additional AGN in the sample based on
either X-ray or IR data.  For the analysis presented here, we exclude
AGN from the sample based on the presence of high ionization UV
emission lines.  The effect of including AGN/QSOs in the rest-frame UV
LF is discussed further in \S~\ref{sec:vvds}.

\break 

\subsection{Spectroscopic Completeness}
\label{sec:speccomp}

Assessing photometric and spectroscopic completeness is a key
ingredient in determining the total completeness of our survey.  The
photometric completeness (i.e., the fraction of galaxies at redshifts
$1.9\la z \la 3.4$ that satisfy either the BX or LBG color selection)
is discussed in \S~\ref{sec:ic}.  Here we focus on the extent to which
the redshift distribution of the spectroscopic sample reflects that of
the photometric sample as a whole.  There are several observations
that suggest that the redshift selection functions for the
spectroscopic samples reflect the overall redshift selection functions
had we obtained spectroscopic redshifts for every single candidate.
First, the success of measuring redshifts is primarily a function of
the weather conditions (e.g., cirrus, seeing) at the time of
observation, with a $90\%$ success rate in the best conditions.
Repeat observations of objects for which we were unable to secure a
redshift initially indicate that the redshift distribution of
spectroscopic failures is similar to that of spectroscopic successes.
In other words, our failure to measure a redshift is generally not
attributable to the redshift being far from what one would expect from
the color selection criteria.

\begin{figure}[hbt]
\plotone{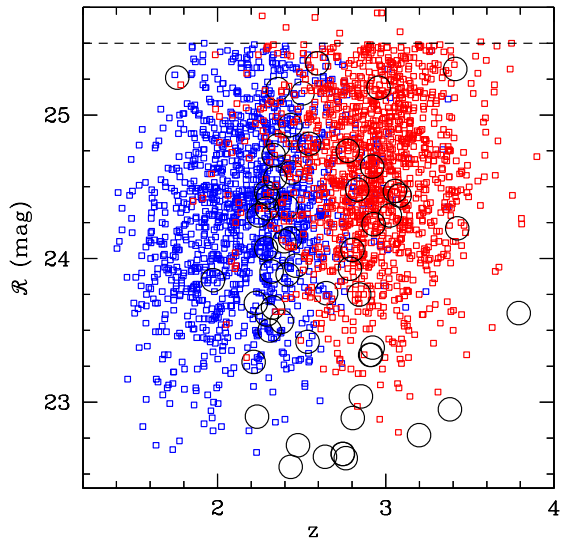}
\caption{Apparent magnitude versus redshift for spectroscopically
confirmed BX objects (in blue) and LBGs (in red) in the redshift range
$1.4<z<4.0$.  AGN/QSOs in the BX and LBG samples are denoted by the
large open circles.  The dashed horizontal line indicates the
self-imposed $\rs=25.5$ limit to the photometric (and hence
spectroscopic) samples.  The few objects shown with magnitudes fainter
than this limit are from the Q1422 field.
\label{fig:rvz}}
\end{figure}

Second, Figure~\ref{fig:rvz} demonstrates that optical apparent
magnitude is independent of redshift for the BX and LBG samples,
keeping in mind that the $\rs=25.5$ limit is applied to the
photometric (and hence also spectroscopic) sample.  This is important
because if the redshifts of objects were correlated with their optical
apparent magnitude, then we might expect the redshift distribution of
spectroscopically identified candidates to differ from candidates in
general given that our mask prioritization scheme gives more weight to
candidates with magnitudes in the range $23.5\la \rs \la 24.5$
(\S~\ref{sec:specfollow}).  Given these results, we proceed under the
assumption that spectroscopic selection does not significantly bias
the recovered redshift distribution relative to that of the underlying
photometric sample.

It is instructive to note that, with respect to the redshift
distribution, there are two forms of completeness we must concern
ourselves with.  The first is how well the redshift selection function
for the {\it spectroscopic} sample reflects the underlying selection
function for the {\it photometric} sample.  We have just argued that
the spectroscopic and photometric samples must have similar redshift
distributions.  The second is how well the {\it photometric} selection
function reflects the underlying redshift distribution of {\it all}
star-forming galaxies.  As discussed in \S~\ref{sec:photoz}, the
number density of galaxies is roughly constant as a function of
redshift in the redshift ranges $1.9\le z<2.7$ and $2.7\le z<3.4$.
The modulation of this intrinsic (roughly constant) redshift
distribution into the Gaussian distribution of
Figure~\ref{fig:zhistall} can be modeled with great precision by way
of Monte Carlo simulations, as we demonstrate below.  Readers who wish
to skip directly to the results may proceed to \S~\ref{sec:results1}.

\section{METHOD: INCOMPLETENESS CORRECTIONS}
\label{sec:ic}

A primary aim of this analysis is to connect the observed properties
of BX galaxies and LBGs to the underlying population of {\it all}
star-forming galaxies at redshifts $z\sim 2-3$.  To this end, we have
constructed a plausible population of galaxies with a range of
redshifts ($1.4\la z\la 4.0$), luminosities, and reddening, and
determined the fraction of these galaxies that would satisfy our color
criteria.  As is typically done, inverting these fractions and
applying them to the observed counts allows one to estimate the
underlying distribution of galaxies.  In this section we discuss in
detail the procedure used to reconstruct the intrinsic population of
galaxies at redshifts $1.9\la z\la 3.4$.

\break

\subsection{Monte Carlo Simulations}
\label{sec:mc}

We employed a Monte Carlo approach both to (1) determine the
transformation between the intrinsic properties of a galaxy (e.g., its
luminosity, reddening, and redshift) and its observed rest-UV colors
and (2) quantify the effects of photometric errors in their measured
rest-UV colors, similar to the method used in \citet{shapley01},
\citet{adel00}, and \citet{steidel99}.  Template galaxies with
intrinsic sizes of $0\farcs 05$ to $0\farcs 8$ and exponential light
profiles were convolved with the average PSF (typically $1''$) of the
optical images.  Variations in the light profile used (e.g.,
exponential, de Vaucouleurs) have a negligible effect on the
simulation results; the intrinsic size of the light emitting region
($\sim 0\farcs{5}-0\farcs{8}$ based on {\it HST} ACS observations;
\citealt{law07}) is almost always smaller than the seeing disk.

The expected rest-UV colors of a galaxy with a particular redshift and
reddening are computed by assuming a \citet{bruzual96} template galaxy
with constant star formation for 1~Gyr and a \citet{calzetti00}
extinction law.\footnote{Note that because our selection, and hence
simulations, are concerned with the rest-UV colors, adopting a
\citet{maraston06} model (where most of the difference with the
\citet{bruzual96} model is in the rest-optical) should minimally
affect our results.}  The BX selection criteria were designed to
select $z\sim 2$ galaxies with a range of SEDs similar to those found
for LBGs at higher redshifts \citep{adelberger04}.  Spectral synthesis
modeling and external multi-wavelength information indicates that most
UV-selected $z\sim 2-3$ galaxies can be described by long duration
($>100$~Myr) starbursts and the constant star formation model
described above should reproduce this behavior to the extent required
by the simulations (e.g., \citealt{shapley05}).  In particular, the
rest-UV colors of galaxies are essentially constant after $10^{8}$
years of star formation, once the mix of O and B stars
stabilizes\footnote{There is considerable leeway in the best-fit star
formation histories for the optical/IR SEDs of UV-selected $z\sim 2$
galaxies, but external constraints point to burst timescales of $>
100$~Myr \citep{shapley05, erb06b}.}.  The \citet{calzetti00}
reddening law reproduces the {\it average} expected star formation
rates of $z\sim 2-3$ galaxies based on extinction free stacked X-ray
and radio estimates (e.g., \citealt{reddy04}) and further reproduces
the average dust obscuration of galaxies with bolometric luminosities
in the range $10^{11}\la L_{\rm bol}\la 10^{12.2}$~L$_{\odot}$ where
the bulk of our sample lies \citep{reddy06a}.  The use of a constant
star forming model and the Calzetti reddening law should therefore
adequately parameterize the SEDs of most optically-bright star-forming
galaxies at $z\sim 2-3$.  An advantage of spectroscopic followup of
photometrically selected BX galaxies and LBGs is that we can also
constrain the effects of IGM opacity and Ly$\alpha$
absorption/emission (\S~\ref{sec:lyas}), both of which are redshift
dependent.  All of these perturbing effects will result in a wide
range of spectral shapes and should account for any galaxies that are
not exactly described by a \citet{calzetti00} attenuated constant star
forming SED.

A large distribution of galaxy colors was then computed assuming a
particular luminosity function (LF) and the observed $\ebmv$
distribution for spectroscopically confirmed galaxies.  Small
variations in the assumed Schechter parameters of the LF do little to
change the results, since our main goal is to sufficiently populate
redshift space and rest-UV color space with a realistic distribution of
objects.  The results are also insensitive to small variations in the
assumed $\ebmv$ distribution as long as the range of $\ebmv$ chosen
reflects that expected for the galaxies.  A by-product of the
luminosity function analysis is that we also compute the best-fit
underlying $\ebmv$ distribution.  The validity of the assumed LF and
$\ebmv$ distributions can be tested by comparing with the inferred LF
and $\ebmv$ distributions.  Significant differences between the
assumed and inferred distributions imply that the initial assumptions
for the LF and $\ebmv$ distribution were different from their true
values.  The colors were corrected for opacity due to the
intergalactic medium (IGM) assuming a \citet{madau95} model, and
corrected for filter and CCD responses and airmasses appropriate for
the each field of the survey.

The intrinsic rest-UV colors are randomly assigned to simulated
galaxies that are then added to the images in increments of 200
galaxies at a time.  This ensures that the image including all added
(simulated) galaxies has confusion statistics similar to the observed
image, since this will affect the photometric uncertainties and
systematics due to blending.  We then attempt to recover these
simulated galaxies using the same software used to recover the real
data, and record whether a simulated galaxy is detected and what its
observed magnitude and colors are.  We repeated this procedure until
approximately $2\times 10^{5}$ simulated galaxies were added to each
of the $U_{\rm n}$, $G$, and ${\cal R}$ images of each field.  This
large number of simulated galaxies is necessary in order to
sufficiently populate each bin of luminosity, reddening, and redshift.

The end product of the simulations are sets of transformations for
each field that give the probabilities that galaxies with intrinsic
luminosities ($L'$), reddenings ($E'$), and redshifts ($z'$) will be
observed to have luminosities $L$, reddenings $E$, and redshifts $z$
(or alternatively, the probabilities that galaxies with true
properties $L'E'z'$ will be measured with a particular set of rest-UV
colors).

\subsection{Photometric Uncertainties}
\label{sec:photerrsec}

We have used the results of the Monte Carlo simulations
(\S~\ref{sec:mc}) to estimate the photometric errors and determine
optimal bin sizes for subsequent analysis.  For each simulated galaxy
that is detected, we have recorded the true and measured rest-UV
colors.  As the uncertainties may vary depending on magnitude or
color, we have binned the detected galaxies in magnitude and color for
each field and have only considered galaxies that would be detected as
candidates since these are the only objects that are relevant to our
analysis.  We used bin sizes of 0.5~mag in $\rs$ and 0.2~mag in
$U_{\rm n}-G$ and $G-\rs$ color to determine the uncertainties in the
recovered magnitudes and colors of objects in each field.  Systematic
bias in the $G-\rs$ color was estimated by computing the quantity
$\Delta[G-\rs] = (G-\rs)_{\rm meas} - (G-\rs)_{\rm true}$ which was
typically $\la 0.04$~mag with uncertainty estimated to be
$\sigma(\Delta[G-\rs])\sim 0.09$~mag.  The typical random
uncertainties in $U_{\rm n}-G$ and $\rs$ are $\sim 0.15$~mag and $\sim
0.13$~mag, respectively.  These quantities were determined using the
same method as presented in \citet{shapley05}, \citet{steidel03}, and
\citet{shapley01}.  The uncertainties were generally larger for
objects faint in $\rs$ \citep{steidel03}.  The field-to-field results
were consistent with each other (i.e., the typical biases and
uncertainties from field-to-field were within $0.1$~mag of each
other).  The photometric errors are slightly smaller in size than the
bin sizes ($0.2$~mag) used to estimate the reddening and luminosity
distribution.  A more refined method discussed in \S~\ref{sec:tpf}
will correct for any systematic scattering of objects into adjacent
bins due to photometric error and/or Ly$\alpha$ perturbations to the
colors.

\subsection{Ly$\alpha$ Equivalent Width ($W_{\rm Ly\alpha}$) Distribution}
\label{sec:lyas}

\begin{figure}[bht]
\plotone{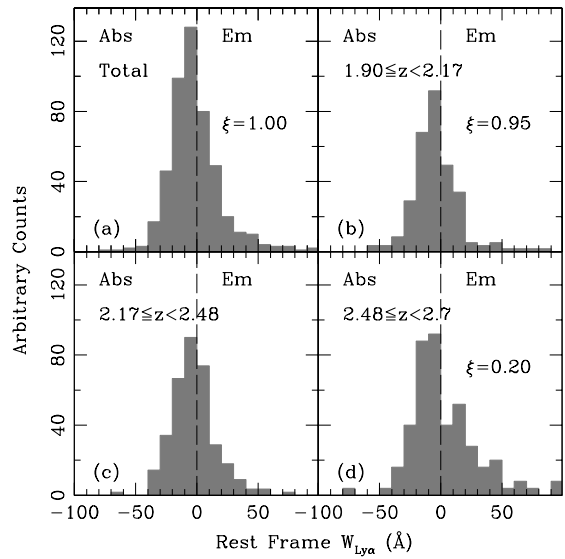}
\caption{(a) Rest frame Ly$\alpha$ equivalent width ($W_{\rm
Ly\alpha}$) distribution for $482$ spectroscopically observed $z\sim
2$ galaxies.  Panels (b)-(d) show the $W_{\rm Ly\alpha}$ distribution
for subsets in redshift.  We use the convention that $W_{\rm
Ly\alpha}>0$ implies emission.  The distributions are absorption
dominated in all cases.  The $\xi$ values indicate the probability
that the distributions are drawn from the same parent population as
the $W_{\rm Ly\alpha}$ distribution for galaxies at $2.17\le z <2.48$,
the redshift range where Ly$\alpha$ does not affect the rest-UV
colors.
\label{fig:lya}}
\end{figure}

\begin{figure}[hbt]
\plotone{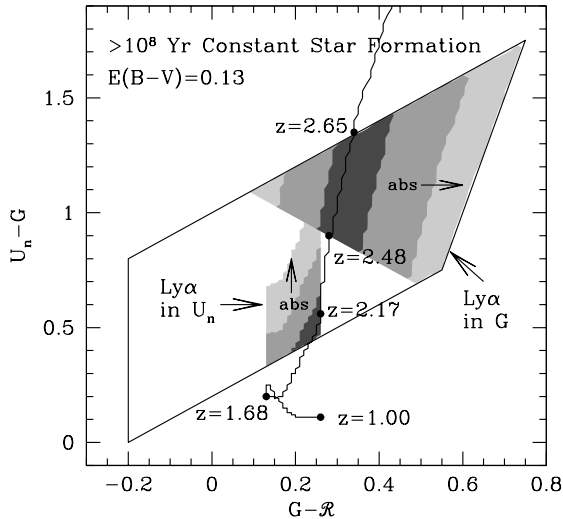}
\caption{Perturbation of $\ugr$ colors from Ly$\alpha$ absorption and
emission.  The trapezoid is the BX selection window defined by
Equation~\ref{eq:bx}.  The $\ugr$ colors of a template galaxy with
constant star formation for $>100$~Myr (after which the UV colors are
essentially constant) and $\ebmv = 0.13$ (the mean for the $z\sim 2$
sample) assuming a \citet{calzetti00} extinction law is shown by
the solid curve, proceeding from redshift $z=1$ to $3$.  The lower and
upper shaded regions correspond to redshift ranges where the
Ly$\alpha$ line falls in the $U_{\rm n}$ and $G$-bands, respectively.
In the absence of photometric errors and assuming all galaxies can be
described by the SED assumed here, galaxies with redshifts $1.68\la
z\la 2.17$ and $2.48\la z \la 2.93$ will fall in the dark gray regions
with a probability of $64\%$ based on the $W_{\rm Ly\alpha}$
distribution in Figure~\ref{fig:lya}.  The medium and light gray
regions correspond to scattering probabilities of $30\%$ and $6\%$,
respectively.  Arrows labeled ``abs'' indicate the direction in which
the colors will be perturbed with increasing Ly$\alpha$ absorption.
\label{fig:ewdist}}
\end{figure}

The presence of Lyman alpha absorption and/or emission can perturb the
observed rest-UV colors of $z\sim2-3$ galaxies by up to ~0.75 mag
depending on the redshift and intrinsic (rest-frame) Ly$\alpha$
equivalent width $W_{\rm Ly\alpha}$.  To investigate these effects, we
measured the $W_{\rm Ly\alpha}$ for 414 spectroscopically confirmed BX
galaxies with redshifts $1.9\le z<2.7$ in 7 different fields of the BX
survey. The resulting distribution for all galaxies with $1.9\le
z<2.7$ is shown in Figure~\ref{fig:lya}a, with the characteristic
asymmetric shape.  The photometric scattering probability associated
with this $W_{\rm Ly\alpha}$ distribution is shown in
Figure~\ref{fig:ewdist}.  The two shaded ``zones'' in
Figure~\ref{fig:ewdist} reflect the redshift ranges where Ly$\alpha$
falls within the $U_{\rm n}$ and $G$-bands.  This figure demonstrates
how galaxies that are targeted by the BX criteria can be shifted out
of the BX selection window due to Ly$\alpha$ emission or absorption.

Since our ultimate goal is to determine how the $W_{\rm Ly\alpha}$
distribution perturbs the intrinsic colors of galaxies (i.e., the
colors we would measure in the absence of absorption and/or emission),
we must first determine whether the measured $W_{\rm Ly\alpha}$
distribution reflects the intrinsic distribution for the parent
population of galaxies.  In other words, we must check if our color
selection criteria introduces significant biases into the measured
$W_{\rm Ly\alpha}$ distribution.  We can begin by examining some
characteristics of the measured $W_{\rm Ly\alpha}$ for BX galaxies and
LBGs, summarized in Table~\ref{tab:lyaew}.  The BX distribution has
$\langle W_{\rm Ly\alpha}\rangle \sim -1$~\AA, somewhat lower than the mean
$W_{\rm Ly\alpha}$ for LBGs.  While the measurements of $W_{\rm
Ly\alpha}$ for {\it individual} galaxies may be highly uncertain, the
difference in the {\it average} $W_{\rm Ly\alpha}$ suggests that the
high redshift (LBG) population has a higher incidence of Ly$\alpha$ in
emission than the low redshift (BX) population.  This disparity
between the lower and higher redshift populations can be better
appreciated by examining column (3) of Table~\ref{tab:lyaew} that
shows that the fraction of galaxies with $W_{\rm Ly\alpha}\ge 20$~\AA\,
($f20$) is almost twice as high among LBGs as it is for BX galaxies.

The change in $f20$ is even more apparent when we consider BX galaxies
in different redshift ranges: $f20$ for BX galaxies with redshifts
between $z\approx 2.5$ and $z=2.7$ is twice that of BX galaxies with
redshifts between $z=1.9$ and $z\approx 2.5$ (Table~\ref{tab:lyaew}).
These results suggest that star-forming galaxies exhibit a higher
incidence of Ly$\alpha$ emission at higher redshifts, but is this
trend intrinsic to high redshift galaxies, or is it introduced as a
result of color selection bias, as Figure~\ref{fig:ewdist} suggests?  

\begin{deluxetable}{lcc}
\tabletypesize{\footnotesize}
\tablewidth{0pc}
\tablecaption{Measured $W_{\rm Ly\alpha}$ Distributions}
\tablehead{
\colhead{Sample} &
\colhead{$\langle W_{\rm Ly\alpha}\rangle$\tablenotemark{a}} &
\colhead{$f(W_{\rm Ly\alpha})\ge 20$~\AA\tablenotemark{b}}}
\startdata
BX (ALL: $1.90\le z<2.70$) & $-1$~\AA & 0.12 \\
BX ($1.90\le z<2.17$) & $-1$~\AA & 0.08 \\
BX ($2.17\le z<2.48$) & $-2$~\AA & 0.11 \\
BX ($2.48\le z<2.70$) & $2$~\AA & 0.20 \\
LBG ($2.70\le z<3.4$) & $9$~\AA & 0.23 \\
\enddata
\tablenotetext{a}{Mean rest-frame $W_{\rm Ly\alpha}$.}
\tablenotetext{b}{Fraction with $W_{\rm Ly\alpha} \ge 20$.}
\label{tab:lyaew}
\end{deluxetable}

We can test for systematics induced by the color criteria by examining
$f20$ for BX galaxies at redshifts $2.17\le z<2.48$, where Ly$\alpha$
lies outside the $U_{\rm}$ and $G$-bands.  These galaxies have a
similar $f20$ to that of $z<2.17$ galaxies (Table~\ref{tab:lyaew}),
implying that the fraction of absorption versus emission line systems
culled by the BX criteria is similar between the $z<2.17$ and the
$2.17\le z<2.48$ samples (this assumes that there is little evolution
in the $W_{\rm Ly\alpha}$ distribution between the $z<2.17$ and
$2.17\le z<2.48$ subsamples).  Focusing on the high redshift subsample
with $2.48\le z<2.70$, Figure~\ref{fig:ewdist} suggests that these
galaxies are more likely to satisfy the BX criteria if they have
Ly$\alpha$ in absorption, yet their $f20$ is similar to that of LBGs.
In other words, the $2.48\le z<2.70$ subsample has an $f20$ value that
does not indicate a preferential selection of absorption over emission
line galaxies relative to that of the lower redshift subsamples.
Rather, the $f20$ value is {\it larger} than those for the lower
redshift subsamples and is similar to that of the LBGs.  These
conclusions are supported by a Kolmogorov-Smirnov test, the results of
which are summarized in Figure~\ref{fig:lya}.  Namely, $\xi$ in the
figure indicates the probability that the $W_{\rm Ly\alpha}$
distributions for the total sample, the sample with $1.90\le z<2.17$,
and the sample with $2.48\le z<2.70$, are drawn from the same parent
population as the sample with $2.17\le z<2.48$, where Ly$\alpha$ does
not effect the $\ugr$ colors.  Galaxies with $2.48\le z<2.70$ have a
$W_{\rm Ly\alpha}$ distribution that deviates significantly from the
one at $2.17\le z<2.48$.

Assuming that the UV properties of galaxies are independent of their
Ly$\alpha$ line profiles would then suggest that the BX color criteria
do not significantly modulate the intrinsic $W_{\rm Ly\alpha}$
distribution of $z\sim 2$ galaxies.\footnote{\citet{shapley03} have
demonstrated that $W_{\rm Ly\alpha}$ is in fact dependent upon the
rest-frame UV colors and magnitudes of galaxies.  However, the small
biases that these trends may have on the observed $W_{\rm Ly\alpha}$
do not have a significant impact on the derived LFs at $z\sim 2$ and
$z\sim 3$.}  For the purposes of our simulations, we make the
approximation that the observed $W_{\rm Ly\alpha}$
distribution for BX galaxies can be applied to our simulated galaxies
to obtain the average perturbation of their rest-UV colors.

Since the Ly$\alpha$ line falls in the $G$-band for galaxies in the
entire redshift range $2.7\le z<3.4$, we cannot examine trends in the
$W_{\rm Ly\alpha}$ distribution for the LBGs in the same way we did
for the BXs.  However, in \S~\ref{sec:lyadist} we justify why the $W_{\rm
Ly\alpha}$ distribution for LBGs should approximately reflect the
intrinsic $W_{\rm Ly\alpha}$ distribution for $z\sim 3$ galaxies.
Also in \S~\ref{sec:lyadist}, the incompleteness corrections are
used to test whether our initial assumptions of the $W_{\rm Ly\alpha}$
distributions are correct.

\subsection{Quantifying Incompleteness}
\label{sec:cc}

\subsubsection{Effective Volume ($V_{\rm eff}$) Method}
\label{sec:veff}

The fraction of galaxies with a given set of binned properties that
satisfy the color criteria can be computed directly from the results of
the Monte Carlo simulations.  These binned properties might be the
optical luminosity ($L$), redshift ($z$), and reddening
($\ebmv$) of a galaxy.  Under the assumption that these properties are
independent of each other, and if we let the indices $i$, $j$, and $k$
run over the range of values of $L$, $z$, and $\ebmv$, then the
true number of galaxies in the $ijk$th bin can be approximated as
\begin{eqnarray}
n_{ijk}^{true} \simeq n_{ijk}^{obs} / \bar{p}_{ijk}
\label{eq:truenum}
\end{eqnarray}
where $\bar{p}_{ijk}$ are the mean probabilities that a galaxy in the
$ijk$th bin is (a) detected and (b) satisfies the color criteria
(e.g., \citealt{adel02t}).  These probabilities $\bar{p}_{ijk}$ are simply
\begin{eqnarray}
\bar{p}_{ijk} = {1\over n_{ijk}}\sum^{n} p_{ijkn}
\label{eq:sumprob}
\end{eqnarray}
where $p_{ijkn}$ is the probability that the $n$th simulated galaxy in
the $ijk$th bin will be detected as a candidate, and $n_{ijk}$ is the
total number of simulated galaxies in the $ijk$th bin.  The values
$p_{ijkn}$ take into account the probability that the colors of the
$n$th simulated galaxy will be perturbed by the $W_{\rm Ly\alpha}$
distribution of Figure~\ref{fig:lya} and still be selected as a BX
object.  They also fold in the probability that a non-candidate
simulated galaxy will fall in the BX selection window.  Dividing by
$n_{ijk}$ normalizes the mean probabilities $\bar{p}_{ijk}$ and
accounts for both the fraction of galaxies whose photometric errors
scatter them out of the BX selection window and galaxies that are not
detected in the simulations.  If the true comoving volume
corresponding to the $j$th bin in redshift is $V_j$, then the
effective volume associated with the $j$th bin in $z$ is
\begin{eqnarray}
V^{\rm eff}_j \equiv V_j \sum^{ik}\bar{p}_{ijk} = V_j \times \xi_j,
\end{eqnarray}
where $\xi_j$ are commonly referred to as ``completeness coefficients'':
\begin{eqnarray}
\xi_j \equiv \sum^{ik}\bar{p}_{ijk}.
\end{eqnarray}
The photometric properties of each field are unique due to differences
in the observing conditions, and this will affect the computed
$\xi_j$.  We can then determine the completeness coefficients for each
field and then perform a weighted-average of them (i.e., weighted
according to the field size) to obtain mean completeness coefficients,
$\bar{\xi}_j$.

\subsubsection{Maximum Likelihood Method ($V_{\rm lik}$)}
\label{sec:tpf}

While the procedure just described can be used to make an initial
guess as to the shape of the reddening and luminosity distributions,
it can lead to spurious results, particularly for objects whose true
colors are such that they lie outside of or close to the edges of the
BX selection window.  Equation~\ref{eq:truenum} is approximately true
only if the average measured properties of a galaxy are the same as
the true (simulated) properties, and this will certainly not be the
case for galaxies that are preferentially scattered into the BX window
due to photometric errors or the presence of Ly$\alpha$
absorption/emission (e.g., \citealt{adel02t}).  The approach described
above will also not take into account photometric bias and the
preferential scattering of objects from one bin to another if the bin
sizes are comparable to (or smaller than) the photometric errors
\citep{adel02t}.  

\begin{figure}[hbt]
\plotone{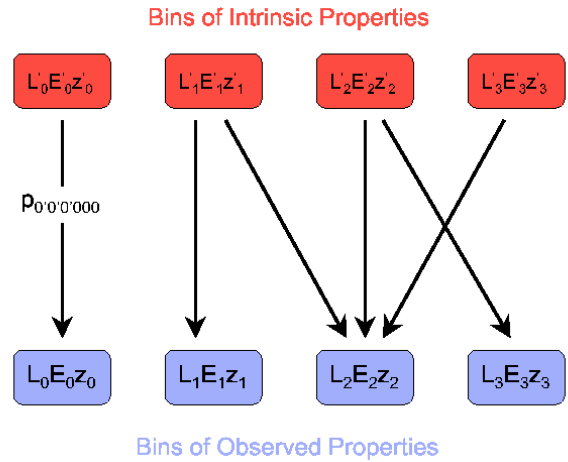}
\caption{Cartoon illustration of how the probability that a galaxy with
intrinsic (true) properties $L'E'z'$ may not have a one-to-one correspondence
with bins of observed (measured) properties $LEz$ (or measured colors $ugr$).
\label{fig:transprob}}
\end{figure}

Figure~\ref{fig:transprob} further illustrates these issues.  In the
simplest case, galaxies that fall within a particular bin of true
properties (say, $L_{\rm 0}'E_{\rm 0}'z_{\rm 0}'$) will, on average,
have measured properties corresponding to bin $L_{\rm 0}E_{\rm
0}z_{\rm 0}$.  In this case, we can use the approach of
\S~\ref{sec:veff} (i.e., the $V_{\rm eff}$ method) to simply divide
the observed number of galaxies in bin $L_{\rm 0}E_{\rm 0}z_{\rm 0}$
by the probability that galaxies in bin $L_{\rm 0}'E_{\rm 0}'z_{\rm
0}'$ will be observed in bin $L_{\rm 0}E_{\rm 0}z_{\rm 0}$ (call that
probability $p_{0'0'0' \rightarrow 000}$), as shown by the leftmost
arrow in Figure~\ref{fig:transprob}.  However, we can point to several
examples that suggest that there may not be a one-to-one
correspondence between bins of intrinsic and observed properties, as
illustrated by the remaining arrows in Figure~\ref{fig:transprob}.
First, in order to accurately compute the luminosity and reddening
distributions at $z\sim 2$, we cannot make our bin sizes much larger
than the typical photometric errors since the observed range of $\umg$
and $\gmr$ colors for galaxies at a single redshift ($\la 0.8$~mag) is
only $\la 4$ times the typical photometric error in $\umg$ and $\gmr$.
Therefore, galaxies that ought to fall within a particular bin of
measured properties will be scattered into adjacent bins.  This would
not be a problem if each bin of measured properties gained and lost an
equal number of galaxies, but since the luminosity and reddening
distributions are peaked, photometric errors will scatter galaxies
away from the peak and into the wings of the distributions.  Second,
the distribution of errors in colors is not symmetric with respect to
the true values such that there is a systematic tendency to scatter
galaxies into redder bins more often than bluer ones
\citep{steidel03}.  Third, the presence of Ly$\alpha$ absorption in a
galaxy's spectrum will, depending on the redshift, cause us to
overestimate the reddening.  Finally, there will be some galaxies
whose true properties are such that on average they lie outside the
selection windows, and only get scattered into the sample because of
photometric errors, such as might be the case for galaxies lying close
to the color selection boundaries.

Because of these systematic effects, the number of galaxies within a
particular bin of measured properties will be some weighted
combination of the numbers of galaxies within intrinsic bins that
contribute to that measured bin of properties.  The weights are simply
the ``transitional'' or scattering probabilities indicated by the
arrows in Figure~\ref{fig:transprob}, and these can be determined from
the Monte Carlo simulations.  Formally, the number of galaxies we
expect to observed in a bin of measured luminosity $L$, reddening $E$,
and redshift $z$ within a field of size $\Delta \Omega$ is
\begin{eqnarray}
\bar{n}(L,E,z){dV \over d\Omega dz} = \nonumber \\
\mu \int dL'dE'dz'f(L')g(E')h(z')p_{\rm L'E'z' \rightarrow LEz} {dV \over d\Omega dz},
\label{eq:fulleq}
\end{eqnarray}
where $\mu$ is related to the total comoving number density of
galaxies; $p_{\rm L'E'z' \rightarrow LEz}$ is the (transitional)
probability that galaxies with intrinsic $L'E'z'$ will be measured
to have $LEz$; and f(L'), g(E'), and h(z') are the intrinsic
distributions of luminosity, reddening, and redshift, respectively,
normalized such that
\begin{eqnarray}
\int_{L_{\rm min}}^{L_{\rm max}}dLf(L)=\int_{E_{\rm min}}^{E_{\rm
max}}dEg(E) = \nonumber \\ \int_{z_{\rm min}}^{z_{\rm max}}dzh(z) = 1
\end{eqnarray}
\citep{adel02t}.  Our goal is to determine the intrinsic distributions
$f(L)$, $g(E)$, and $h(z)$, but inverting Eq.~\ref{eq:fulleq} to solve
for these distributions is intractable.  One alternative is to compute
the likelihood (${\cal L}$) of observing our data, which is expressed
as a list of galaxies with observed $L_{i}E_{i}z_{i}$, for a given set
of $fgh$ distributions:
\begin{eqnarray}
{\cal{L}}({L_{i}E_{i}z_{i}})\propto \exp \left[-\mu\Delta\Omega\int dLdEdz \bar{n}(L,E,z){dV \over d\Omega dz} \right] \nonumber \\
\prod_i \bar{n}(L_{i}E_{i}z_{i}).
\label{eq:contlik}
\end{eqnarray}
The discrete form of Eq.~\ref{eq:contlik}, extended to incorporate each of $l$ different
fields, can be expressed as
\begin{eqnarray}
{\cal{L}}({n_{ijkl}})\propto \exp \left[-\sum_{ijkl} \bar{n}_{ijkl}\right] \prod_{ijkl}\bar{n}_{ijkl}^{n_{ijkl}},
\label{eq:dislik}
\end{eqnarray}
where $\bar{n}_{ijkl}$ is the mean number of galaxies in the $i^{\rm
th}$ bin of luminosity, $j^{\rm th}$ bin of reddening, and $k^{\rm
th}$ bin of redshift in the $l^{\rm th}$ field that the assumed values of $f_i$, $g_j$, and
$h_k$ imply; and $n_{ijkl}$ is the observed number of galaxies in the
same bin (e.g., \citealt{adel02t}).  The discrete version of Eq.~\ref{eq:fulleq} is
\begin{eqnarray}
\bar{n}_{ijkl} = \mu \Delta\Omega_{l}\sum_{i'j'k'}f_{i'}g_{j'}h_{k'}V_{k'}p_{l,i',j',k'\rightarrow ijk},
\end{eqnarray}
where $\Delta\Omega_{l}$ is the size of the $l^{\rm th}$ field,
$V_{k'}$ is the comoving volume in Mpc$^3$~arcmin$^{-2}$ corresponding
to bin $k'$ in redshift, and $p_{l,i',j',k'\rightarrow ijk}$ is the
probability that a galaxy in the $l^{th}$ field in the $i'j'k'$ bin of
luminosity, reddening, and redshift, will have measured properties
corresponding to bin $ijk$.  Assuming that the data quality does not vary
significantly from field-to-field, we can simplify the probabilities such that
\begin{eqnarray}
\bar{p}_{i'j'k'\rightarrow ijk} \equiv \sum_{l}\Delta\Omega_{l}p_{li'j'k'\rightarrow ijk}/
\sum_{l}\Delta\Omega_{l}.
\end{eqnarray}
Maximizing the likelihood as expressed in Eq.~\ref{eq:dislik} is
equivalent to minimizing
\begin{eqnarray}
-\ln{\cal{L}} \propto \sum_{ijk} \bar{n}_{ijk} - \sum_{ijk} n_{ijk}\ln \bar{n}_{ijk},
\label{eq:minlik}
\end{eqnarray}
and is more amenable to computation than Eq.~\ref{eq:dislik}.

\subsubsection{Implementation of the Maximum-Likelihood Method}
\label{sec:implementation}

We first used the Monte Carlo simulations to determine the
transitional probabilities that relate the true luminosities,
reddenings, and redshifts of galaxies to their observed rest-UV colors.
Following the discussion of \S~\ref{sec:lyas}, the colors of galaxies
were perturbed by randomly assigning a $W_{\rm Ly\alpha}$ according to
the distributions shown in Figure~\ref{fig:lya} for the BX sample and
the distribution shown in Figure~8 of \citet{shapley03} for LBGs (see
also Figure~\ref{fig:lyacompall}).  We took advantage of both the
$U_{\rm n}-G$ and $G-{\cal R}$ colors in our analysis of the $z\sim 2$
sample to provide more stringent constraints on the $\ebmv$
distribution, something that was not possible at $z\sim 3$ where most
galaxies only had limits in $U_{\rm n}$ either due to severe
blanketing by the Ly$\alpha$ forest or the suppression of continuum
flux shortward of the Lyman limit.  

\begin{figure}[bth]
\plotone{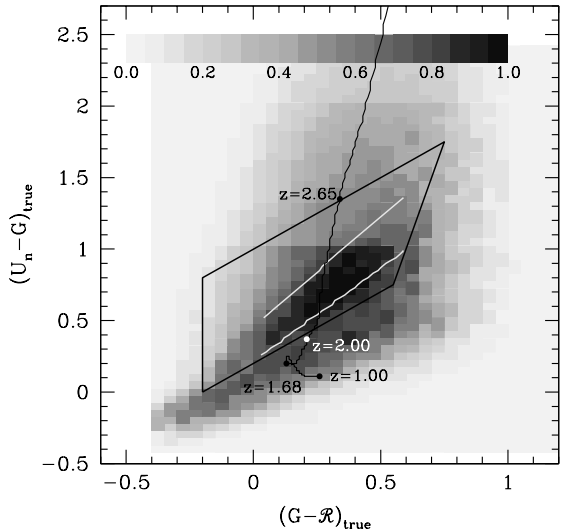}
\caption{Relative probability distribution for galaxies with intrinsic
colors $(U_{\rm n}-G)_{\rm true}$ and $(G-{\cal R})_{\rm true}$ to be
detected and selected as BX objects (solid line same as in
Figure~\ref{fig:ewdist}).  The distribution is weighted according to
the incidence of galaxies with a particular set of intrinsic colors as
determined from the LF and $\ebmv$ distributions used to compute the
transformation between intrinsic and observed colors.  The
distribution is non-zero exterior to the BX window (trapezoid) as a
result of photometric error and Ly$\alpha$ line perturbations of the
colors.  Galaxies with expected (or intrinsic) $U_{\rm n}-G$ colors
bluer than required to satisfy BX criteria are particularly prone to
selection as discussed in \S~\ref{sec:lyas}.  The region between the
white curves denotes the swath of color space where galaxies with
redshifts $2.17< z \le 2.48$ are expected to lie.  These galaxies'
colors are unaffected by Ly$\alpha$ line perturbations.
\label{fig:simdist}}
\end{figure}

Figure~\ref{fig:simdist} is useful in visualizing the transitional
probabilities, in this case for the BX sample, where we show the
relative probability distribution for galaxies between $1.0<z<3.0$ to
be selected by the BX criteria.  The probability distribution is
weighted by the incidence of galaxies with intrinsic colors as
determined from the LF and $\ebmv$ distributions assumed in computing
the transitional probabilities.  This distribution reflects both
photometric error and Ly$\alpha$ perturbation of the expected rest-UV
colors.  One noticeable feature of Figure~\ref{fig:simdist} is the
divergent behavior of the selection function for low ($z\la 2.0$) and
high ($z\ga 2.7$) redshift galaxies, where higher redshift galaxies
have redder $U_{\rm n}-G$ colors for a given SED.  This can be
understood, in part, by examining Figure~\ref{fig:ewdist}.  If
$z\sim2$ galaxies can be reasonably described by the SED and reddening
assumed above then we would expect that galaxies with $z>2.7$ would
only be scattered into the BX window if there were large changes in
their colors, either due to photometric errors or Ly$\alpha$
perturbation.  First, we find no evidence that photometric errors
increase for galaxies at higher redshifts.  Second, the $(1+z)$
dependence of the observed $W_{\rm Ly\alpha}$ will result in a larger
color change (for a fixed rest-frame $W_{\rm Ly\alpha}$) for higher
redshift galaxies than for lower redshift galaxies, such that the
scattering probability distribution covers a larger area in color
space, making it less likely for a particular source to fall within
the BX selection window.  Finally, the $U_{\rm n}-G$ color changes
more rapidly for higher redshift galaxies where Ly$\alpha$ forest
absorption begins to increasingly affect the $U_{\rm n}$-band.  All of
these effects could explain the relatively small number of $z>2.7$
galaxies singled out with the BX criteria.  The advantage of rest-UV
selection is that the drop off in BX efficiency for $z>2.7$ can be
compensated for by adopting the $z\sim 3$ LBG criteria whose selection
function begins to rise for $z>2.7$ and which use exactly the same
filter set, negating the need for additional observations
\citep{steidel03}.

Unlike the $z>2.7$ galaxies discussed above, $1.0\la z\la 2.0$
galaxies are crowded into a narrower region of color space as is
evident from Figure~\ref{fig:ewdist}.  Small variations in colors as a
result of photometric errors or Ly$\alpha$ absorption can shift a
large numbers of such galaxies into the BX selection window.  This
effect can be viewed in Figure~\ref{fig:simdist}, where there is a
high relative probability for galaxies with blue $U_{\rm n}-G$ colors
(the ``BM'' galaxies; e.g., Figure~10 of \citealt{adelberger04}) to
satisfy BX selection, partly due to the effect of Ly$\alpha$
absorption in these systems (cf., Figure~\ref{fig:lya}b).  The highest
density region in this figure (between the two white curves of
Figure~\ref{fig:simdist}) occurs in the same color space expected to
be occupied by galaxies at redshifts where the Ly$\alpha$ line does
not affect the $\ugr$ colors ($2.17< z\le 2.48$).
Figure~\ref{fig:simdist} also demonstrates the fallacy of the
assumption in Equation~\ref{eq:truenum}, where the true and observed
rest-UV colors may be significantly and, more importantly,
systematically different for galaxies lying in particular regions of
color space.  Figure~\ref{fig:simdist} is meant to be purely
illustrative and, in reality (as in our simulations), the probability
distribution will be ``smeared'' out when one considers galaxies with
a range of spectral shapes.

The effects of IGM opacity, Ly$\alpha$ absorption/emission, and
photometric error (\S~\ref{sec:mc},\ref{sec:lyas},
\ref{sec:photerrsec}) imply that simple boxcar approximations to the
redshift selection function (even in photometric surveys) are unrealistic,
irrespective of the wavelengths used to select galaxies.  The
advantage of our combined Monte Carlo, photometric, and spectroscopic
approach is that even complicated selection functions can be
quantified relatively easily and thus be corrected for in the final
analysis.

\begin{figure}[hbt]
\plottwo{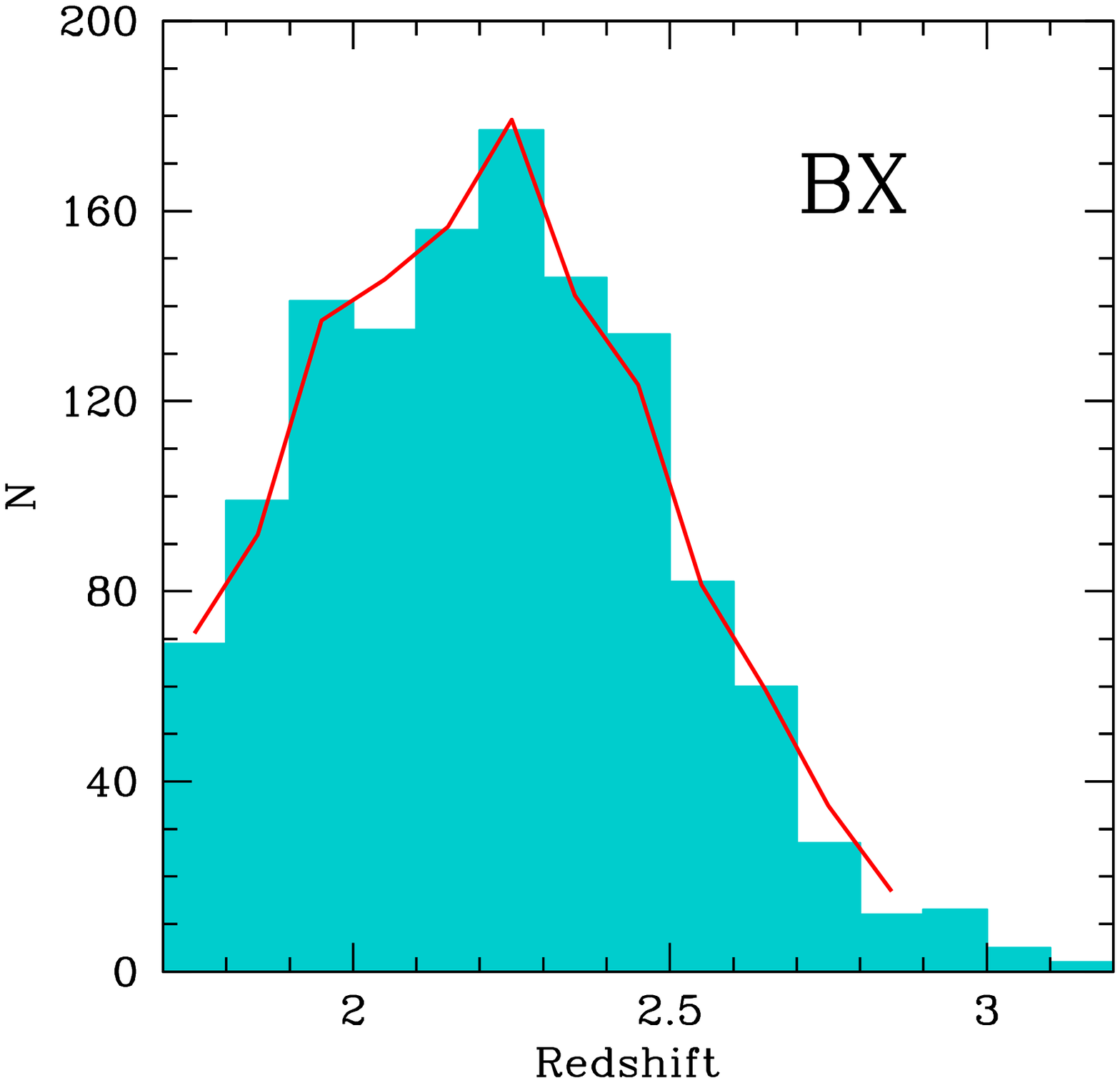}{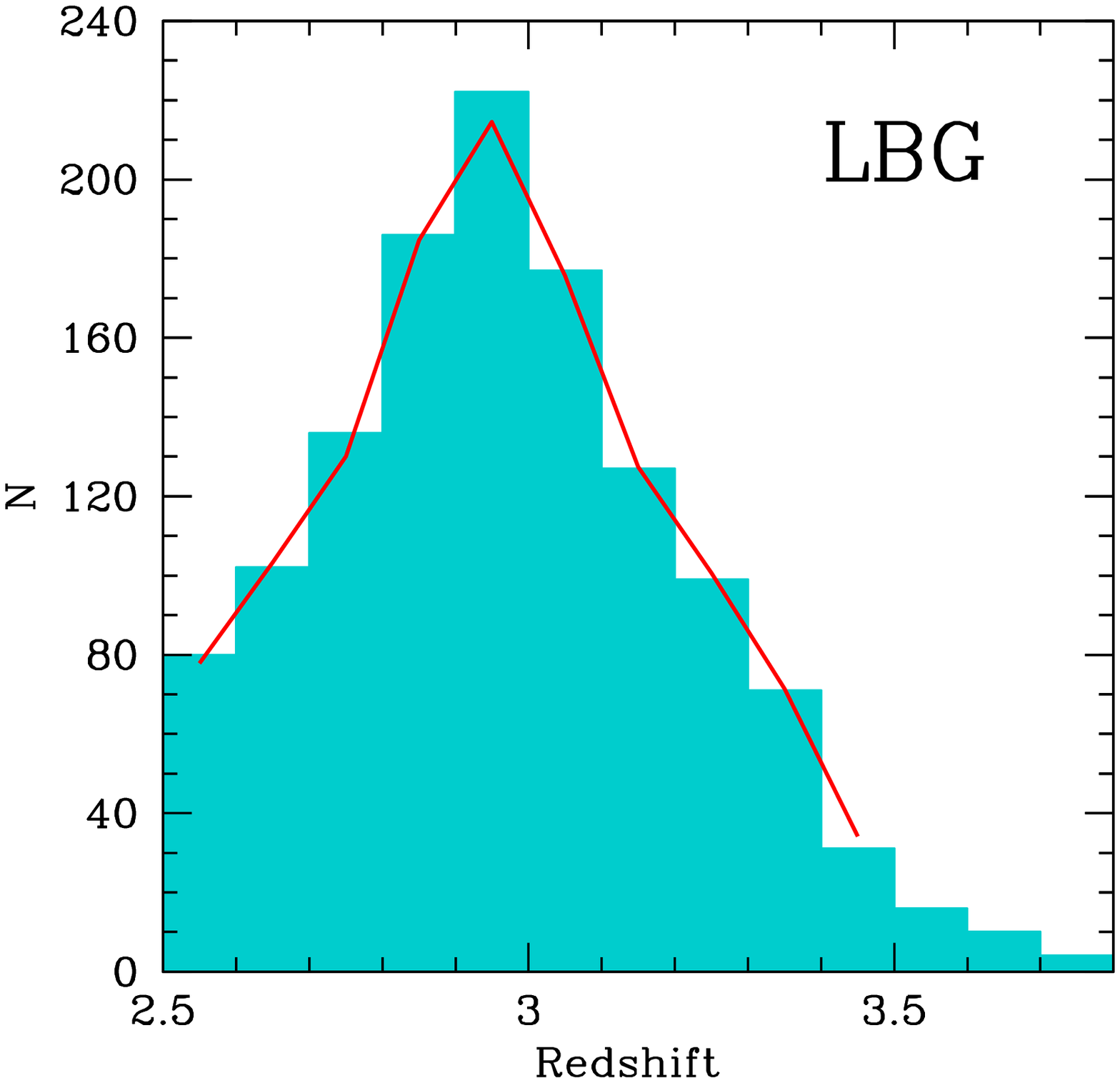}
\caption{Expected redshift distributions (lines) given our
best-fit reddening and luminosity distributions, compared with the
observed redshift distributions of BX galaxies (left panel) and
LBGs (right panel), indicated by the shaded histograms.
\label{fig:bxlbgexp}}
\end{figure}

\begin{figure*}[hbt]
\plotone{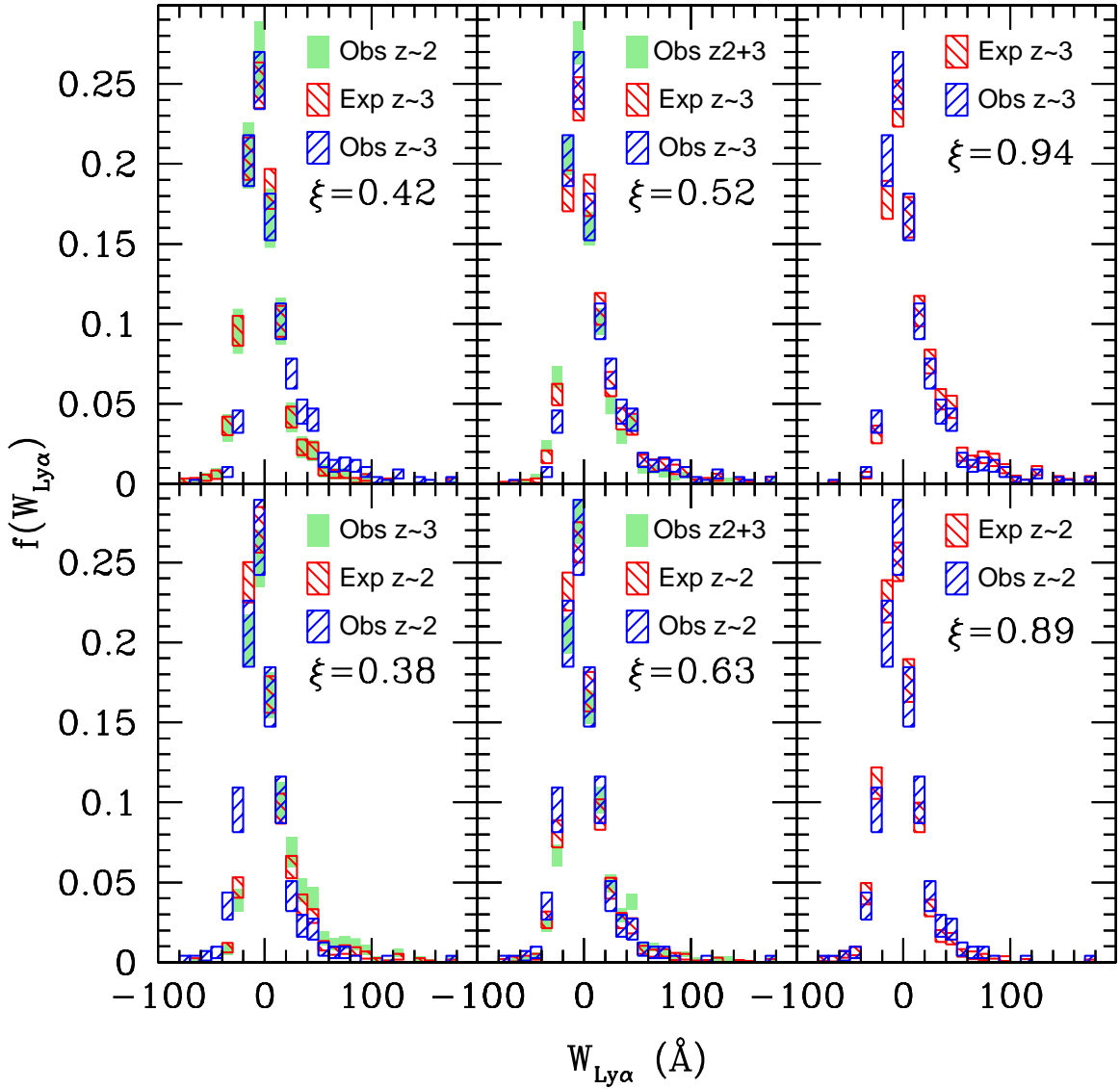}
\caption{Comparisons between expected and observed $W_{\rm Ly\alpha}$
distributions for different assumptions of the intrinsic $W_{\rm
Ly\alpha}$ distributions between redshifts $2.7\le z<3.4$ (top panels)
and $1.9\le z<2.7$ (bottom panels).  The assumed intrinsic
distributions are denoted by green rectangles in the left and middle
panels.  The assumed intrinsic distributions are the same as the
observed distributions (blue rectangles) in the right panels.  The
expected and observed distributions are indicated in red and blue,
respectively, in all six panels.  The length of the bars represent the
dispersion in values of the $W_{\rm Ly\alpha}$ distribution derived
assuming many realizations of the LF and $\ebmv$ distribution (see
text).  The parameter $\xi$ denotes the likelihood that the observed
and expected distributions are drawn from the same distribution.  We
use the convention that $W_{\rm Ly\alpha} > 0$ implies Ly$\alpha$
emission.
\label{fig:lyacompall}}
\end{figure*}

In practice, minimizing Eq.~\ref{eq:minlik} while simultaneously
varying the intrinsic distributions of luminosity, reddening, and
redshift ($fgh$) can lead to spurious results given the large
parameter space and possibility of numerous local minima in likelihood
space.  A reasonable approach is to then make some simplifying
assumptions, such as fixing the redshift distribution to be constant
and assuming an LF computed using the method of \S~\ref{sec:veff}.
One can then minimize Eq.~\ref{eq:minlik} with respect to the
distribution of spectral shapes ($g$) as parameterized by the $\ebmv$
color excess, using values of $p_{i'j'k'\rightarrow ijk}$ relevant for
the spectroscopic sample.  In other words, $p_{i'j'k'\rightarrow ijk}$
will give the probability that a galaxy with true properties in the
$i'j'k'^{\rm th}$ bin will be measured with properties in the
$ijk^{\rm th}$ bin and be spectroscopically observed.  The probability
that a candidate lying within a particular bin of $\rs$ magnitude will
be spectroscopically observed is approximated using the spectroscopic
fractions listed in Table~\ref{tab:specfrac}.  These spectroscopic
fractions are then multiplied by the probability that an object is a
star-forming galaxy (i.e., not an AGN/QSO) using the AGN/QSO fractions
in the relevant magnitude range (Table~\ref{tab:specfrac}).  At this
stage, we must rely on the spectroscopic sample since we can only
estimate $\ebmv$ for galaxies with redshifts.  Then, keeping $g$ fixed
to the best-fit $\ebmv$ distribution, we take advantage of the full
photometric sample to minimize Eq.~\ref{eq:minlik} with respect to the
luminosity distribution ($f$).  The revised estimate of $f$ can then
be held fixed to refine our estimate of $g$.  The process goes through
several iterations where at the last stage we vary $f$, $g$, and $h$
simultaneously.  The results of this procedure indicate that our
initial assumption of a constant redshift distribution (i.e., number
of galaxies in each of the redshift bins is roughly constant) is a
reasonable one to make.  The redshift distributions predicted for BX
galaxies and LBGs given the maximum-likelihood $fgh$ distributions are
excellent matches to the observed redshift distributions of BX
galaxies and LBGs, as shown in Figure~\ref{fig:bxlbgexp}.

Uncertainties in the luminosity and $\ebmv$ distribution were
estimated by generating many fake realizations of our observed data
from the catalogs of simulated galaxies, and recomputing the best-fit
$fgh$.  The dispersion in measurements of $fgh$ are taken to be the
$1$~$\sigma$ errors.  It is important to note that the errors in our
estimates are due to a combination of Poisson noise and field-to-field
dispersion.  Unlike all other previous estimates of the $z\sim 2-3$ UV
LF, our determination incorporates the largest spectroscopic sample of
galaxies at these redshifts and automatically takes into account the
{\it systematic} effects mentioned in \S~\ref{sec:tpf}.

\section{RESULTS: INTRINSIC $W_{\rm Ly\alpha}$ AND $\ebmv$ DISTRIBUTIONS}
\label{sec:results1}

\subsection{Validity of Assumed $W_{\rm Ly\alpha}$ Distributions}
\label{sec:lyadist}

An important question is whether the distribution of Ly$\alpha$
emission and absorption profiles of galaxies changes as a function of
redshift.  Such trends with redshift may indicate fundamental
differences in the ISM of galaxies and/or changing large-scale
environments as a function of redshift.  Can we do better job of
determining whether the intrinsic $W_{\rm Ly\alpha}$ distribution
of galaxies changes as a function of redshift?  Ideally, we would have
liked to include the $W_{\rm Ly\alpha}$ distribution as another free
parameter in the maximum-likelihood method discussed in
\S~\ref{sec:tpf},\ref{sec:implementation}, so that instead of
maximizing three functions ($fgh$), we would be maximizing four.
However, this would needlessly complicate our ability to determine the
maximum-likelihood $fgh$ distributions, especially since the
luminosity distribution ($f$) is insensitive to small changes in the
$W_{\rm Ly\alpha}$ distribution.  As a compromise, we {\it can}
investigate how different assumptions of the intrinsic $W_{\rm
Ly\alpha}$ distributions of galaxies affect the distributions that we
expect to measure.

Figure~\ref{fig:ewdist} illustrates how the color selection criteria
can modulate the observed $W_{\rm Ly\alpha}$ distribution of
galaxies, such that the observed distribution may be different than
the intrinsic distribution.  The Monte Carlo simulations discussed in
\S~\ref{sec:ic} allow us to directly compare the measured (observed)
$W_{\rm Ly\alpha}$ distributions for the BX and LBG samples with those
expected based on the transitional probabilities.  The results of this
comparison are summarized in Figure~\ref{fig:lyacompall}, which shows
$W_{\rm Ly\alpha}$ for various assumptions of the input $W_{\rm
Ly\alpha}$ distribution.  We consider three cases.  In the first case,
we assume that the intrinsic $W_{\rm Ly\alpha}$ distribution of
galaxies at $z\sim 3$ is identical to that measured at $z\sim 2$.  In
the second case, we assume that the intrinsic $W_{\rm Ly\alpha}$
distribution at $z\sim 3$ is an equally weighted combinations of the
measured distributions at $z\sim 2$ and $z\sim 3$.  In the third case,
we assume that the intrinsic $W_{\rm Ly\alpha}$ distribution at $z\sim
3$ is identical to the measured distribution at $z\sim 3$.  Three
analogous cases are considered for the $z\sim 2$ sample.

For example, the top left panel of Figure~\ref{fig:lyacompall} shows
the $W_{\rm Ly\alpha}$ distribution for $2.7\le z<3.4$ galaxies that
one would expect (red rectangles) if the {\it intrinsic} distribution
at $2.7\le z<3.4$ is identical to the measured $W_{\rm Ly\alpha}$
distribution for lower redshift ($1.9\le z<2.7$) galaxies (green
rectangles, labeled ``Obs $z\sim 2$'').  The validity of assuming a
particular {\it intrinsic} distribution can be tested by comparing the
expected distribution (red rectangles) with the actual measured
distribution (blue rectangles).  The vertical sizes of the rectangles
for the observed distributions reflect Poisson errors.  Uncertainties
in the expected distributions (red rectangles) are computed by
constructing many samples of galaxies drawn randomly from the
maximum-likelihood luminosity and $\ebmv$ distributions
(\S~\ref{sec:ebmv},\S~\ref{sec:uvlf}), and fixing the vertical bar
size to the dispersion in the $W_{\rm Ly\alpha}$ distributions
measured for each of these simulated samples.

The top left panel of Figure~\ref{fig:lyacompall} shows that assuming
an intrinsic distribution of $W_{\rm Ly\alpha}$ for galaxies at
redshifts $2.7\le z<3.4$ that is identical to the measured
distribution of $W_{\rm Ly\alpha}$ for BX-selected ($1.9\le z<2.7$)
galaxies results in an expected distribution at $2.7\le z<3.4$ that
deviates significantly from the distribution that we actually {\it
measured}.  In this case, the expected distribution exhibits a larger
fraction of galaxies with absorption and a deficit of emission-line
galaxies when compared with the measured $W_{\rm Ly\alpha}$
distribution.  Therefore, the intrinsic $W_{\rm Ly\alpha}$ for $2.7\le
z<3.4$ galaxies must have lower and higher fractions, respectively, of
absorption and emission-line galaxies than what is observed among
lower redshift ($1.9\le z<2.7$) galaxies.  The bottom left panel of
Figure~\ref{fig:lyacompall} tells a similar story.  Assuming $1.9\le
z<2.7$ galaxies have an intrinsic $W_{\rm Ly\alpha}$ identical to that
measured for LBGs ($2.7\le z<3.4$) results in an expected distribution
for $1.9\le z<2.7$ galaxies that has a lower frequency of
absorption-line systems than what is actually observed at $1.9\le
z<2.7$.  Therefore, the intrinsic $W_{\rm Ly\alpha}$ distribution for
$1.9\le z<2.7$ galaxies must include a larger fraction of galaxies
with Ly$\alpha$ in absorption than what is observed for higher
redshift ($2.7\le z<3.4$) galaxies.  

The middle panels of Figure~\ref{fig:lyacompall} show what happens if
we assume that the intrinsic $W_{\rm Ly\alpha}$ distribution at
$1.9\le z<2.7$ and $2.7\le z <3.4$ is simply an equally-weighted
combination of the distributions measured in these two redshift
ranges.\footnote{Since the comoving number densities of galaxies at
redshifts $1.9\le z<2.7$ and $2.7\le z <3.4$ are similar, it is
reasonable to approximate the combined $W_{\rm Ly\alpha}$ distribution
as an equally weighted sum of the measured distributions in these two
redshift ranges.}  In the bottom middle panel, the expected and
observed distributions are not significantly different, so it is
plausible that the intrinsic distribution of $W_{\rm Ly\alpha}$ for
redshift $1.9\le z<2.7$ galaxies resembles an equally-weighted
combination of the distributions measured for BX galaxies and LBGs.
However, as the top middle panel indicates, such an intrinsic
distribution overpredicts the number of galaxies at $2.7\le z<3.4$
with Ly$\alpha$ in absorption.  Decreasing the fraction of galaxies
with absorption in the intrinsic distribution (e.g., by assuming some
non-equally weighted combination of $W_{\rm Ly\alpha}$ distributions
at low and high redshifts) may result in a better match for the
observed $z\sim 3$ distribution, but would lose agreement with the
observed $z\sim 2$ distribution.

Finally, if we assume that the intrinsic $W_{\rm Ly\alpha}$
distributions at $1.9\le z<2.7$ and $2.7\le z <3.4$ are identical to
those we actually measure in these two redshift ranges, then the
expected distributions are very close to what is actually measured
(right panels of Figure~\ref{fig:lyacompall}).  In fact, the expected
distributions are only marginally different from the {\it assumed}
intrinsic distributions, even if the assumptions are erroneous
(compare red and green rectangles in left and middle panels of
Figure~\ref{fig:lyacompall}; a Kolmogorv-Smirnov (KS) test indicates a
$\ga 50\%$ probability that the intrinsic and expected distributions
are drawn from the same populations).  These observations suggest that
the BX and LBG color selection criteria do not alter significantly the
parent $W_{\rm Ly\alpha}$ distribution of galaxies.  We already
discussed in \S~\ref{sec:lyas} why this must be the case for BX
galaxies, since there is a redshift range covered by BX selection
where the $\ugr$ colors are unaffected by Ly$\alpha$; we have just
shown it to be true for LBGs also.

In all cases shown in Figure~\ref{fig:lyacompall}, we have quantified
the disparity in the observed and expected $W_{\rm Ly\alpha}$
distributions by computing the statistic $\xi$ as follows.  We
generated 10000 realizations of the expected $W_{\rm Ly\alpha}$
distribution (in the same way as we did to compute the uncertainties
in the expected distribution; see above).  We then performed a KS test
to determine the probability ($p_{\rm KS}$) that each of these
realizations are drawn from the same distribution as the observed
$W_{\rm Ly\alpha}$ distribution.  The quantity $\xi$ is then defined
as the ratio of the number of realizations where $p_{\rm KS}<0.5$ to
the total number of realizations (10000).  Low values of $\xi$
indicate that the expected and observed distributions are less likely
to have been drawn from the same parent distribution.  The values of
$\xi$ are given in each panel of Figure~\ref{fig:lyacompall}, and
support our conclusion that the color criteria do not significantly
alter the intrinsic $W_{\rm Ly\alpha}$ distributions.  In summary, we
find evidence that the fraction of emission-line galaxies ($f20$)
appears to increase with redshift (Table~\ref{tab:lyaew}) and that
such a trend is most likely not due to selection bias, as demonstrated
by the differences in the expected and observed $W_{\rm Ly\alpha}$
distributions for galaxies at lower and higher redshift
(Figure~\ref{fig:lyacompall}).

\subsection{$\ebmv$ Distributions}
\label{sec:ebmv}

\begin{figure*}[hbt]
\plottwo{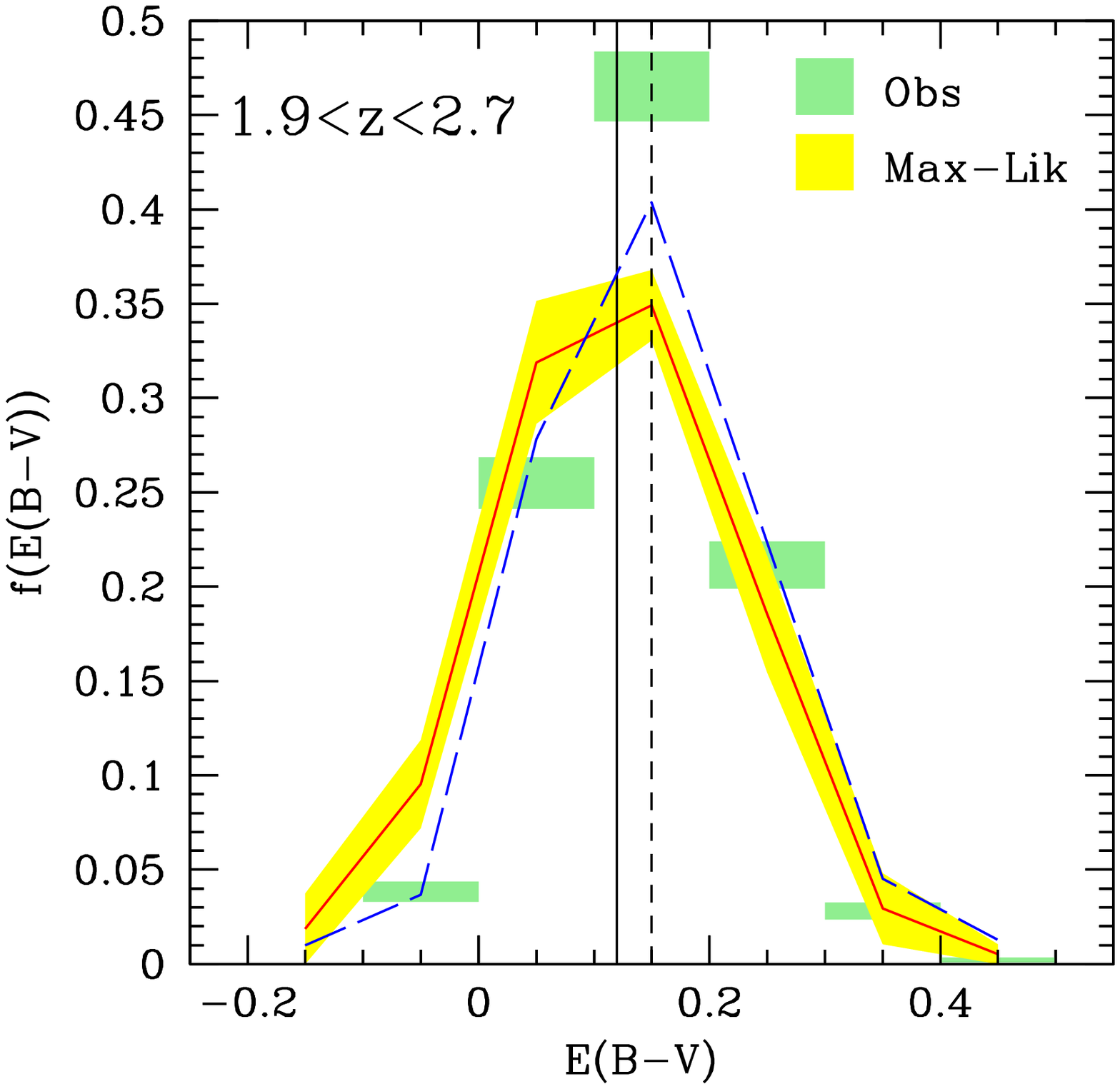}{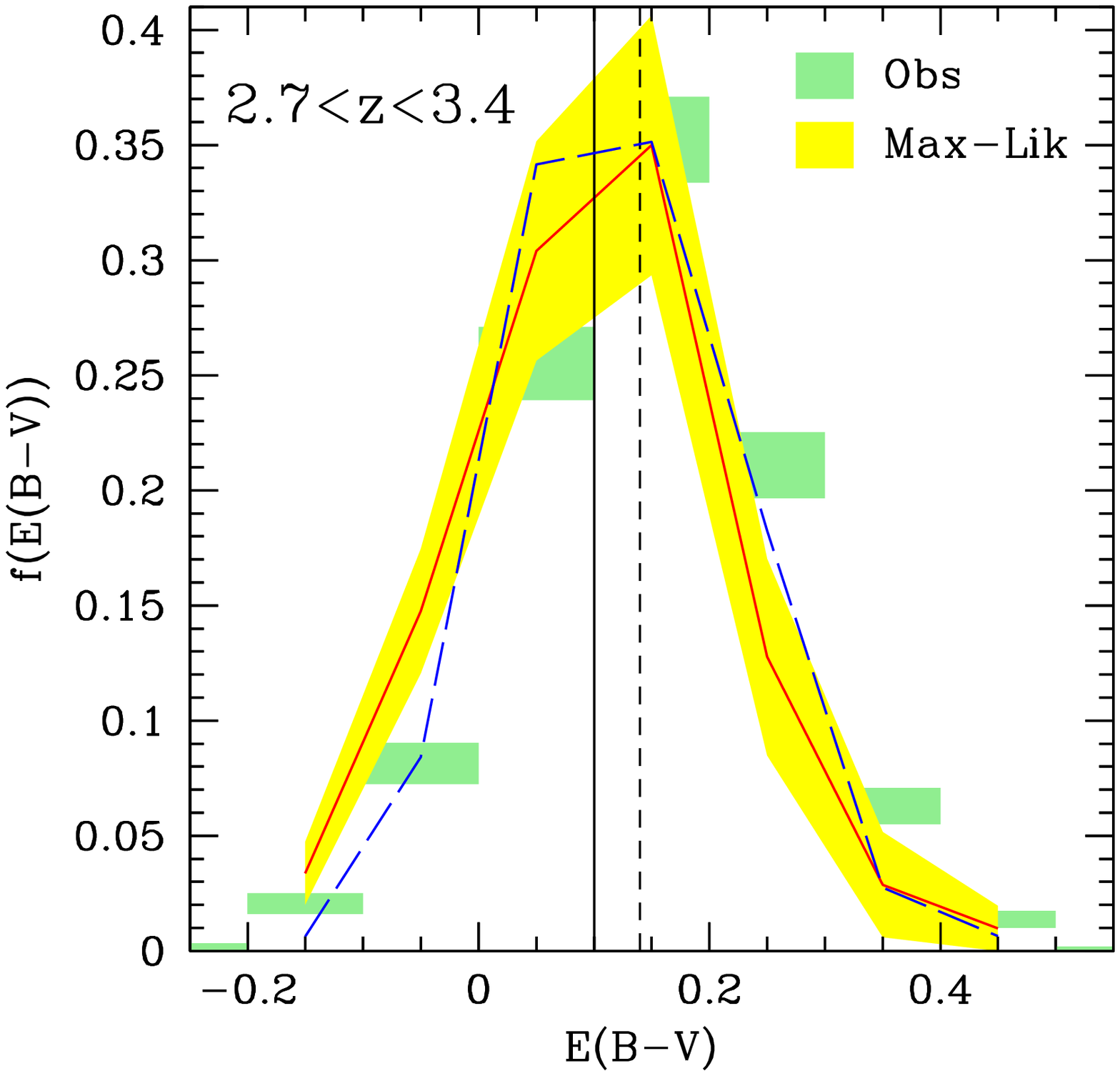}
\caption{Comparison of best-fit and observed $\ebmv$ distributions for
galaxies at redshifts $1.9\le z<2.7$ (left panel) and $2.7\le z<3.4$
(right panel).  Histograms denote the observed distributions, computed
from the $\gmr$ colors, and assuming a constant star formation model
attenuated by the \citet{calzetti00} law.  The width of each bar
reflects the Poisson error in the corresponding bin.  The red lines
and yellow shaded regions indicate the mean and $1$~$\sigma$ errors on
the maximum-likelihood (best-fit) distributions.  The blue dashed
lines indicate the distribution without correcting for Ly$\alpha$
perturbation to the observed colors.  Dashed and solid vertical lines,
respectively, denote the average $\ebmv$ for the observed and best-fit
distributions. The $\ebmv$ distribution data are summarized in
Table~\ref{tab:ebmvtab}.
\label{fig:ebmvfig}}
\end{figure*}

A useful by-product of the maximum-likelihood method
(\S~\ref{sec:tpf},\ref{sec:implementation}) is the distribution of
galaxy spectral shapes, parameterized by the color excess $\ebmv$,
corrected for incompleteness.  Figure~\ref{fig:ebmvfig} shows the
best-fit $\ebmv$ distributions, compared with the observed
distributions, for galaxies at redshifts $1.9\le z<2.7$ and $2.7\le
z<3.4$ (also tabulated in Table~\ref{tab:ebmvtab}).  Using the $V_{\rm
eff}$ method (\S~\ref{sec:veff}) results in $\ebmv$ distributions that
are within $10\%$ of the observed distributions and therefore deviate
significantly from our best-fit distributions at $z\sim 2$ and $z\sim
3$.  The analysis indicates that the true $\ebmv$ distributions are
slightly bluer, on average, than observed.  Table~\ref{tab:ebmvtab}
lists the mean and dispersion of $\ebmv$ for the observed and
maximum-likelihood distributions for galaxies at redshifts $1.9\le
z<2.7$ and $2.7\le z<3.4$.

\begin{deluxetable*}{lcccc}
\tabletypesize{\footnotesize}
\tablewidth{0pc}
\tablecaption{Normalized $\ebmv$ Distributions}
\tablehead{
\colhead{$\ebmv$} &
\colhead{BX (Measured)} &
\colhead{$1.9\le z<2.7$ (Max-Lik)} &
\colhead{LBG (Measured)} &
\colhead{$2.7\le z<3.4$ (Max-Lik)}}
\startdata
$-0.2$ --- $-0.1$\tablenotemark{a} & $<0.01$ & $0.02\pm0.02$ & $0.02\pm0.01$ & $0.03\pm0.01$ \\
$-0.1$ --- 0.0\tablenotemark{a} & $0.04\pm0.01$ & $0.10\pm0.02$ & $0.08\pm0.01$ & $0.15\pm0.03$ \\
0.0 --- 0.1 & $0.26\pm0.01$ & $0.32\pm0.03$ & $0.26\pm0.02$ & $0.30\pm0.05$ \\
0.1 --- 0.2 & $0.47\pm0.02$ & $0.34\pm0.02$ & $0.35\pm0.02$ & $0.35\pm0.06$ \\
0.2 --- 0.3 & $0.21\pm0.01$ & $0.19\pm0.03$ & $0.21\pm0.01$ & $0.13\pm0.04$ \\
0.3 --- 0.4 & $0.03\pm0.01$ & $0.03\pm0.02$ & $0.06\pm0.01$ & $0.03\pm0.02$ \\
0.4 --- 0.5 & $<0.01$ & $0.01\pm0.01$ & $0.01\pm0.01$ & $0.01\pm0.01$ \\
$\langle\ebmv\rangle$ & $0.15\pm0.07$ & $0.12\pm0.12$ & $0.14\pm0.07$ & $0.10\pm 0.09$ \\
\enddata
\tablenotetext{a}{We measure a non-negligible number of galaxies with $\ebmv<0$ since
the $\ebmv$ of dust-free and/or very young galaxies is dominated by intrinsic variations
in the SED (see text).}
\label{tab:ebmvtab}
\end{deluxetable*}

The measured $\ebmv$ distribution for the BX sample is expected to be
slightly biased toward redder spectral shapes than the intrinsic
values because our photometric method makes the colors appear slightly
redder than they really are --- and thus $\ebmv$ is redder ---
particularly for fainter galaxies (\S~\ref{sec:tpf}).  In addition,
the presence of Ly$\alpha$ absorption in a galaxy's spectrum will,
depending on the redshift, cause the $G$-band magnitude to appear
fainter than the true broadband magnitude (corrected for line
effects), such that $\ebmv$ will be overestimated.  This latter effect
can be visualized for $1.9\le z<2.7$ galaxies in the left panel of
Figure~\ref{fig:ebmvfig}: the distribution uncorrected for the effects
of Ly$\alpha$ (dashed blue line) is systematically redder than the
corrected distribution (solid red line), owing to the presence of
Ly$\alpha$ absorption among, and the low $f20$ value of, the majority
of galaxies at $1.9\le z<2.7$ (e.g., Figure~\ref{fig:lya},
\ref{fig:lyacompall}).  The systematic effects induced by Ly$\alpha$
are less apparent in the $\ebmv$ distribution for $2.7\le z<3.4$
galaxies, primarily because the LBG selection window spans a region of
color space that is significantly larger than the typical color change
induced by Ly$\alpha$ perturbations.

Before proceeding with a discussion of $\ebmv$ as an indicator of dust
reddening and the variation of $\ebmv$ with redshift and apparent
magnitude, we remind the reader that we can only correct for the
incompleteness of objects whose colors are such that they are
scattered into the selection windows.  In other words, there are
undoubtedly galaxies at these redshifts that will never scatter into
the BX/LBG selection windows, for example, those galaxies that are
optically-faint either because they have little star formation or are
very dusty starbursts (e.g., DRGs and SMGs).  Therefore, the $\ebmv$
for such dusty galaxies will not be reflected in the distributions
shown in Figure~\ref{fig:ebmvfig}.  Typically, such very dusty
galaxies would have $\ebmv > 0.45$, although a significant fraction
also show bluer $\ebmv$ comparable to those of BX/LBGs
\citep{chapman05}.  Because these dusty star-forming and quiescent
galaxies are in large part optically-faint, not accounting for them in
our analysis should minimally affect our $\ebmv$ distribution for {\it
optically-bright} galaxies (Figure~\ref{fig:ebmvfig}).  Further, as we
show in \S~\ref{sec:vvds}, comparison of our UV LF with those derived
from magnitude limits surveys suggests that we must be reasonably
complete for UV-bright ($\rs<25.5$) galaxies at $z\sim 2-3$.

\subsubsection{$\ebmv$ as a Proxy for Dust Reddening}
\label{sec:ebmvdust}

Up until now, we have been using $\ebmv$ (the rest-frame UV slope) to
parameterize the range of spectral shapes observed among high redshift
galaxies.  A number of studies have shown that $\ebmv$ also has a
physical interpretation: it correlates very well with the reddening,
or dust obscuration, of most high redshift galaxies (e.g.,
\citealt{calzetti00,adel00,reddy06a}).  Here we define reddening as
the attenuation of luminosity by dust which can be parameterized, for
example, by the quantity $L_{\rm bol}/L_{\rm UV}$.  Combining {\it
Spitzer} MIPS data for a sample of {\it spectroscopically-confirmed}
redshift $1.5\la z\la 2.6$ galaxies where {\it K}-corrections could be
computed accurately, \citet{reddy06a} showed that $\ebmv$ not only
correlates with $L_{\rm bol}$ for galaxies with $L_{\rm bol} \la
10^{12.3}$~L$_{\odot}$, but that the correlation is identical to that
established for local galaxies \citep{calzetti00,meurer99}.

The $\ebmv$ for relatively dust-free (or very young) galaxies is
dominated by intrinsic variations in the SEDs of high
redshift galaxies, and so $\ebmv$ is not a direct indicator of
reddening for these galaxies (which is why we measure a non-negligible
number density of galaxies with $\ebmv< 0$ when assuming a single
SED).  Further, there is a significant presence of very dusty galaxies
with $L_{\rm bol}\ga 10^{12}$~L$_{\odot}$ that are optically-bright
($\rs<25.5$) and satisfy the rest-UV color criteria, but have $\ebmv$
that severely underpredict their attenuations and bolometric
luminosities \citep{reddy06a}.  Finally, we remind the reader that we
cannot account for the $\ebmv$ of objects that have a zero probability
of being scattered into our sample.  Nonetheless, our
completeness-corrected estimates of the $\ebmv$ distributions suggest
an average attenuation between dust-corrected and uncorrected UV
luminosity, $L_{\rm UV}^{cor}/L_{\rm UV}$, of $\sim 4-5$.  This is
similar to the value measured from (1) stacked X-ray data for BX
galaxies and LBGs \citep{nandra02,reddy04}, and (2) MIPS luminosities
and dust-corrected UV and H$\alpha$ luminosities
\citep{reddy06a,erb06c}.  It is also the same value advocated by
\citet{steidel99} in correcting observed UV luminosities for dust
extinction among $z\sim 3$ LBGs.

\subsubsection{Comparison of Reddening Distributions with Redshift}

\begin{figure}[hbt]
\plotone{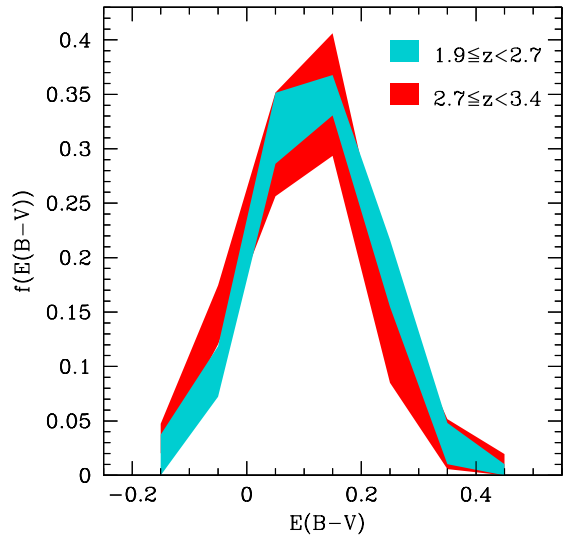}
\caption{Comparison of maximum-likelihood $\ebmv$ distributions for
galaxies at redshifts $1.9\le z<2.7$ and $2.7\le z<3.4$.
\label{fig:ebmvcomp}}
\end{figure}

Remarkably, we find very little evolution in the reddening
distribution between redshifts $1.9\la z\la 3.4$, despite the roughly
$730$~Myr timespan between the mean redshifts for the low ($\langle
z\rangle=2.30$) and high ($\langle z\rangle=3.05$) redshift samples,
as shown in Figure~\ref{fig:ebmvcomp}.  It is not surprising that the
two distributions should span a similar range of $\ebmv$ since the BX
criteria were designed to selected galaxies with a similar range of
spectral properties as LBGs.  Even so, the incompleteness corrections
modulate two very different observed $\ebmv$ distributions for the lower
and higher redshift samples to the point where they are virtually
identical.  The difference in the fraction of large $W_{\rm Ly\alpha}$ emission
systems between the two samples (\S~\ref{sec:lyadist}), and the possibility that
such large $W_{\rm Ly\alpha}$ emission systems could be young and relatively
dust-free galaxies (\S~\ref{sec:discussion}), does little to
modulate the overall reddening distributions since such galaxies constitute a
small fraction of the overall population.

The similarity in $\ebmv$ and dust attenuation between $z\sim 2$ and
$z\sim 3$ galaxies agrees with the stacked X-ray studies of BXs and
LBGs \citep{reddy04, nandra02}.  As we discuss in
\S~\ref{sec:discussion}, the lack of evolution in $\ebmv$ implies that
the extinction properties of bright ($\rs\le 25.5$) star-forming
galaxies are not changing significantly between redshift $z\sim 3$ and
$z\sim 2$, unlike the situation at lower ($z\la 2$) and higher ($z\ga
3$) redshifts (e.g., \citealt{reddy06a, bouwens06}).

\subsubsection{Reddening Distribution as a Function of Rest-Frame UV Magnitude}
\label{sec:ebmvtor25}

Before turning to a discussion of the LF, we must first determine
whether the best-fit $\ebmv$ distribution shows any systematic changes
as a function of rest-frame UV magnitude, since such changes can, in
principle, affect the shape and normalization of the LF.  As a first
test, we restricted the maximum-likelihood analysis
(\S~\ref{sec:tpf},\ref{sec:implementation}) to particular magnitudes in the
range $22.0\le\rs\le 25.5$, and we did not find any significant trend
in the $\ebmv$ distribution as a function of magnitude to $\rs=25.5$.

\begin{figure}[hbt]
\plotone{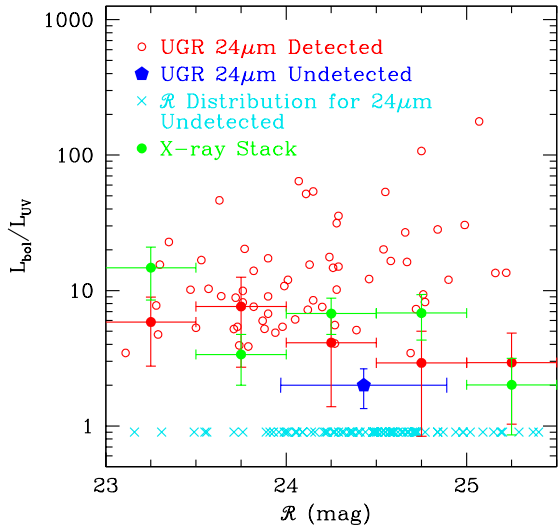}
\caption{Distribution of attenuation factors, parameterized as $L_{\rm
bol}/L_{\rm UV}$, as inferred from {\it Spitzer} MIPS data, as a
function of apparent optical magnitude $\rs$ for rest-UV-selected
galaxies with redshifts $1.5\la z\la 2.6$.  Also indicated is the
stacked average for $48$ galaxies undetected at $24$~$\mu$m (large
blue pentagon) and unconfused with brighter sources; the distribution
in $\rs$ magnitude for a larger sample of $73$ galaxies undetected at
$24$~$\mu$m is shown by the arbitrarily normalized crosses.  Total
stacked $24$~$\mu$m and X-ray averages are indicated by the solid red
circles and green points, respectively, that include all galaxies.
\label{fig:irxvr}}
\end{figure}

Because $\ebmv$ can be used as a proxy for the reddening of galaxies
(\S~\ref{sec:ebmvdust}), we have further investigated whether the
reddening distribution varies as a function of observed apparent
magnitude by exploiting the multi-wavelength data in several survey
fields.  To accomplish this, we relied on our interpretation of the
{\it Spitzer} MIPS data for a sample of BX-selected galaxies in the
GOODS-N field; these data give us an independent probe of the dust
emission in $z\sim 2$ galaxies.  Figure~\ref{fig:irxvr} shows the dust
obscuration factors, parameterized as $L_{\rm bol}/L_{\rm UV}$, where
$L_{\rm bol}\equiv L_{\rm IR} + L_{\rm UV}$ (infrared plus UV
luminosity),\footnote{Here we define $L_{\rm IR}$ as the total
luminosity between $8$ and $1000$~$\mu$m.} as a function of observed
optical magnitude, from the MIPS analysis of the GOODS-North field by
\citet{reddy06a}.  The open red circles indicate rest-UV-selected
objects at $1.5\le z\le 2.6$, most of which are BX galaxies, detected
at $24$~$\mu$m, and the large pentagon and crosses denote the average
stack and distribution in $\rs$ magnitude, respectively, for galaxies
undetected at $24$~$\mu$m.  The total $24$~$\mu$m stacked averages
including both detected and undetected galaxies at $24$~$\mu$m are
shown by the solid red circles.  Similarly, the $L_{\rm bol}/L_{\rm
UV}$ inferred from X-ray stacked averages (computed in manner similar
to that presented in \citealt{reddy04} where $L_{\rm bol}$ is
determined from the X-ray flux) including all galaxies, irrespective
of direct detection in the {\it Chandra} 2~Ms data \citep{alexander03}
are shown by the green points.  While there is some evidence that the
{\it dispersion} in attenuation factor increases towards fainter
magnitudes, as evidenced by the larger spread of $24$~$\mu$m detection
galaxies and as would be expected if optically-faint galaxies have
contribution from both heavily dust-obscured objects as well as those
with intrinsically low star formation rates, the results of
Figure~\ref{fig:irxvr} suggest that the {\it average} extinction
correction (based on the stacked points) is approximately constant
over the range in $\rs$ magnitude considered here.\footnote{We note
that \citet{reddy06a} excluded objects from their analysis that were
directly detected in the {\it Chandra} 2~Ms data in the GOODS-N field
\citep{alexander03} of which almost all are AGN.}  These results
confirm the trends noted by \citet{adel00}, who used local templates
to deduce that the observed UV luminosities of galaxies at redshifts
$z=0$, $z\sim 1$, and $z\sim 3$, are insensitive to dust obscuration,
$L_{\rm bol}/L_{\rm UV}$ (e.g., Figure~17 of \citealt{adel00}).  We
have confirmed this trend explicitly at redshifts $1.5\la z\la 2.6$.
The observed (unobscured) UV luminosity (i.e., the emergent luminosity
after attenuation by dust) to $\rs=25.5$ will also be insensitive to
bolometric luminosity since dust obscuration is tightly correlated
with bolometric luminosity \citep{reddy06a}.  While the attenuation
factors and bolometric luminosities of $z\sim 2-3$ galaxies are
insensitive to the {\it unobscured} UV luminosity, at least to
$\rs=25.5$, there is a very strong dependence of the {\it
dust-corrected} UV luminosity (or IR or bolometric luminosity) on the
attenuation factors of galaxies (\citealt{reddy06a, adel00}; see also
\S~\ref{sec:irevol}).  For the purposes of the present analysis, we
will assume that the reddening distribution of galaxies is constant to
$\rs=25.5$.  We return to the issue of how a varying reddening
distribution affects our calculation of the total luminosity density
in \S~\ref{sec:pahirlf}.

\section{RESULTS: UV LUMINOSITY FUNCTIONS}
\label{sec:uvlf}

\subsection{Preferred LFs}

To provide the closest match between rest-frame wavelengths, and thus
avoid cosmological {\it K}-corrections, we used $\rs$-band as a tracer
of rest-frame $1700$~\AA\, emission at the mean redshift of the LBG
sample, $\langle z\rangle\sim 3.05$.  Similarly, we used a
``composite'' magnitude between $G$ and $\rs$-band ($m_{\rm G\rs}$) as
a tracer of rest-frame $1700$~\AA\, at the mean redshift of the BX
sample, $\langle z\rangle \sim 2.30$, where $m_{\rm G\rs}$ is simply
the magnitude corresponding to the average of the $G$ and $\rs$
fluxes.  Absolute magnitudes were computed using the standard
relation:
\begin{eqnarray}
M_{\rm AB}(1700\AA) = m - 5\log(d_{\rm L}/10\,{\rm pc}) + 2.5\log(1+z),
\label{eq:absappmag}
\end{eqnarray}
where $M_{\rm AB}(1700\AA)$ is the absolute magnitude at rest-frame
$1700$~\AA, $d_{\rm L}$ is the luminosity distance, and $m$ is the
apparent magnitude at $\rs$-band at $z\sim 3$ or at the composite
$G\rs$-band at $z\sim 2$.  We have made the reasonable assumption that
the SED {\it K}-correction is approximately zero for the average
rest-UV SED of BX-selected galaxies after a star formation age of
$100$~Myr for the typical reddening ($\ebmv\sim 0.15$) of galaxies in
our sample.

\begin{figure*}[hbt]
\plotone{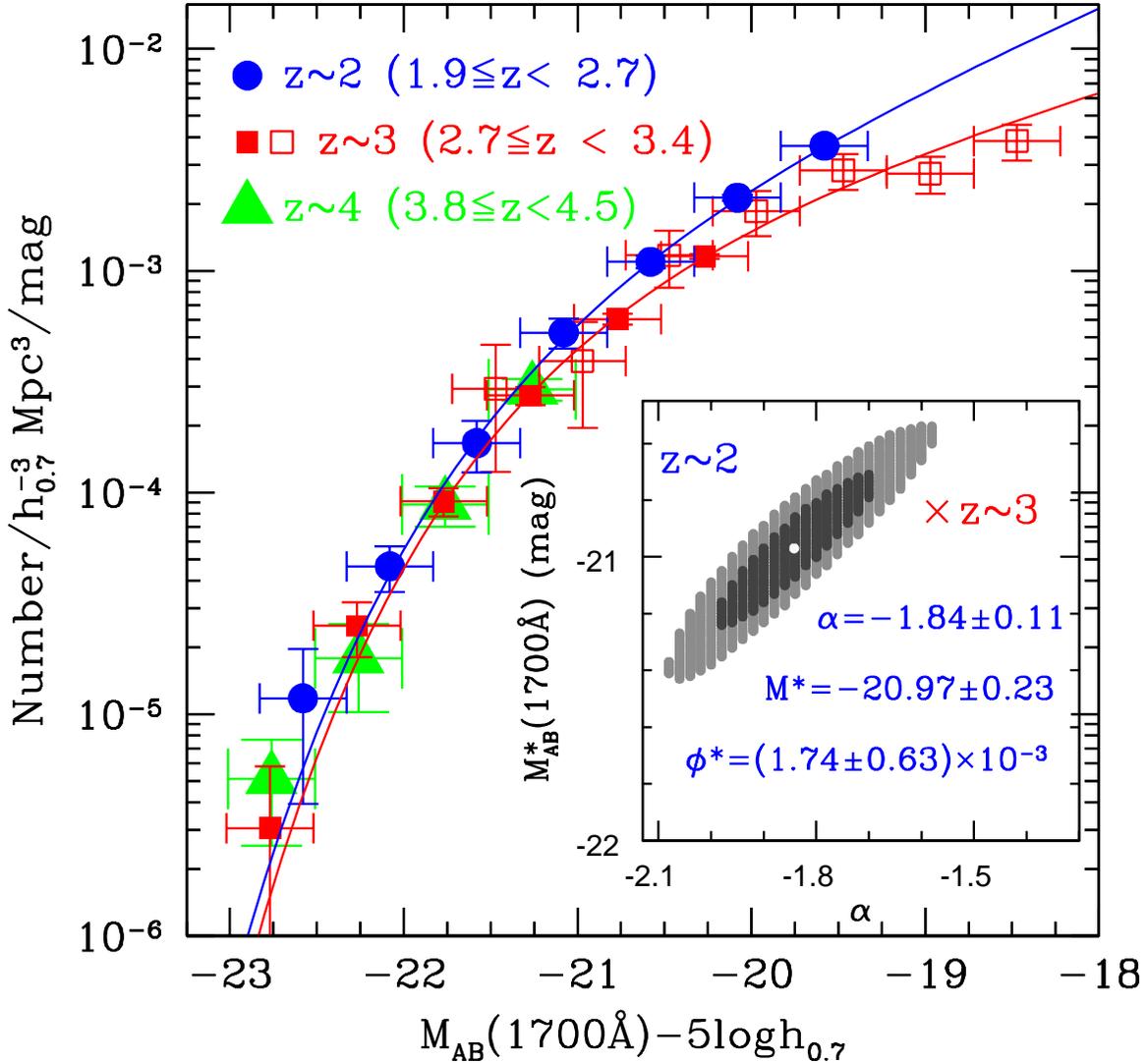}
\caption{Rest-frame UV luminosity functions at $z\sim 2$ (solid
circles) and $z\sim 3$ (solid and open squares for ground-based
observations and {\it HST}, respectively) computed in our analysis,
compared with $z\sim 4$ results (triangles) from \citet{steidel99}.
All data have been recast to the same cosmology used throughout this
paper.  Also indicated are the best-fit \citet{schechter76} functions
for the $z\sim 2$ (blue line) and $z\sim 3$ (red line) LFs.  No shift
in normalization was applied to the LFs.
Confidence contours demonstrate the degeneracy between $\alpha$ and
$M^{\ast}$ for the $z\sim 2$ fit, as shown in the inset.  The red
cross denotes $\alpha$ and $M^{\ast}$ for $z\sim 3$ galaxies.  
\label{fig:rlfg}}
\end{figure*}

The maximum-likelihood rest-frame $1700$~\AA\, luminosity functions
for $z\sim 2$ and $z\sim 3$ galaxies are shown in
Figure~\ref{fig:rlfg}, and listed in Table~\ref{tab:uvlf}.  The LFs
were computed by using the entire photometric sample and holding the
best-fit $\ebmv$ distribution (as determined from the spectroscopic
sample; Figure~\ref{fig:ebmvfig}) fixed.  The extension of the
spectroscopically-determined $\ebmv$ distribution to the photometric
sample is a reasonable approximation given that (a) the spectroscopic
and photometric samples are likely to have the same redshift
distribution (\S~\ref{sec:speccomp}) and (b) the $\ebmv$ distribution
remains unchanged as a function of $\rs$ magnitude to $\rs=25.5$
(\S~\ref{sec:ebmvtor25}).  By nature of the maximum-likelihood method,
our LF computation includes corrections for the systematic effects of
photometric bias and Ly$\alpha$ perturbations.  Errors in the
luminosity functions reflect both Poisson counting statistics and
field-to-field variations; the latter are accounted for by examining
the dispersion in the LF as a function of magnitude for each of the
fields of the survey (see \S~\ref{sec:ftf}).  The best-fit
\citet{schechter76} function and parameters for the $z\sim 2$ LF are
also indicated.  Uncertainties in the faint-end slope $\alpha$,
characteristic absolute magnitude $M^{\ast}$ (or characteristic
luminosity $L^{\ast}$), and characteristic number density
$\phi^{\ast}$ are estimated by simulating many realizations of the LF
as allowed by the errors (assuming the errors follow a normal
distribution), fitting a Schechter function to each of these
realizations, and then determining the dispersion in measured values
for $\alpha$, $M^{\ast}$, and $\phi^{\ast}$ for these realizations.

\begin{deluxetable}{lcc}
\tabletypesize{\footnotesize}
\tablewidth{0pc}
\tablecaption{Rest-Frame UV Luminosity Functions of $1.9\la z\la 3.4$ Galaxies}
\tablehead{
\colhead{} &
\colhead{} &
\colhead{$\phi$} \\
\colhead{Redshift Range} &
\colhead{M$_{\rm AB}(1700\AA)$} &
\colhead{($\times 10^{-3}$~$h^{3}_{0.7}$~Mpc$^{-3}$~mag$^{-1}$)}}
\startdata
$1.9\le z<2.7$ & $-22.83$ --- $-22.33$ & $0.012\pm0.008$ \\
& $-22.33$ --- $-21.83$ & $0.05\pm0.01$ \\
& $-21.83$ --- $-21.33$ & $0.17\pm0.04$ \\
& $-21.33$ --- $-20.83$ & $0.53\pm0.08$ \\
& $-20.83$ --- $-20.33$ & $1.10\pm0.07$ \\
& $-20.33$ --- $-19.83$ & $2.13\pm0.07$ \\
& $-19.83$ --- $-19.33$ & $3.66\pm0.02$ \\
\\
$2.7\le z<3.4$ & $-23.02$ --- $-22.52$ & $0.0031\pm0.0027$ \\
& $-22.52$ --- $-22.02$ & $0.025\pm0.007$ \\
& $-22.02$ --- $-21.52$ & $0.09\pm0.01$ \\
& $-21.52$ --- $-21.02$ & $0.27\pm0.03$ \\
& $-21.02$ --- $-20.52$ & $0.60\pm0.03$ \\
& $-20.52$ --- $-20.02$ & $1.16\pm0.03$ \\
\enddata
\label{tab:uvlf}
\end{deluxetable}

Based on integrating our maximum-likelihood LFs, the fraction of
star-forming galaxies with redshifts $1.9\le z<2.7$ and $M_{\rm
AB}(1700\AA) < -19.33$ (i.e., $\rs=25.5$ at $z=2.3$) that have colors
that satisfy BX criteria is $\approx 58\%$.  Similarly, the fraction
of star-forming galaxies with redshifts $2.7\le z<3.4$ and $M_{\rm
AB}(1700\AA) < -20.02$ ($\rs=25.5$ at $z=3.05$) that have colors that
satisfy the LBG criteria is $\approx 47\%$.  Note that some galaxies
escaping LBG selection will be scattered into the BX window, and vice
versa.  Also, some galaxies that are intrinsically fainter (or
brighter) than $\rs=25.5$ will be scattered into (or out of) the BX
and LBG samples due to photometric error.  The {\it total} fraction of
galaxies with $1.9\le z<3.4$ and $\rs<25.5$ that satisfy {\it either}
the BX or LBG criteria is $0.55$.

In the following sections, we examine various aspects of the
luminosity functions derived here, including the differences in the LF
derived using the $V_{\rm eff}$ versus maximum-likelihood method, the
significance (or lack thereof) of the Schechter parameters, and
field-to-field variations.  We conclude this section by examining how
photometric redshifts can introduce non-trivial biases in the
computation of the LF.

\break

\subsection{Comparison of the $V_{\rm eff}$ and Maximum-Likelihood Methods}

The maximum-likelihood technique was used to derive the LFs presented
here.  However, because many published LFs are derived using the less
accurate $V_{\rm eff}$ method, particularly for dropout samples at
high redshift, it is useful to determine how close (or how far) such
determinations are from reality by comparing with our
maximum-likelihood value.

\begin{figure}[hbt]
\plotone{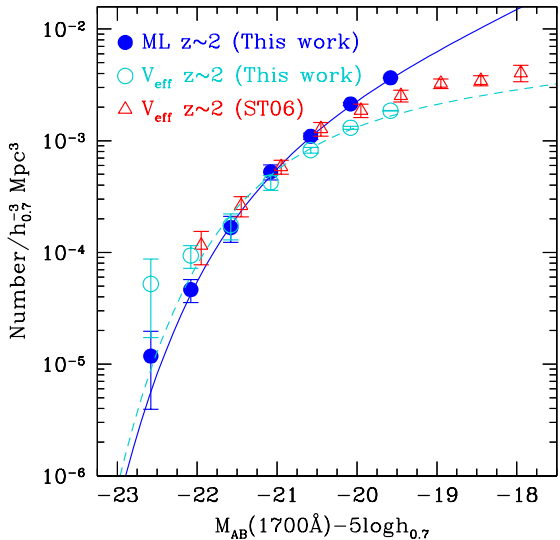}
\caption{Comparison of the maximum-likelihood LF at $z\sim 2$ with
the $V_{\rm eff}$ determinations from our work (open circles)
and \citet{sawicki06a} (open triangles).
\label{fig:lfcomp}}
\end{figure}

Figure~\ref{fig:lfcomp} compares the LFs at $z\sim 2$ computed using
the $V_{\rm eff}$ (\S~\ref{sec:veff}) and maximum-likelihood
(\S~\ref{sec:tpf}) methods, along with a comparison of the $V_{\rm
eff}$ determination at $z\sim 2$ from \citet{sawicki06a}.  The figure
clearly demonstrates the systematic bias at both bright and faint
magnitudes of the $V_{\rm eff}$ LF with respect to the
maximum-likelihood value.  These biases are particularly apparent for
criteria that target a narrow range in color space, such as the BX
criteria, where photometric scatter or perturbations due to Ly$\alpha$
can be as large or larger than the width of the color selection
windows (e.g., Figure~\ref{fig:ewdist}, \ref{fig:simdist}).  For
example, we would have inferred a significantly shallower faint-end
slope of $\alpha = -1.21\pm0.15$ had we relied on the LF derived from
the $V_{\rm eff}$ method.\footnote{Note that our $V_{\rm eff}$
determination is slightly different from that of \citet{sawicki06a},
despite the use of the exact same filter set and color criteria
between the two studies, since our $V_{\rm eff}$ determination
includes the effects of Ly$\alpha$ line perturbations to the rest-UV
colors and incorporates the maximum-likelihood $\ebmv$ distribution in
the LF calculation.}  For the LBG criteria, we find little difference
in the $V_{\rm eff}$ and maximum-likelihood determinations of the LF,
mostly due to the fact that the LBG selection window covers a larger
area of color space and outlier objects (with colors placing them near
the selection boundaries) for which the bias is the largest will make
up a significantly smaller fraction of the sample.

\subsection{Schechter Parameters}
\label{sec:faintendUV}

The spectroscopic sample allows us to accurately constrain the LF,
taking into account sample completeness, interloper fraction, and line
perturbations, for galaxies with $\rs<25.5$.  It is brighter than this limit
that we consider our LF to be most robust.  The results for $z\sim 2$
galaxies fainter than $\rs=25.5$ is less certain given that our
determination of the $z\sim 2$ faint-end slope relies on a
spectroscopic sample that extends only $\sim 4$ times fainter than the
characteristic luminosity of $z\sim 2$ galaxies.  Formally, we find a
faint-end slope at $1.9\le z<2.7$ of $\alpha=-1.84\pm0.11$, with a
characteristic magnitude of $M^{\ast}_{\rm AB}(1700\AA) =
-20.97\pm0.23$.

\begin{deluxetable*}{lcccc}
\tabletypesize{\footnotesize}
\tablewidth{0pc}
\tablecaption{Best-fit Schechter Parameters for UV LFs of $1.9\la z\la 3.4$ Galaxies}
\tablehead{
\colhead{Redshift Range} &
\colhead{$\alpha$} &
\colhead{M$^{\ast}_{\rm AB}(1700\AA)$} &
\colhead{$\phi^{\ast}$} &
\colhead{$\chi^{2}$}}
\startdata
$1.9\le z<2.7$ (ground-based) & $-1.84\pm0.11$ & $-20.97\pm0.23$ & $(1.74\pm0.63)\times 10^{-3}$ & 0.73 \\
$1.9\le z<2.7$ (ground-based) & $-1.60$ (fixed) & $-20.60\pm0.08 $ & $(3.31\pm0.22)\times 10^{-3}$ & 5.81 \\
\\
$2.7\le z<3.4$ (ground-based) & $-1.85\pm0.22$ & $-21.21\pm0.16$ & $(1.02\pm0.52)\times 10^{-3}$ & 0.18 \\
$2.7\le z<3.4$ (ground+space) & $-1.57\pm0.11$ & $-20.84\pm0.12$ & $(1.66\pm0.63)\times 10^{-3}$ & 5.72 \\
\enddata
\label{tab:schechparms}
\end{deluxetable*}

The \citet{steidel99} analysis of the $z\sim 3$ LF included
$U$-dropout galaxies in HDF-N where the redshift distribution was
modeled using the color criteria of \citet{dickinson98} and assuming
the range of intrinsic spectral shapes of LBGs found by \citet{adel00}
(open squares in Figure~\ref{fig:rlfg}).  Based on the combined Keck
spectroscopic and HDF-N $U$-dropout samples, \citet{steidel99} found a
steep faint-end slope $\alpha=-1.60\pm0.13$.  Further refinement of
the incompleteness corrections by \citet{adel00} resulted in a
combined fit to the ground-based spectroscopic and HDF $U$-dropout
samples of $\alpha=-1.57\pm0.11$.  Fitting only the ground-based
(spectroscopically determined) points from our analysis at $z\sim 3$
yields $\alpha=-1.85\pm0.22$ and $M^{\ast}_{\rm
AB}(1700\AA)=-21.21\pm0.16$.  Combining our data with the HDF-N
$U$-drop points, and excluding the faintest HDF point that may suffer
from incompleteness \citep{steidel99}, results in a fit with
$\alpha=-1.57\pm 0.11$ and $M^{\ast}_{\rm AB}(1700\AA)=-20.84\pm0.12$.
Not surprisingly, these values are in excellent agreement with those
found by \citet{adel00}, primarily because the same faint (HDF) data
are used to determine the faint-end slope.  The best-fit Schechter
function at $z\sim 3$ is also indicated in Figure~\ref{fig:rlfg}.

While $\alpha$ and $M^{\ast}$ are useful in parameterizing the general
shape of the LF, we caution against over-interpreting their validity
when accounting for faint galaxies that are beyond current
spectroscopic capabilities.  The absence of spectroscopic constraints
on the asymptotic faint-end slope and a less than exponential fall-off
of bright sources both conspire to make $\alpha$ steeper (i.e., more
negative).  However, these parameters are useful in describing a local
approximation to data points that are not far from $L^{\ast}$.  For
convenience, the best-fit parameter values from our analysis are
listed in Table~\ref{tab:schechparms}.  In the subsequent analysis, we
will assume $\alpha=-1.6$ for the faint-end slope of the UV LF at
$z\sim 2-3$.  Note that if the steeper faint-end slopes inferred from
our shallower ground-based data ($\alpha\sim -1.85$) accurately
reflect reality, then this will serve to increase the total UV
luminosity density of galaxies with SFRs between $0.1$ and
$1000$~M$_{\odot}$~yr$^{-1}$ by $\sim 20\%$ and $\sim 50\%$ at $z\sim
2$ and $z\sim 3$, respectively, relative to the values obtained with
$\alpha=-1.6$.

\subsection{Field-to-Field Variations}
\label{sec:ftf}

\begin{figure}[hbt]
\plotone{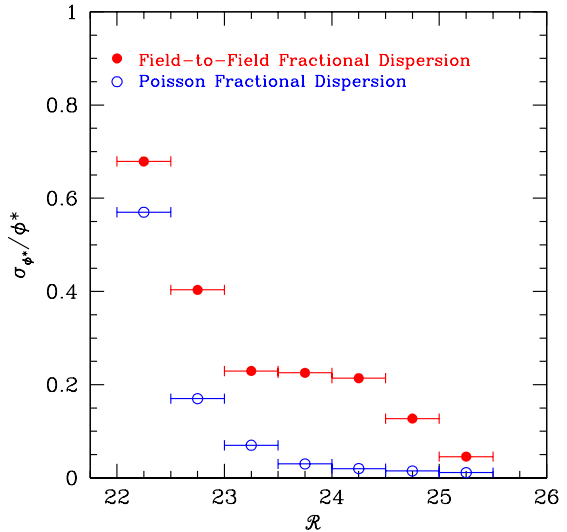}
\caption{Fractional dispersion in normalization of the $z\sim 2$ UV LF
as a function of apparent magnitude, both in terms of field-to-field
(solid circles) and Poisson (open circles) variations.
\label{fig:normdisp}}
\end{figure}

Access to multiple uncorrelated fields allows us to judge the effects
of large scale structure on the derived LF.  The dispersion in
normalization between the luminosity function in bins of $\rs$ derived
in individual fields is a strong function of $\rs$, as illustrated in
Figure~\ref{fig:normdisp} for the $z\sim 2$ sample.  The points in
this figure are determined using the following steps.  First, we
computed the maximum-likelihood LF in each of the $14$ fields of the
$z\sim 2$ survey.  The dispersion in LF values from field-to-field,
within a given magnitude bin, are determined by weighting the LF
values by the field size such that LF determinations from larger
fields are given more weight than LF determinations from smaller
fields.  The fractional dispersion in normalization is then defined as
the ratio of the dispersion of these weighted values and the weighted
mean value of the LF in each bin.  This fractional dispersion in
normalization is much larger at the bright-end for $\rs<22.5$ and
decreases significantly for galaxies with fainter magnitudes.  This
trend results from statistical fluctuations at the bright end due to
the smaller number of galaxies and the fact that the clustering
correlation function is a strong function of magnitude
\citep{adelberger05a}.  We further note that at least 4 of the 14
$z\sim 2$ survey fields show significant redshift-space over-densities
(e.g., HS1700 field; \citealt{steidel05}).  The effect of such
over-densities on the derived LF will of course depend on the redshift
of the over-densities with respect to the BX selection function.  An
over-density at $z=2.8$ is unlikely to affect the LF derived for
$1.9\le z<2.7$ galaxies to the same extent as an over-density at
$z=2.3$ (placing it in the middle of the BX selection function).  One
option when working in single fields is to use the available
spectroscopy and known selection function to model the effects of such
over-densities on the derived LFs, or use Monte Carlo simulations to
estimate uncertainties in the normalization of the derived LF
\citep{bouwens06}.  Because our analysis includes many uncorrelated
fields (14 and 29 for the $z\sim 2$ and $z\sim 3$ samples,
respectively) spread throughout the sky, we assume that the average
LFs are representative of $1.9\la z \la 3.4$ galaxies.  Any remaining
uncertainty in normalization of the average LF (i.e., the uncertainty
reflected in the field-to-field fractional dispersion shown in
Figure~\ref{fig:normdisp}) is added in quadrature with Poisson
counting error (shown as open circles in Figure~\ref{fig:normdisp}) to
determine the total error bars shown in Figure~\ref{fig:rlfg}.  We
remind the reader that the {\it systematic} effects of photometric
bias and Ly$\alpha$ perturbations are already reflected in the derived
LFs.

\subsection{Effect of Photometric Redshifts}
\label{sec:photoz}

In light of recent literature regarding photometric estimates of the
UV LFs at redshifts $z\sim 2-3$ (e.g., \citealt{gabasch04}), it is
worthwhile to briefly examine how our derived LFs would change in the
absence of spectroscopic information, instead relying on photometric
redshifts ($z_{\rm phot}$).  We derived the photometric redshift
error, defined as
\begin{eqnarray}
\Delta z \equiv z_{\rm phot} - z_{\rm spec},
\end{eqnarray}
for a sample of $925$ star-forming galaxies with spectroscopic
redshifts $1.4<z_{\rm spec}<3.5$ that lie in fields with sufficient
multi-wavelength data to warrant SED analysis.  This is by far the
largest spectroscopic sample at these redshifts, and it enables us to
investigate how $\Delta z$ varies as a function of both redshift and
magnitude since, in principle, the error will depend both on the
relative placement of spectral breaks across the photometric filters
(i.e., the redshift of the galaxy) and on the quality of the
photometry and significance of the detection (i.e., the apparent
magnitude of the galaxy).

\begin{figure*}[hbt]
\plottwo{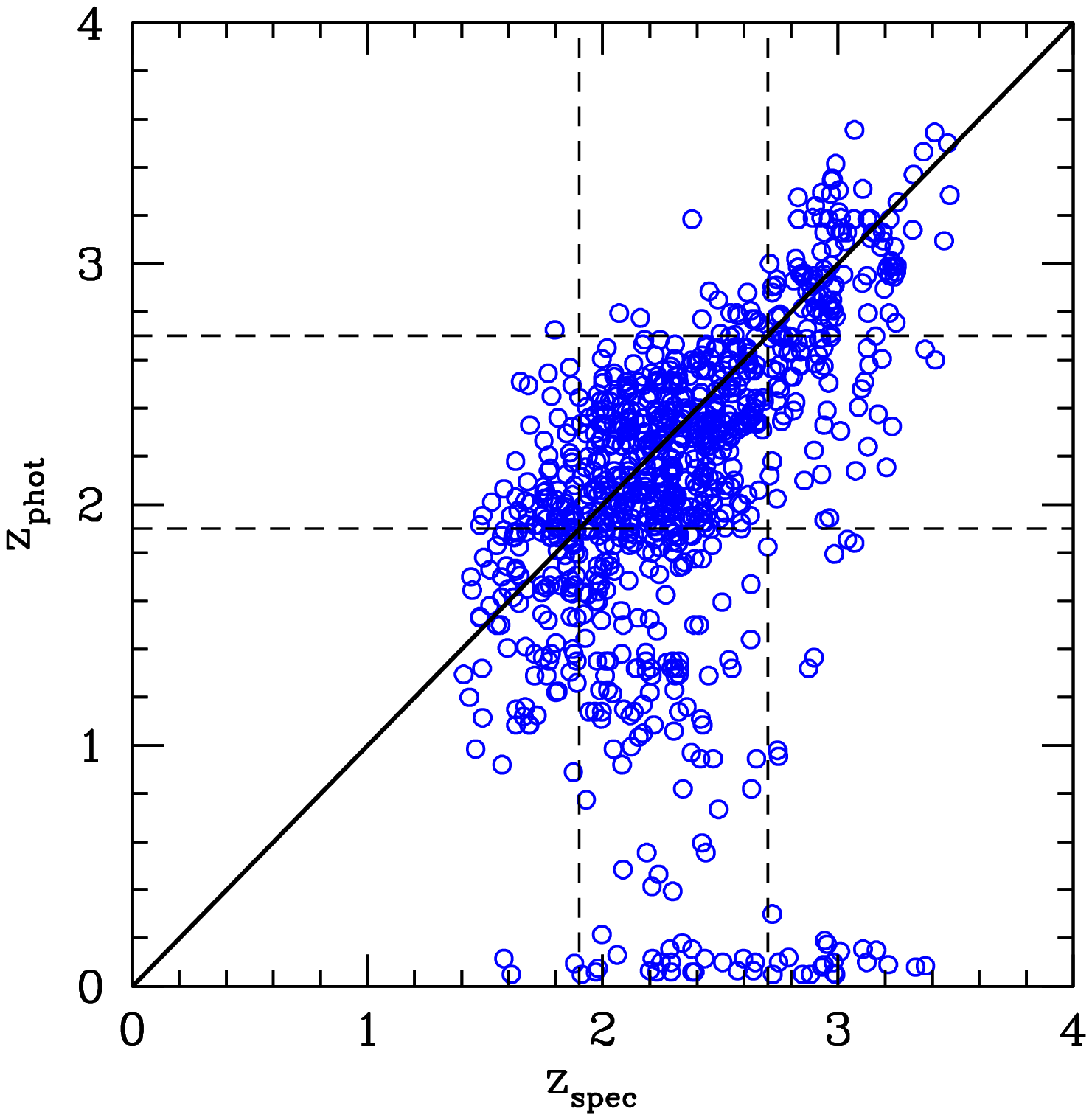}{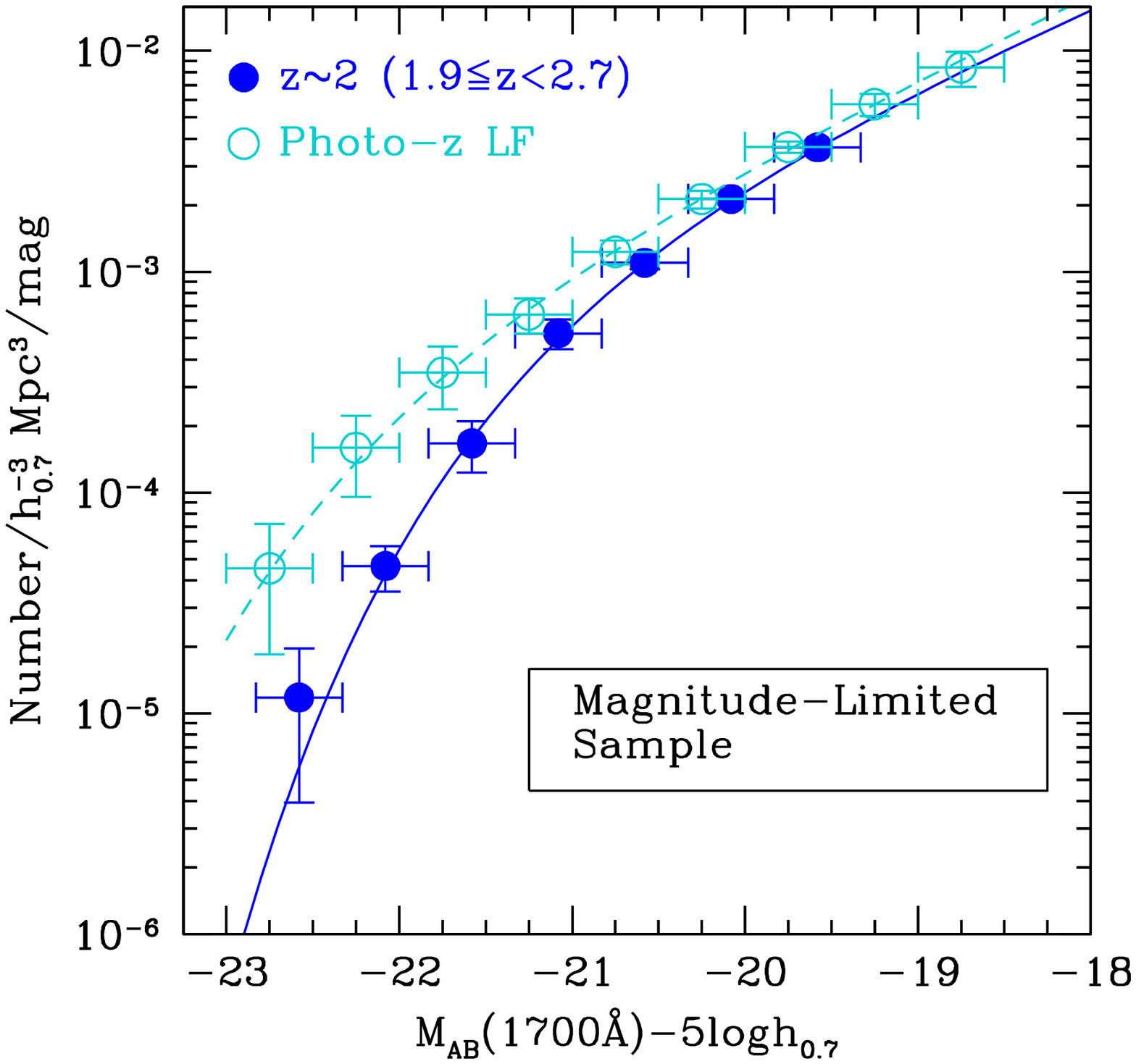}
\caption{({\it Left:}) Comparison between the photometric redshifts derived
using HyperZ \citep{bolzonella00} and spectroscopic redshifts from our
ground-based survey for a sample of $925$ star-forming galaxies.  The solid
line denotes the unity relationship, and dashed lines demarcate the region
over which the UV LF is computed.  ({\it Right:}) Comparison of the UV LF
derived using the photometric redshift distribution in the left panel
(open circles) with our spectroscopic determination (solid circles).
\label{fig:photoz}}
\end{figure*}

\begin{deluxetable*}{lcc}
\tabletypesize{\footnotesize}
\tablewidth{0pc}
\tablecaption{Biases and Dispersions in Photometric Redshift Errors ($\Delta z = z_{\rm phot}-z_{\rm spec}$) for Star-Forming Galaxies at $1.4\le z\le 3.5$\tablenotemark{a}}
\tablehead{
\colhead{} &
\colhead{All Galaxies} &
\colhead{Excluding $z_{\rm phot} \le 0.6$ Galaxies} \\
\colhead{Redshift-Range} &
\colhead{$M>-20.97$\,\,\,\,\,$M\le -20.97$} &
\colhead{$M>-20.97$\,\,\,\,\,$M\le -20.97$}}
\startdata
$1.4\le z<1.9$ & $0.01\pm0.40$\,\,\,\,\,$0.07\pm0.48$ & $0.04\pm0.35$\,\,\,\,\,$0.07\pm0.33$ \\
$1.9\le z<2.7$ & $-0.26\pm0.58$\,\,\,\,\,$-0.28\pm0.62$ & $-0.18\pm0.42$\,\,\,\,\,$-0.14\pm0.40$ \\
$2.7\le z<3.5$ & $-0.50\pm0.93$\,\,\,\,\,$-0.43\pm0.86$ & $-0.20\pm0.42$\,\,\,\,\,$-0.21\pm0.39$ \\
\enddata
\tablenotetext{a}{$\Delta z \equiv z_{\rm phot}-z_{\rm spec}$.}
\label{tab:photoz}
\end{deluxetable*}

Photometric redshifts were estimated using the HyperZ code of
\citet{bolzonella00}.  We only considered galaxies with detections in
at least the following bands: $G$,$\rs$, and $\ks$.  At least half of
the resulting $925$ objects also have detections in either the
$J$-band and/or {\it Spitzer} IRAC bands.  All but 52 of the 925
objects are detected at $U_{\rm n}$; the remaining 52 are all ``C''
candidates (\S~\ref{sec:colorselection}) at redshifts $z>2.7$.  We
considered a variety of star formation histories, reddening, and
redshifts when fitting the data using HyperZ.  Figure~\ref{fig:photoz}
compares the photometric and spectroscopic redshifts for the sample of
925 galaxies.  The biases and dispersions in photometric redshifts for
galaxies fainter and brighter than M$^{\ast}=-20.97$ in different
redshift ranges are listed in Table~\ref{tab:photoz}, both including
and excluding catastrophic outliers with $z_{\rm phot} < 0.6$.  Even
excluding outliers with $z_{\rm phot} < 0.6$ results in significant
redshift error dispersions of $\sigma(\Delta z)\sim 0.33-0.42$.
Further, in all cases over the redshift ranges where we compute the
LFs, $1.9\le z<2.7$ and $2.7\le z<3.4$, we find significant
photometric biases ranging from $\Delta z \sim 0.2-0.5$, in the sense
that $z_{\rm phot}$ is systematically under-estimated (i.e.,
luminosity is over-estimated).

A simulation was constructed to examine the effect of these biases and
dispersions on the LF, similar to the method presented in
\citet{marchesini07}, but modified to (a) allow for many realizations
of the intrinsic LF and (b) account for photometric redshift errors
using the empirical data of Figure~\ref{fig:photoz} and
Table~\ref{tab:photoz}.  To accomplish this, we first generated many
realizations of the UV LF as allowed by the (normally-distributed)
errors of the spectroscopically determined LF.  We then drew
magnitudes randomly from a Schechter distribution determined by
fitting a Schechter function to each of these realizations of the
intrinsic LF.  Since the probability of a galaxy lying at redshift
$z$,
\begin{eqnarray}
p(z) \propto {{dV}\over{dz}} \propto {{d_{\rm L}^{2}(z)}\over{(1+z)^{2}}}{1 \over {\sqrt{\Omega_{\Lambda}+\Omega_{\rm M}(1+z)^{3}}}},
\end{eqnarray}
is roughly constant over the redshift interval $1.4\le z\le 3.5$, we
drew redshifts from a uniform distribution.  The result is a list of
simulated redshift and magnitude pairs, $(z_{\rm spec},M)$, for
galaxies.  A photometric redshift was assigned to each galaxy by
randomly drawing a redshift from the distribution of $z_{\rm phot}$
(in left panel of Figure~\ref{fig:photoz}) within a box of width
$\delta z_{\rm spec}=0.4$ centered at $z_{\rm spec}$, using the
$z_{\rm phot}$ distribution for galaxies either brighter or fainter
than $M^{\ast}=-20.97$ depending on the magnitude $M$ of the simulated
galaxy.  The absolute magnitude of the galaxy was then recomputed
assuming the photometric redshift.  We then reconstructed the LF at
$1.9\le z<2.7$ for each realization assuming the photometric
redshifts.  The average LF from these many realizations is indicated
by the open circles in the right panel of Figure~\ref{fig:photoz}.
Because there are more galaxies scattered into, rather than out of,
the redshift range $1.9\le z<2.7$, the net result is that we would
have over-estimated the intrinsic LF had we relied on photometric
redshifts.  While the difference in the photometric and spectroscopic
LFs on the faint-end is small, it becomes quite significant for
galaxies brighter than $M^{\ast}$.  This systematic difference arises
from the fact that the change in absolute magnitude ($\Delta M$) for a
fixed $\Delta z$ and apparent magnitude will be larger for galaxies
scattered from low to high redshift than for galaxies scattered from
high to low redshift.  For example, a galaxy at redshift $z_{\rm
spec}=1.6$ scattering to redshift $z_{\rm phot}=2.0$, implying $\vert
\Delta z\vert =0.4$, results in $\Delta M\approx 0.44$.  However, a
galaxy at redshift $z_{\rm spec}=3.0$ scattering to redshift $z_{\rm
phot}=2.6$ (i.e., the same $\vert \Delta z\vert$ as above) results in
$\Delta M\approx 0.26$.  The net effect is that the bright-end of the
LF is systematically inflated with respect to the faint-end.

There are three further issues to note.  First, it has become common
in the literature to estimate photometric redshift errors independent
of fitting the stellar populations of galaxies by simply shifting
prescribed fixed templates until a best-fit redshift is reached.
Redshift errors derived in this manner will underestimate the true
error in redshift obtained by marginalizing over the uncertainties of
fitting those templates to the broadband photometry.  Second, the
simulation performed here benefited from the {\it a priori} knowledge
that all the galaxies truly lie at the correct redshifts $1.9\le
z<2.7$.  Photometric redshift scatter (e.g., Figure~\ref{fig:photoz})
will generally be larger than that presented here since there will
undoubtedly be some very low redshift galaxies ($z<1.4$) that are
scattered into the range $1.9\le z<2.7$.  Third, we remind the reader
that the photometric redshift errors derived here are for
optically-bright ($\rs<25.5$) objects with spectroscopic redshifts.
It is likely that the photometric redshift errors will be larger than
assumed here for very faint galaxies where the photometric
uncertainties may be larger.  This, in turn, may bias the faint-end
slope more severely than reflected in our simulations.  We stress that
the results of the photometric redshift simulation presented here
(Figure~\ref{fig:photoz}) are {\it unique} to our sample.  As a
result, the biases in the LF may be different for surveys that
incorporate a different number of photometric filters with differing
photometric data quality, although the photometric redshift accuracy
found here is similar to that presented in \citet{shapley05} and
\citet{reddy06a} using more (different) bands.  In any case, this
section illustrates how photometric redshifts can induce non-trivial
biases in the LF.

\subsection{UV LF Summary}

To summarize, this section has focused on our measurements of the UV
LF at $z\sim 2$ and $z\sim 3$.  Our method for computing the LFs takes
into account a number of systematic effects including contamination
from low redshift interlopers and AGN, Ly$\alpha$ line perturbations
to the observed colors of galaxies, and photometric scatter.  A large
number of independent fields allows us to control for sample variance.
Further, spectroscopic redshifts enable us to precisely correct for
the effect of IGM opacity on the rest-UV colors.  Our method for
computing the LFs uses a maximum-likelihood technique to account for
the systematic scattering of galaxies in parameter (e.g., luminosity,
reddening, and redshift) space.  Given this detailed treatment, we
consider the UV LF at $z\sim 2$ and $z\sim 3$ derived here to be the
most robust measurements yet.  Comparison of our UV LF with the
(corrected) determination from a magnitude limited sample suggests
that our determination must be reasonably complete for galaxies with
$\rs<25.5$ (see \S~\ref{sec:vvds}).

In the following sections, we will discuss how we can combine our
determinations of the UV LFs at $z\sim 2-3$ with what we know about
the extinction properties of high redshift galaxies to infer LFs at
other wavelengths.  We will primarily focus on inferences of the IR
and bolometric LFs at $z\sim 2-3$, but also present H$\alpha$ LFs at
similar redshifts, the latter of which may be useful for current and
future emission line studies.

\section{RESULTS: REST-FRAME $8$~$\mu$m, INFRARED, and BOLOMETRIC LUMINOSITY FUNCTIONS}
\label{sec:irlf}

As suggested in the previous analysis, correcting the rest-UV LF for
the effects of dust extinction is a key component in recovering the
star formation rate density.  Aside from our knowledge of the $\ebmv$
distribution at high redshift (\S~\ref{sec:ebmv}), extensive
multi-wavelength data enable us to independently determine the
extinction corrections relevant for the same sample of
spectroscopically confirmed galaxies that we used to compute the UV
LF.

Before the advent of panchromatic galaxy surveys, it was common to
simply apply an average correction for extinction, typically a factor
of $4-5$ \citep{steidel99}.  Subsequently, extensive multi-wavelength
data have placed our extinction corrections on a much more solid
footing.  For instance, initial X-ray and radio stacking analyses
(e.g., \citealt{nandra02, reddy04}) indicated that high redshift
UV-selected populations with $\rs<25.5$ have average obscuration
factors ($L_{\rm IR}/L_{\rm UV}$) around $4-5$, supporting the average
correction advocated by \citet{steidel99}.  Further progress was made
by taking advantage of the unique sensitivity of the {\it Spitzer}
MIPS instrument, allowing us to directly detect for the first time the
dust emission from $L^{\ast}$ galaxies at $z\ga 1.5$ \citep{reddy06a}.
The \citet{reddy06a} analysis confirmed the {\it average} trends
established by previous X-ray stacking studies, and further
demonstrated that moderate luminosity galaxies ($10^{11}\la L_{\rm
bol}\la 10^{12.3}$) at $z\sim 2$ follow the \citet{meurer99}
attenuation law found for local UV-selected starburst galaxies.  The
importance of this analysis for the present study is that we can
directly relate the $\ebmv$ distribution of most $z\sim 2$ galaxies
(\S~\ref{sec:ebmv}) with their distribution in obscuration,
$L_{\rm IR}/L_{\rm UV}$.  

In this section, we present our constraints on the $8$~$\mu$m,
infrared, and bolometric luminosity functions of $z\sim 2-3$ galaxies,
as derived from our UV LFs and the known extinction properties of
galaxies at these redshifts.  We present IR LFs based on two different
recipes (our $\ebmv$ distribution and the distribution of $24$~$\mu$m
fluxes) for evaluating the dust attenuation of galaxies.  In
\S~\ref{sec:bolmeas}, we combine our UV LF with the $\ebmv$
distribution to infer the IR LF.  In subsequent sections, we combine
our UV LF with the observed $24$~$\mu$m properties of galaxies to
infer the IR LF.  The two methods are compared in detail in
\S~\ref{sec:pahirlf}.

\subsection{Extinction-Corrected Measures of the Luminosity Function}
\label{sec:bolmeas}

As a first step, we can use the \citet{meurer99} relation to recover the
dust-corrected LF.  The method proceeded with the following steps:

1.  We first generated many realizations of the maximum-likelihood UV
LF and $\ebmv$ distribution at $z\sim 2$, assuming normal LF and
$\ebmv$ distribution errors.  We randomly chose an LF and $\ebmv$
distribution from these many realizations, to create an LF/$\ebmv$
pair, $\{{\cal L},{\cal E}\}$.

2.  Because the LF and $\ebmv$ distributions do not change
significantly over the redshift range $1.9\le z<2.7$ and because the
$\ebmv$ distribution is insensitive to absolute magnitude down to our
spectroscopic limit (\S~\ref{sec:ebmvtor25}), we can assume that the
{\it intrinsic} redshift $z$, magnitude $M$, and reddening $\ebmv$ of
a galaxy are independent variables.  Redshifts were drawn randomly
from a uniform distribution.  Magnitudes were drawn from the range
$-23\la M(1700\AA) \la -15.5$ according to a Schechter distribution
that described the luminosity function ${\cal L}$ from the $\{{\cal
  L},{\cal E}\}$ pair.  The faint limit of $M(1700\AA)=-15.5$
corresponds to an unobscured SFR of $\sim 0.1$~M$_{\odot}$~yr$^{-1}$
using the \citet{kennicutt98} calibration.  We drew galaxies down to
this low limit of unobscured luminosity because such galaxies can be
scattered to bins of higher luminosity after correcting for
extinction.  Similarly, $\ebmv$ values were drawn randomly from the
$\ebmv$ distribution ${\cal E}$ from the $\{{\cal L},{\cal E}\}$ pair,
excluding negative $\ebmv$ values that reflect unphysical reddening
values.  The result is a list of galaxies associated with a triplet
($z$,$M$,$\ebmv$).

3. The rest-frame $1700$~\AA\, specific luminosity of each galaxy is
calculated as
\begin{eqnarray}
L_{\rm 1700} = {{4\pi d^{2}_{\rm L}}\over{(1+z)}} 10^{-0.4(48.60+m_{\rm 1700})},
\end{eqnarray}
where $d_{\rm L}$ is the luminosity distance at redshift $z$ and
$m_{\rm 1700}$ is the apparent magnitude of the galaxy with absolute
magnitude $M$ at redshift $z$ (Eq.~\ref{eq:absappmag}).  We then
calculate $\nu L_{\nu}$ at $1700$~\AA\, to yield the UV luminosity.
The $\ebmv$ for the galaxy is used in conjunction with the
\citet{calzetti00} relation to derive the dust-corrected UV luminosity.

4. To determine the IR luminosity corresponding to this dust-corrected
UV luminosity, we assumed that the UV and IR emission are tied
directly to the SFR of the galaxy.  The IR luminosity is assumed to be
the luminosity which, when added to the unobscured UV luminosity,
yields the same SFR that would have been obtained from the
dust-corrected UV luminosity, assuming the \citet{kennicutt98}
relations.  These IR luminosities are then perturbed by a normal
distribution with sigma of $0.3$~dex to account for the dispersion
between dust-corrected UV and IR luminosity (or, alternatively, the
dispersion between $\ebmv$ and IR luminosity; e.g.,
\citealt{meurer99}).

5. These IR luminosities are then binned to produce an IR LF.  This is
the IR LF corresponding to the $\{{\cal L},{\cal E}\}$ pair selected
in step (1).

Steps (1)-(5) are repeated many times, each time randomly drawing
different $\{{\cal L},{\cal E}\}$ pairs.  Aside from uncertainties in
the rest-frame UV faint-end slope, there are two other systematics
that can bias our determination of the IR LF: (1) a change in the faint-end
slope of the rest-frame UV LF and (2) a change in the attenuation of
UV-faint galaxies.  We now discuss these two systematic effects in detail.

First, we must determine how changing the number density of such faint
objects, determined by the faint-end slope $\alpha$, affects the IR
LF.  In principle, we could simply fix $\alpha=-1.6$ when fitting the
different realizations of the UV LF at $z\sim 2$ and compare with the
results obtained by allowing $\alpha$ to vary freely in the Schechter
fits to the realizations.  However, this method will cause us to
underestimate the errors on the faint-end of the IR LF.  To obtain a
truer estimate at the faint-end, we allowed $\alpha$ to vary freely
around a normal distribution with a mean of $\langle \alpha\rangle
=-1.6$ and standard deviation of $\sigma(\alpha) = 0.11$, similar to
the dispersion in $\alpha$ that we measure when fitting the UV LF
(Figure~\ref{fig:rlfg} and Table~\ref{tab:schechparms}).  For
simplicity, we assume that $\alpha$ varies according to
$\langle\alpha\rangle = -1.6\pm0.11$ at $z\sim 2$ in the subsequent
discussion.

A second systematic effect that can bias the determination of
the IR LF is the distribution of extinction among $\rs > 25.5$
galaxies.  For the calculation of the IR LF from $\ebmv$ we considered
two cases.  In the first case, we assume that the $\ebmv$ distribution
is constant to arbitrarily faint rest-UV magnitudes.  In the second
case, we assume that the $\ebmv$ distribution is constant to
$\rs=25.5$ (e.g., Figure~\ref{fig:irxvr} and \S~\ref{sec:ebmvtor25}),
but then suddenly changes to have a mean $\langle \ebmv \rangle =
0.04$ (with same dispersion) for galaxies fainter than $\rs=25.5$ (we
refer to this second case as a discontinuous $\ebmv$ distribution).
We have assumed this value of $\ebmv=0.04$ because it is similar to
the $\ebmv$ observed for very faint ($\la 0.1$~L$^{\ast}$) UV-selected
galaxies inferred from dropout samples at higher redshifts (Bouwens
et~al.  2007, submitted).  Because $\rs=25.5$ is an arbitrary limit
dictated by efficient spectroscopic followup, it is highly unlikely
that the $\ebmv$ distribution will suddenly change fainter than this
limit.  Rather, the distribution is likely to gradually fall towards
lower $\ebmv$, or bluer rest-frame continuum spectral slopes
($\beta$), proceeding to fainter galaxies, assuming that such fainter
galaxies have lower star formation rates and are less dusty than
$\rs<25.5$ galaxies.  Therefore, the true $\ebmv$ distribution will
very likely fall between the two extremes assumed above.  We will
return to this point shortly.\footnote{We make the reasonable
assumption that very dusty ULIRGs with faint UV luminosities make up a
small fraction of the total number of UV-faint galaxies.  Assuming
otherwise would imply a significantly larger number of IR luminous
galaxies than are presently observed in shallow IR surveys (e.g.,
\citealt{perez05,caputi07}).}

\begin{figure}[hbt]
\plotone{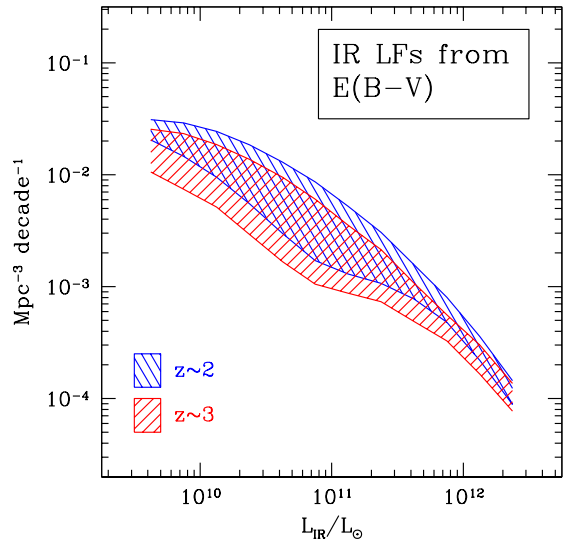}
\caption{Infrared luminosity functions at $z\sim 2$ and $z\sim 3$,
calculated assuming the \citet{meurer99} and \citet{kennicutt98}
relations to convert unobscured UV luminosity and $\ebmv$ to infrared
luminosities.  The width of the shaded regions reflect the uncertainty
in the rest-frame UV-slope and the attenuation distribution for $\rs>25.5$
galaxies (see text).
\label{fig:calzlf}}
\end{figure}

\begin{deluxetable*}{lccc}
\tabletypesize{\footnotesize}
\tablewidth{0pc}
\tablecaption{$8$~$\mu$m, IR, and Bolometric Luminosity Functions of $1.9\la z\la 3.4$ Galaxies\tablenotemark{a}}
\tablehead{
\colhead{} &
\colhead{} &
\colhead{$1.9\le z<2.7$} &
\colhead{$2.7\le z<3.4$ (Predicted)} \\
\colhead{} &
\colhead{} &
\colhead{(h$_{0.7}^{3}$~Mpc$^{-3}$~decade$^{-1}$)} &
\colhead{(h$_{0.7}^{3}$~Mpc$^{-3}$~decade$^{-1}$)}}
\startdata
$\log [\nu L_{\nu}(8\mu m)]$ & 8.25 --- 8.50 & $(3.01\pm1.81)\times 10^{-2}$ & $(2.36\pm1.79)\times 10^{-2}$ \\
& 8.50 --- 8.75 & $(3.08\pm1.92)\times 10^{-2}$ & $(2.25\pm1.69)\times 10^{-2}$ \\
& 8.75 --- 9.00 & $(2.73\pm1.70)\times 10^{-2}$ & $(1.89\pm1.40)\times 10^{-2}$ \\
& 9.00 --- 9.25 & $(2.20\pm1.39)\times 10^{-2}$ & $(1.45\pm1.07)\times 10^{-2}$ \\
& 9.25 --- 9.50 & $(1.59\pm1.02)\times 10^{-2}$ & $(1.02\pm0.72)\times 10^{-2}$ \\
& 9.50 --- 9.75 & $(1.10\pm0.69)\times 10^{-2}$ & $(6.55\pm4.50)\times 10^{-3}$ \\
& 9.75 --- 10.00 & $(7.26\pm4.21)\times 10^{-3}$ & $(4.13\pm2.38)\times 10^{-3}$ \\
& 10.00 --- 10.25 & $(4.47\pm2.31)\times 10^{-3}$ & $(2.64\pm1.33)\times 10^{-3}$ \\
& 10.25 --- 10.50 & $(2.72\pm1.08)\times 10^{-3}$ & $(1.60\pm0.55)\times 10^{-3}$ \\
& 10.50 --- 10.75 & $(1.54\pm0.54)\times 10^{-3}$ & $(8.71\pm2.37)\times 10^{-4}$ \\
& 10.75 --- 11.00 & $(7.75\pm1.92)\times 10^{-4}$ & $(4.62\pm1.02)\times 10^{-4}$ \\
& 11.00 --- 11.25 & $(3.39\pm0.71)\times 10^{-4}$ & $(2.09\pm0.48)\times 10^{-4}$ \\
\\
$\log L_{\rm IR}$\tablenotemark{b} & 9.50 --- 9.75 & $(2.68\pm1.67)\times 10^{-2}$ & $(1.89\pm1.40)\times 10^{-2}$ \\
& 9.75 --- 10.00 & $(2.24\pm1.40)\times 10^{-2}$ & $(1.50\pm1.11)\times 10^{-2}$ \\
& 10.00 --- 10.25 & $(1.69\pm1.09)\times 10^{-2}$ & $(1.09\pm0.78)\times 10^{-2}$ \\
& 10.25 --- 10.50 & $(1.21\pm0.78)\times 10^{-2}$ & $(7.49\pm5.30)\times 10^{-3}$ \\
& 10.50 --- 10.75 & $(8.23\pm4.93)\times 10^{-3}$ & $(4.78\pm3.03)\times 10^{-3}$ \\
& 10.75 --- 11.00 & $(5.52\pm3.01)\times 10^{-3}$ & $(3.16\pm1.68)\times 10^{-3}$ \\
& 11.00 --- 11.25 & $(3.40\pm1.61)\times 10^{-3}$ & $(2.01\pm0.91)\times 10^{-3}$ \\
& 11.25 --- 11.50 & $(2.13\pm0.76)\times 10^{-3}$ & $(1.23\pm0.40)\times 10^{-3}$ \\
& 11.50 --- 11.75 & $(1.18\pm0.38)\times 10^{-3}$ & $(6.86\pm1.86)\times 10^{-4}$ \\
& 11.75 --- 12.00 & $(5.94\pm1.31)\times 10^{-4}$ & $(3.58\pm0.58)\times 10^{-4}$ \\
& 12.00 --- 12.25 & $(2.88\pm0.59)\times 10^{-4}$ & $(1.69\pm0.44)\times 10^{-4}$ \\
& 12.25 --- 12.50 & $(1.10\pm0.34)\times 10^{-4}$  & $(6.55\pm2.37)\times 10^{-5}$ \\
\\
$\log L_{\rm bol}$\tablenotemark{b} & 9.50 --- 9.75 & $(3.69\pm1.78)\times 10^{-2}$ & $(2.57\pm1.60)\times 10^{-2}$ \\
& 9.75 --- 10.00 & $(3.23\pm1.27)\times 10^{-2}$ & $(2.06\pm1.15)\times 10^{-2}$ \\
& 10.00 --- 10.25 & $(2.47\pm0.96)\times 10^{-2}$ & $(1.49\pm0.82)\times 10^{-2}$ \\
& 10.25 --- 10.50 & $(1.61\pm0.83)\times 10^{-2}$ & $(9.53\pm5.88)\times 10^{-3}$ \\
& 10.50 --- 10.75 & $(1.04\pm0.57)\times 10^{-2}$ & $(6.00\pm3.61)\times 10^{-3}$ \\
& 10.75 --- 11.00 & $(6.89\pm3.41)\times 10^{-3}$ & $(3.90\pm1.86)\times 10^{-3}$ \\
& 11.00 --- 11.25 & $(4.25\pm1.84)\times 10^{-3}$ & $(2.49\pm1.05)\times 10^{-3}$ \\
& 11.25 --- 11.50 & $(2.51\pm0.80)\times 10^{-3}$ & $(1.48\pm0.45)\times 10^{-3}$ \\
& 11.50 --- 11.75 & $(1.37\pm0.42)\times 10^{-3}$ & $(7.81\pm1.98)\times 10^{-4}$ \\
& 11.75 --- 12.00 & $(6.67\pm1.32)\times 10^{-4}$ & $(4.05\pm0.74)\times 10^{-4}$ \\
& 12.00 --- 12.25 & $(3.12\pm0.65)\times 10^{-4}$ & $(1.85\pm0.52)\times 10^{-4}$ \\
& 12.25 --- 12.50 & $(1.17\pm0.38)\times 10^{-4}$  & $(7.67\pm2.12)\times 10^{-5}$ \\
\enddata
\label{tab:irlf}
\tablenotetext{a}{Errors include systematic uncertainty in attenuation distribution
for $\rs > 25.5$ galaxies, as described in the text.}
\tablenotetext{b}{The values listed in this table are derived assuming
the \citet{caputi07} calibration between $\nu L_{\nu}(8\mu m)$ and
L$_{\rm IR}$.}
\end{deluxetable*}

For now, our IR LFs estimated from the $\ebmv$ distributions are shown
in Figure~\ref{fig:calzlf} and tabulated in Table~\ref{tab:irlf}.  The
uncertainty in the LFs include uncertainty in the rest-frame UV
faint-end slope and the uncertainty in the $\ebmv$ distribution for
$\rs>25.5$ galaxies.  The upper limit of each LF corresponds to the
first case where $\ebmv$ is held constant.  The lower limit of each LF
corresponds to the second case where $\ebmv$ suddenly decreases to
have a mean of $\langle \ebmv\rangle = 0.04$ for galaxies fainter than
$\rs=25.5$.  In general, the systematic uncertainties related to a
changing attenuation distribution for UV-faint galaxies will dominate
the uncertainties in the faint-end slope (\S~\ref{sec:attfaint}).

\subsection{Distribution of Dust Attenuation Factors}
\label{sec:attendist}

As alluded to above, the distribution of rest-frame $5-8.5$~$\mu$m
luminosities ($L_{\rm 5-8.5\mu m}$) of $z\sim 2$ galaxies observed by
{\it Spitzer}/MIPS can be used to assess the infrared luminosity
function independent of any assumption regarding the relationship
between rest-frame UV slope and extinction, as per the previous
discussion.  To this end, we must quantify the distribution of dust
attenuation among $z\sim 2$ galaxies.  For the subsequent discussion,
we will define the mid-IR (${\cal A}_{\rm MIR}$), far-IR (${\cal
A}_{\rm IR}$), and bolometric attenuation (${\cal A}_{\rm bol}$)
factors as the ratio between $\nu L_{\nu}(8\mu m)$, $L_{\rm IR}$, and
$L_{\rm bol}$, respectively, and $L_{\rm 1700}$.  Following the
calibration of \citet{reddy06a}, we can relate these ${\cal A}$
factors to each other:
\begin{eqnarray}
{\cal A}_{\rm IR} & \approx  12.9\, {\cal A}_{\rm MIR}\nonumber \\
{\cal A}_{\rm bol} & \equiv  {{L_{\rm IR}+L_{\rm 1700}}\over{L_{\rm 1700}}} &  \equiv {\cal A}_{\rm IR}+1.
\label{eq:atteq}
\end{eqnarray}
Note that these attenuation factors, ${\cal A}$, are distinguished
from the rest-frame UV attenuation factor which is the ratio between
the dust-corrected and unobscured UV luminosities.

\begin{figure}[hbt]
\plotone{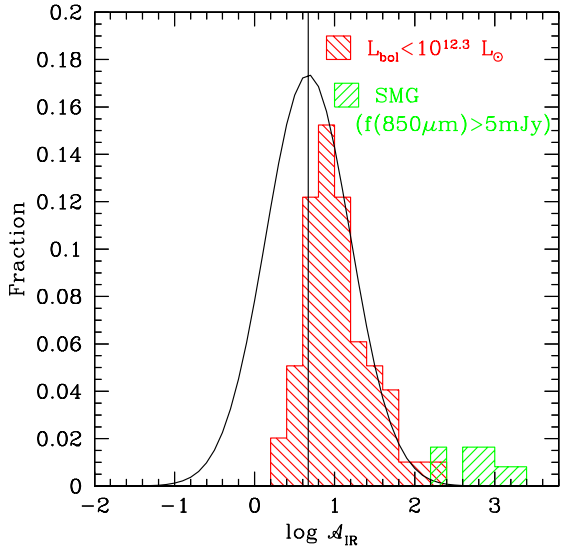}
\caption{Distribution of measured $\log {\cal A}_{\rm IR}$ for
$24$~$\mu$m detected galaxies with $\rs<25.5$ and luminosities $L_{\rm
bol}< 10^{12.3}$~L$_{\odot}$, indicated by the red hashed histogram.
The solid curve denotes the Gaussian fit to the inferred distribution
of all $\ugr$-selected galaxies with $\rs<25.5$, irrespective of
$24$~$\mu$m detection limit, and the vertical line indicates the mean
of the distribution, $\langle \log {\cal A}_{\rm IR}\rangle\approx
0.67$.  Log ${\cal A}_{\rm IR}$ for bright SMGs from the analysis of
\citet{reddy06a} is also shown.
\label{fig:att}}
\end{figure}

The normal distribution of $\ebmv$ for galaxies at $z\sim 2-3$ (e.g.,
Figures~\ref{fig:ebmvfig}, \ref{fig:ebmvcomp}) implies that the
attenuation factors ${\cal A}$ will abide by a log-normal
distribution.  We modeled the shape of the $\log {\cal A}$
distribution by considering the measured $\log {\cal A}$ of
rest-UV-selected galaxies with bolometric luminosities $L_{\rm
bol}<10^{12.3}$~L$_{\odot}$.  From \citet{reddy06a}, $\langle \log {\cal
A}_{\rm IR} \rangle \approx 0.67$, implying $L_{\rm IR}/L_{\rm
1700}\approx 4.7$, for the combined sample of $24$~$\mu$m detected and
undetected rest-UV-selected galaxies to $\rs=25.5$.  This mean
attenuation implies the Gaussian fit shown in Figure~\ref{fig:att},
compared with the distribution of $\log {\cal A}$ for $24$~$\mu$m
detected galaxies.

To construct a fair representation of the attenuation factors of high
redshift galaxies, there is another issue that is pertinent.  Namely,
the distribution above does not take into account the non-negligible
fraction of $z\sim 2-3$ galaxies that have attenuation factors much
larger, {\it on average}, than those of typical galaxies at these
redshifts (e.g., \citealt{chapman03,reddy05a,vandokkum03}).  Virtually
all of these galaxies have luminosities $L_{\rm bol} \ga
10^{12.3}$~L$_{\odot}$ \citep{reddy06a} and $\approx 50\%$ of those
that are also bright at submillimeter wavelengths, $f_{\rm 850\mu
m}\ga 5$~mJy, also satisfy the BX/LBG criteria.  Because our data are
most sensitive to galaxies with luminosities $L_{\rm bol}\la
10^{12.3}$~L$_{\odot}$ and because most galaxies with the largest
attenuation factors have $L_{\rm bol} \ga 10^{12.3}$~L$_{\odot}$ , we
will only consider the $L_{\rm bol} < 10^{12}$~L$_{\odot}$ regime when
computing the IR LF.  We combine our spectroscopically constrained
estimate of the IR LF with higher luminosity data from the literature
in order to compute the total luminosity and star formation rate
densities in \S~\ref{sec:uvhalfevol}, \ref{sec:irevol}, and
\ref{sec:sfrd}.

\subsection{Attenuation of Rest-UV Faint Galaxies}
\label{sec:attfaint}

\begin{figure}[hbt]
\plotone{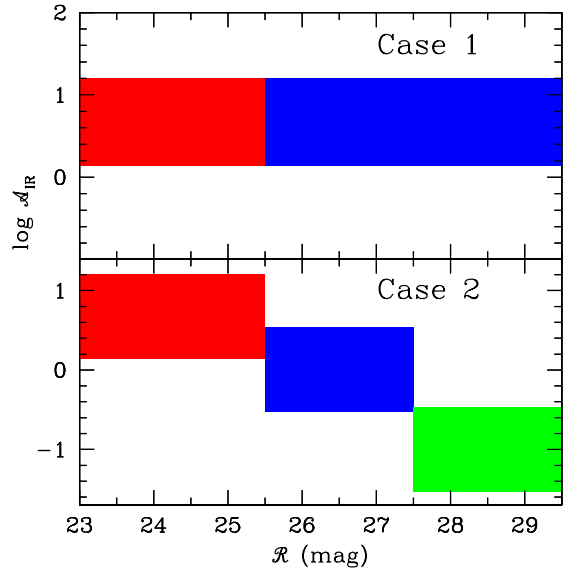}
\caption{Illustration of the two cases we consider for the attenuation
distribution of galaxies as a function of rest-frame UV magnitude.  In
both cases, the height of the bars reflect a $1$~$\sigma$ dispersion of
$0.53$~dex.
\label{fig:attfaintfig}}
\end{figure}

As discussed in \S~\ref{sec:bolmeas}, the attenuation distribution of
galaxies fainter than our $\rs$-band limit for spectroscopy can affect
significantly our inferences of the faint-end of the IR LFs.  To
remind the reader, in constructing the IR LFs that we would have
inferred based on the UV continuum slope, we considered two cases.  In
the first, the distribution of $\ebmv$ is held fixed to the
maximum-likelihood value (e.g., Figure~\ref{fig:ebmvcomp})
irrespective of optical magnitude.  In the second, we assumed the
maximum-likelihood value for those galaxies with $\rs < 25.5$; for
those fainter than this limit, we assumed a mean $\langle\ebmv\rangle
= 0.04$ (corresponding to $\beta \sim -2.0$ using the
\citealt{calzetti00} relation).  To construct the $8$~$\mu$m, IR, and
bolometric LFs based on MIPS $24$~$\mu$m data, we assumed two cases
similar to the ones considered above.  In the first, we assume that
all galaxies can be ascribed to the attenuation distribution shown in
Figure~\ref{fig:att}.  In the second, we assume that the attenuation
distribution shifts to a mean of $\langle \log {\cal A}_{\rm
IR}\rangle = 0$, with the same dispersion as before, for galaxies with
$25.5\le \rs < 27.5$, then shifts again to a lower mean of $\langle
\log {\cal A}_{\rm IR}\rangle = -1$ for galaxies fainter than
$\rs=27.5$.  The latter corresponds to an attenuation such that $90\%$
of the bolometric luminosity of the galaxy emerges in the UV.
Changing the distribution about the specified means does not
significantly affect our conclusions.  The two cases are illustrated
in Figure~\ref{fig:attfaintfig}.

It is very likely that the attenuation distribution does not remain
constant to arbitrarily faint UV magnitude, as is assumed in case 1.
Alternatively, because our $\rs=25.5$ spectroscopic limit is
arbitrary, we do not expect the attenuation properties of galaxies
fainter than this limit to drastically change.  It is more likely that
the attenuation distribution gradually shifts to lower mean values for
galaxies fainter in the UV.  Case 2 can therefore be considered a
reasonable lower extreme to the gradient of attenuation as a function
of rest-UV magnitude.  The true IR LF derived from a realistic
attenuation distribution will likely lie between the IR LFs derived
assuming case 1 and case 2.

\begin{figure*}[hbt]
\plottwo{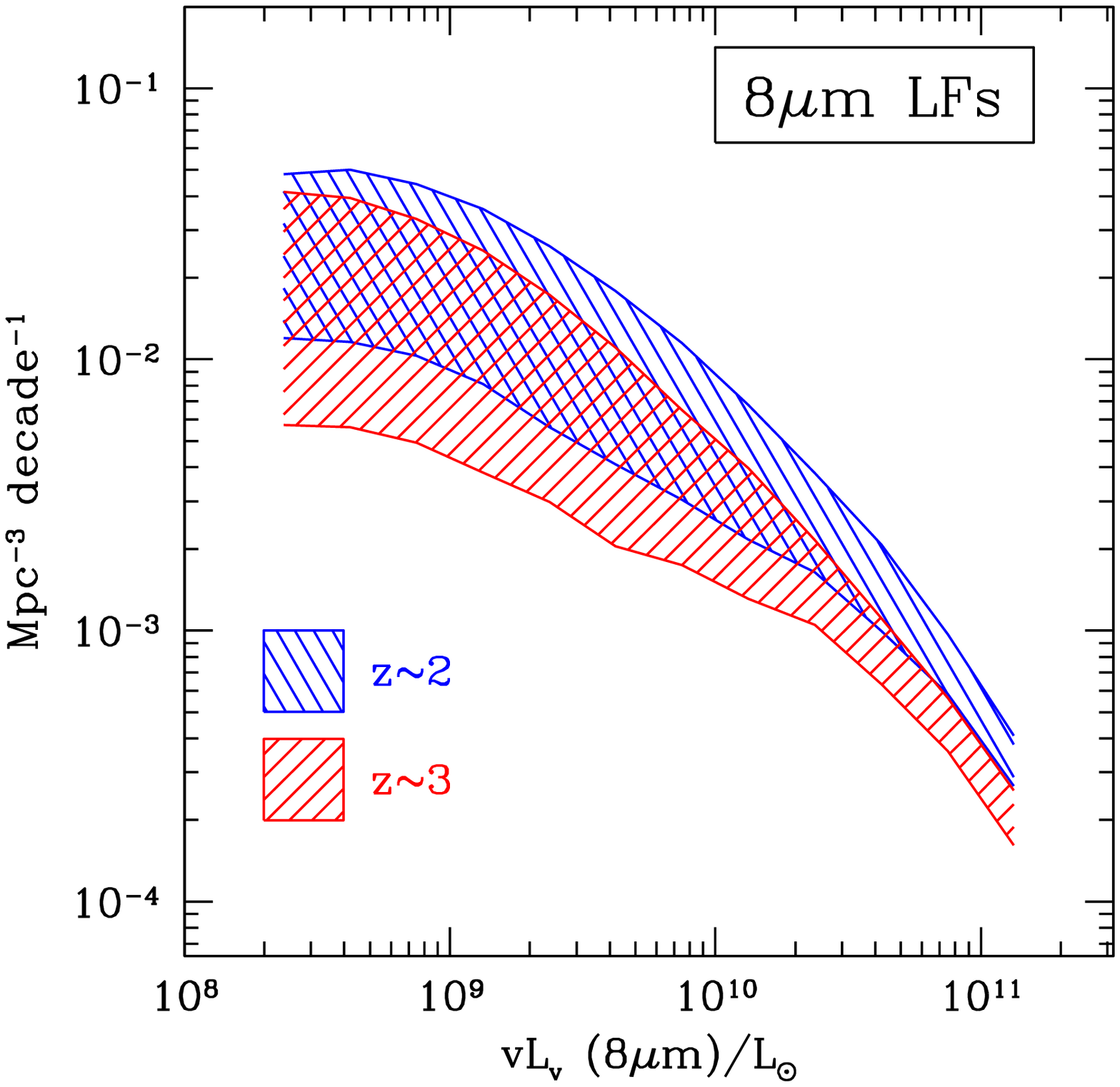}{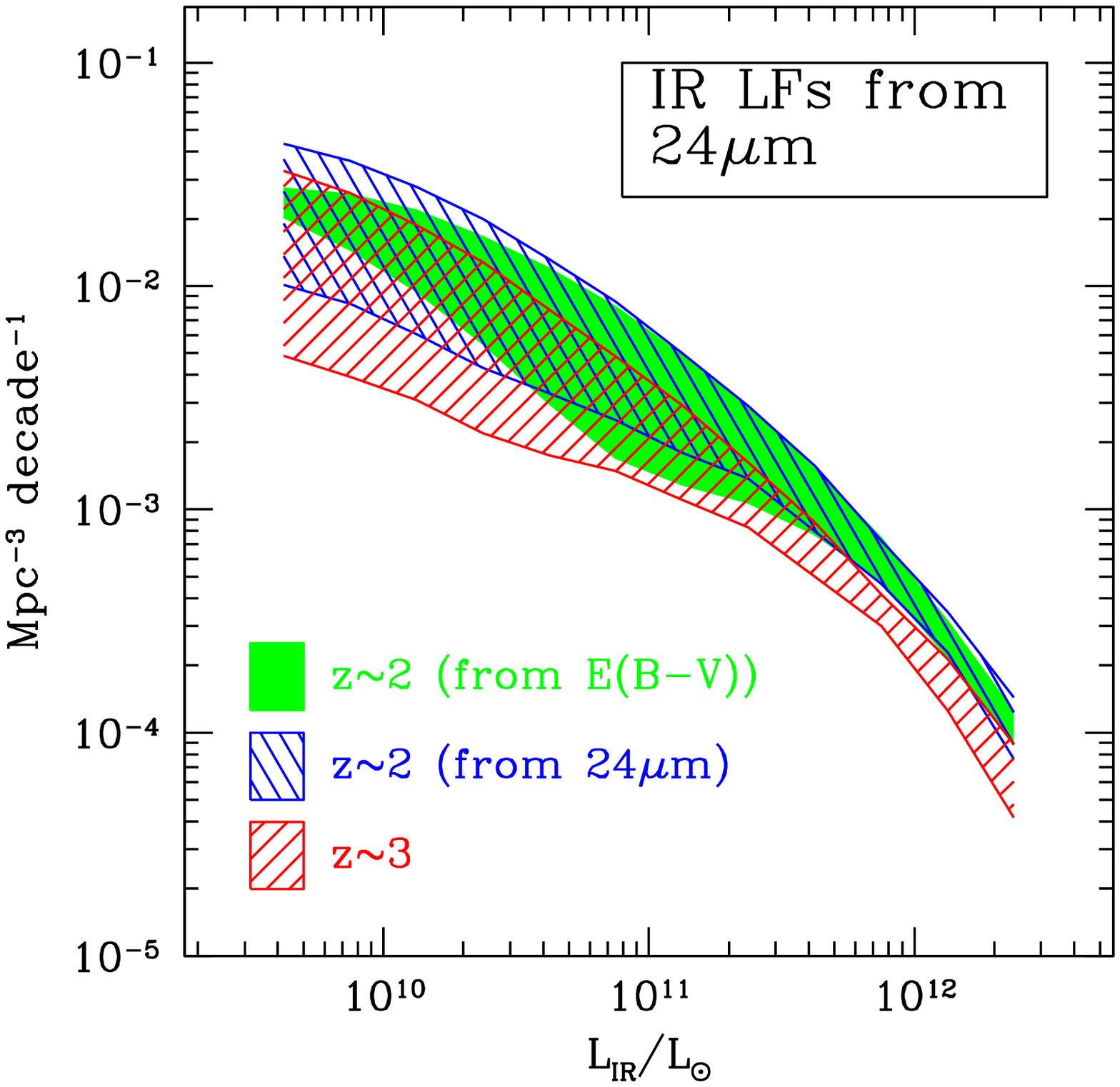}
\caption{$8$~$\mu$m and IR luminosity functions at redshift $1.9\le
z<2.7$, compared with predictions in the higher redshift range $2.7\le
z<3.4$.  The right panel also demonstrates the similarity in the IR LF
derived from the rest-frame UV slope (Figure~\ref{fig:calzlf}) with
that derived using the attenuation factors calculated from {\it
Spitzer}/MIPS observations of $1.5\la z\la 2.6$ galaxies in the
GOODS-N field \citep{reddy06a}
\label{fig:irlf}}
\end{figure*}

Is there any reason to believe that the attenuation has a much steeper
gradient than ``case 2'' (i.e., bottom panel of
Figure~\ref{fig:attfaintfig})?  Future deep stacking analyses and deep
spectroscopic campaigns should resolve this question.  For now we
point out that much more extreme cases for the attenuation of UV-faint
galaxies (i.e., such that they are even less obscured than what we
have considered in case 2) would result in {\it total} IR LD and SFRD
estimates that are comparable, if not lower, than estimates based on
samples of the brightest ($\rs<25.5$ and $\ks(Vega) < 22$) objects at
these redshifts, most of which have luminosities comparable to LIRGs
and ULIRGs \citep{reddy05a}.  Since the total IR LD and SFRD must be
{\it at least as large} as that contributed by LIRGs and ULIRGs, the
attenuation distribution of sub-ULIRG galaxies cannot be much lower
than what we have considered here.  We will return to this point in
\S~\ref{sec:sfrd}.

\subsection{Rest-Frame $8$~$\mu$m and IR Luminosity Functions}
\label{sec:pahirlf}

Combining our UV LFs with the attenuation distribution derived from
MIPS $24$~$\mu$m observations (using the same method described in
\S~\ref{sec:bolmeas}) results in estimates of the IR LFs.
Given the amount of focus in using MIPS $24$~$\mu$m observations to
probe star-forming populations at $z\sim 2$ (e.g., \citealt{caputi07,
reddy06a, papovich06}), it is useful to derive a rest-frame $8$~$\mu$m
luminosity function; we did this using the $8$~$\mu$m attenuation
factors observed for $z\sim 2$ galaxies \citep{reddy06a}.\footnote{For
ease of comparison with previous literature, we express the mid-IR
attenuation factor ${\cal A}_{\rm IR}$ in terms of $\nu L_{\nu}(8\mu
m)$ rather than $L_{\rm 5-8.5\mu m}$ as was used in \citet{reddy06a}.
The relationship between the two for the typical mid-IR SED of
star-forming galaxies is $\nu L_{\nu}(8\mu m) \approx 0.75 L_{\rm
5-8.5\mu m}$.}  We then use the relationship between ${\cal A}_{\rm
IR}$ and ${\cal A}_{\rm MIR}$ (Eq.~\ref{eq:atteq}) to infer the IR
luminosity function.

Our inference of the $8$~$\mu$m and IR LFs at $1.9\le z <2.7$ at faint
and moderate luminosities ($L_{\rm IR}\la 10^{12}$~L$_{\odot}$) are
shown in Figure~\ref{fig:irlf} and listed in Table~\ref{tab:irlf}.
For later comparison, we have assumed the relation between $8$~$\mu$m
and IR luminosity given by \citet{caputi07}.  The upper and lower
limits of the shaded regions in the figure correspond to the two
different cases of attenuation distributions discussed in the previous
section.  We cannot directly measure the rest-frame mid-IR
luminosities of galaxies at $2.7\le z<3.4$, but we show the predicted
$8$~$\mu$m and IR LFs at these redshifts assuming (a) the same
attenuation distribution and (b) the same relationship between mid-IR
and total IR luminosities found for $z\sim 2$ galaxies.
Figure~\ref{fig:irlf} demonstrates that the IR LF we would have
inferred at $10^{9.5}\la L_{\rm IR}\la 10^{12}$~L$_{\odot}$ from the
rest-frame UV slope, or $\ebmv$, is consistent with the one inferred
from the MIPS-determined attenuation factors of these galaxies.  This
similarity reflects the significant correlation between $\ebmv$ and
attenuation for galaxies with moderate luminosities, and such galaxies
are typical of the redshift $z\sim 2-3$ population (i.e., with
luminosities corresponding to $\sim L^{\ast}$).

Because our data are most sensitive to galaxies with $L_{\rm IR}\la
10^{12}$~L$_{\odot}$, we must incorporate direct measurements of the
IR LF for high luminosity objects, such as those from {\it Spitzer}
mid-IR surveys.  There are several published values of the IR LF for
ULIRGs; here we assume the most recent determination from
\citet{caputi07}.  Table~\ref{tab:contributions} lists the
contribution of galaxies in different luminosity ranges to the IR LD,
where we take the total LD to be that of galaxies with $L_{\rm
IR}>6\times 10^8$~L$_{\odot}$.  This limit corresponds to galaxies
with SFRs of $\sim 0.1$~M$_{\odot}$~yr$^{-1}$.  The total LD changes
negligibly by integrating to zero luminosity.
Table~\ref{tab:contributions} shows that --- despite the large
systematic uncertainties at the faint-end induced by variations in the
attenuation distribution --- a significant fraction of the IR LD at
$z\sim 2$ arises from galaxies with sub-ULIRG luminosities.  We will
return this point in \S~\ref{sec:irevol}.

\begin{deluxetable}{lc}
\tabletypesize{\footnotesize}
\tablewidth{0pc}
\tablecaption{Contributions of the IR LD at $z\sim 2$}
\tablehead{
\colhead{$L_{\rm IR}$} &
\colhead{$\log$ IR LD}}
\startdata
$10^{9}$ --- $10^{10}$~L$_{\odot}$ & $8.03\pm0.17$ ($9\%$) \\ 
$10^{10}$ --- $10^{11}$~L$_{\odot}$ & $8.45\pm0.09$ ($23\%$) \\ 
$10^{11}$ --- $10^{12}$~L$_{\odot}$ & $8.68\pm0.06$ ($39\%$) \\ 
$>10^{12}$~L$_{\odot}$ & $8.48\pm0.32$ ($25\%$)\tablenotemark{a} \\ 
{\bf Total ($6\times 10^8 <L_{\rm IR}<\infty$)\tablenotemark{b}:} & {\bf $9.09\pm0.08$} \\ 
\enddata
\tablenotetext{a}{From \citet{caputi07}.}
\tablenotetext{b}{The lower limit of $L_{\rm IR} = 6\times 10^8$~L$_{\odot}$ roughly corresponds
to an SFR of $0.1$~M$_{\odot}$~yr$^{-1}$.}
\label{tab:contributions}
\end{deluxetable}

\subsection{Bolometric Luminosity Functions}

\begin{figure}[hbt]
\plotone{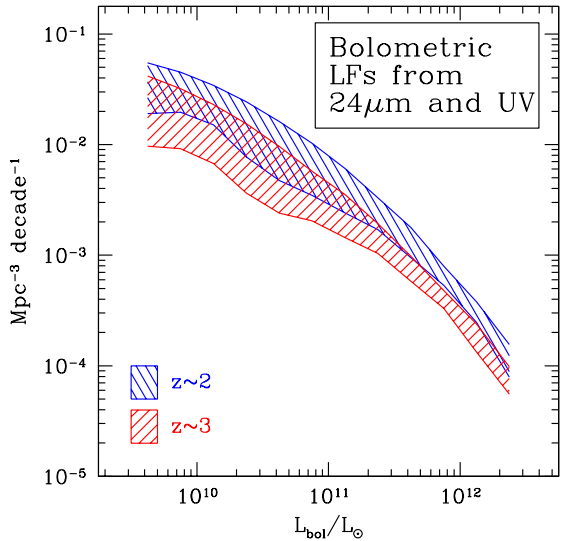}
\caption{The bolometric luminosity function at redshift $1.9\le z<2.7$,
calculated using the sum of the UV (unobscured) and IR
(obscured) luminosities of galaxies.  Our prediction of the bolometric
LF at $2.7\le z<3.4$ is also shown.
\label{fig:bollf}}
\end{figure}

Finally, to gain an accurate picture of the distribution of total
energetics of star-forming galaxies, we must consider the combined
contribution from unobscured (UV) luminosity and obscured (IR)
luminosity.  While the bolometric luminosity should closely follow the
infrared luminosity for luminous galaxies ($L_{\rm bol}\approx L_{\rm
IR}\ga 3\times 10^{11}$~L$_{\odot}$), \citet{reddy06a} show that such
an assumption is no longer valid for galaxies with $L_{\rm bol} \la
3\times 10^{11}$~L$_{\odot}$ (at $z\sim 2$), given the very tight
correlation between dust obscuration and bolometric luminosity.  For
example, a $10^{11}$~L$_{\odot}$ galaxy at $z\sim 2$ will on average
have half of its total luminosity emerging at UV wavelengths;
similarly, a $10^{10.5}$~L$_{\odot}$ galaxy at $z\sim 2$ will on
average have $84\%$ of its bolometric luminosity emerging at UV
wavelengths.\footnote{While the tight correlation between $L_{\rm
bol}$ and attenuation has been observed both locally and at high
redshift, the {\it normalization} of the relationship increases at
higher redshift.  This means, for example, that galaxies at $z\sim 2$
are on average 10 times more luminous for a given dust obscuration (or
are 10 times less dust obscured for a given $L_{\rm bol}$) than $z=0$
galaxies \citep{reddy06a}.  Hence, the fraction of total luminosity
emerging in the UV is larger at higher redshift than it is locally for
galaxies of a given bolometric luminosity.  Note that this observation
is still consistent with the finding that dustier systems dominate the
luminosity density at $z\sim 1-2$ relative to the present-day (e.g.,
\citealt{lefloch05, takeuchi05}), as we discuss in \S~\ref{sec:irevol}.}
Figure~\ref{fig:bollf} shows the bolometric luminosity functions of
star-forming galaxies at redshifts $1.9\le z<3.4$, and values are
listed in Table~\ref{tab:irlf}.  The bolometric LF is larger in all
luminosity bins considered than the IR LFs given that objects will
shift from lower to higher luminosity bins after accounting for the
emergent UV luminosity of high redshift galaxies.  We note that in
computing our prediction for the bolometric LF at $2.7\le z<3.4$, we
have assumed the same distribution of attenuation factors that was
found for $z\sim 2$ galaxies, as was done in computing the IR LFs.
The bolometric LFs are presented here because they give a true picture
as to the total energetic output of galaxies, irrespective of dust
extinction or the fraction of unobscured luminosity.  In section
\S~\ref{sec:sfrd}, we will discuss what the
spectroscopically-constrained bolometric LFs imply for the
contribution of moderate luminosity ($10^{11} \la L_{\rm bol} \la
10^{12}$~L$_{\odot}$) galaxies to the global luminosity density.

\break

\section{RESULTS: H$\alpha$ LUMINOSITY FUNCTION}
\label{sec:half}

We briefly discuss our derivation of the H$\alpha$ LFs here, as they
may be useful for current and future high redshift emission line
studies.  The key ingredient that allows us to convert our UV LF into
an estimate of the H$\alpha$ LF is the correlation between
dust-corrected UV and H$\alpha$ estimates of star formation rates.
\citet{erb06c} found a significant ($6.8$~$\sigma$) correlation with
0.3~dex scatter between the extinction corrected UV estimates and
H$\alpha$ estimates of the SFRs, assuming the \citet{calzetti00} relation,
for a sample of 114 rest-frame UV selected galaxies at $z\sim 2$.  We
can invert the relationship between extinction-corrected SFR and
H$\alpha$ line emission in order to infer the H$\alpha$ LF.  The
spectroscopic H$\alpha$ observations used to establish the correlation
between H$\alpha$ and UV-determined SFRs are described in detail in
\citet{erb06a, erb06c, erb06b}.

\subsection{Method}

\begin{figure*}[hbt]
\plottwo{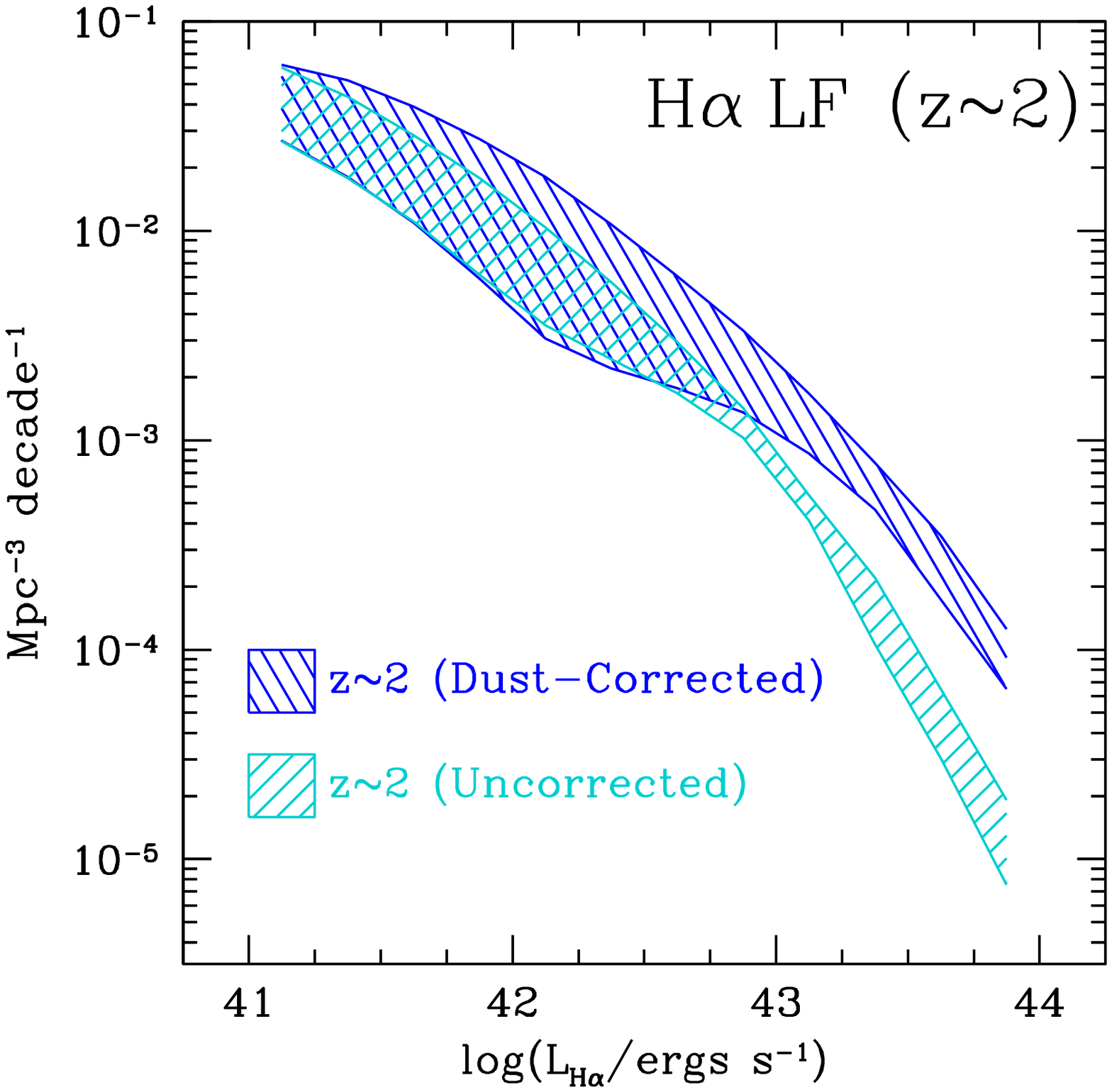}{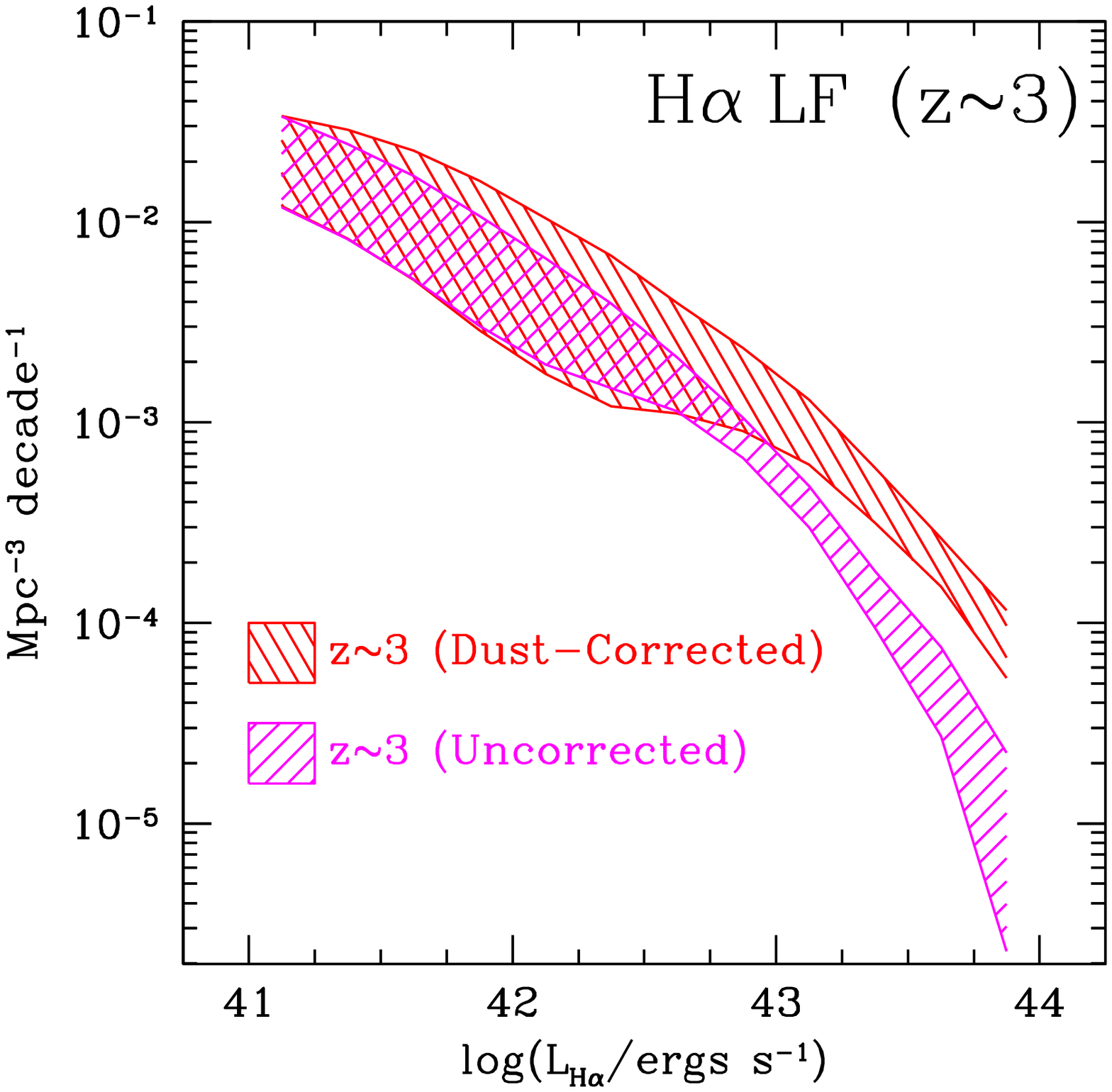}
\caption{Dust-corrected and observed (uncorrected for extinction)
H$\alpha$ LFs inferred for star-forming galaxies with redshifts
$1.9\le z<2.7$ (left) and $2.7\le z<3.4$ (right).  The uncertainty of
each LF, represented by the width of the shaded regions, is dictated
by the uncertainty in (1) the $\ebmv$ distribution for $\rs > 25.5$
galaxies and (2) the rest-frame UV faint-end slope.
Values are tabulated in Table~\ref{tab:half}.
\label{fig:half}}
\end{figure*}

The method used to estimate the H$\alpha$ LF of $z\sim 2$ galaxies is
analogous to that presented in \S~\ref{sec:bolmeas}.  We generated
many realizations of the LF and $\ebmv$ distributions and randomly
selected magnitudes and $\ebmv$.  To determine the H$\alpha$
luminosity corresponding to this dust-corrected UV luminosity, we
assumed that the UV and H$\alpha$ emission are tied directly to the
SFR of the galaxy, where the SFR is calibrated using the
\citet{kennicutt98} relations.  It is then easy to show that
\begin{eqnarray}
L_{\rm H\alpha}[{\rm ergs~s^{-1}}] \approx 1.77\times 10^{13}\,\,L_{\rm 1700}[{\rm ergs~s^{-1}~Hz^{-1}}].
\end{eqnarray}
The resulting H$\alpha$ luminosities are then perturbed by $0.3$~dex
to account for the dispersion in the relation between the
dust-corrected UV and H$\alpha$ estimates \citep{erb06c}.  For
consistency with previous determinations of the H$\alpha$ LF at lower
redshifts, it is useful to derive an H$\alpha$ luminosity function
{\it uncorrected} for extinction at $z\sim 2$.  To accomplish this, we
assume that the $\ebmv$ value, which is derived from the rest-frame UV
colors, reflects the nebular reddening of the galaxy (see also
\citet{erb06c}).  Applying the \citet{calzetti00} relation to the intrinsic
H$\alpha$ luminosity, and assuming the same value of $\ebmv$, yields
an estimate of the observed H$\alpha$
luminosity.\footnote{\citet{erb06c} show that assuming the same
$\ebmv$ in dust-correcting the H$\alpha$ estimates, as opposed to a
smaller nebular reddening as advocated by \citet{calzetti00}, results
in better agreement with dust-corrected UV and stacked X-ray
estimates.}  Note that because the attenuation $A_{\lambda}/\ebmv$ is
a factor of $\approx 3$ smaller at the wavelength of H$\alpha$,
$6563$~\AA, than at $1700$~\AA, it is not correct to simply convert
the observed UV luminosity to a SFR and then back to an observed
H$\alpha$ luminosity: one must take into account the differential
extinction, for a given $\ebmv$, between these two wavelengths.  The
dust-corrected (intrinsic) and uncorrected (observed) H$\alpha$
luminosities are then binned to produce a dust-corrected and observed
H$\alpha$ LF, respectively.  We considered the same two cases for the
$\ebmv$ distribution of UV-faint galaxies as in \S~\ref{sec:bolmeas}.
Our predictions for the uncorrected and dust-corrected H$\alpha$ LFs
at $z\sim 2$ are shown in Figure~\ref{fig:half} and listed in
Table~\ref{tab:half}.

\begin{deluxetable*}{lcc}
\tabletypesize{\footnotesize}
\tablewidth{0pc}
\tablecaption{H$\alpha$ Luminosity Functions of $1.9\la z\la 3.4$ Galaxies\tablenotemark{a}}
\tablehead{
\colhead{} &
\colhead{$1.9\le z < 2.7$} &
\colhead{$2.7\le z < 3.4$ (Predicted)} \\
\colhead{} &
\colhead{Observed $\phi$\,\,\,\,\,Dust-Corrected $\phi$} &
\colhead{Observed $\phi$\,\,\,\,\,Dust-Corrected $\phi$} \\
\colhead{$\log L(H\alpha/ergs~s^{-1})$} &
\colhead{(Mpc$^{-3}$~decade$^{-1}$)} &
\colhead{(Mpc$^{-3}$~decade$^{-1}$)}}
\startdata
41.00 --- 41.25 & $(4.34\pm1.66)\times 10^{-2}$\,\,\,\,\,$(4.44\pm1.75)\times 10^{-2}$ & $(1.72\pm0.55)\times 10^{-2}$\,\,\,\,\,$(2.27\pm1.09)\times 10^{-2}$ \\
41.25 --- 41.50 & $(3.07\pm1.29)\times 10^{-2}$\,\,\,\,\,$(3.52\pm1.71)\times 10^{-2}$ & $(1.33\pm0.48)\times 10^{-2}$\,\,\,\,\,$(1.63\pm0.81)\times 10^{-2}$ \\
41.50 --- 41.75 & $(1.98\pm0.88)\times 10^{-2}$\,\,\,\,\,$(2.51\pm1.41)\times 10^{-2}$ & $(9.18\pm3.83)\times 10^{-3}$\,\,\,\,\,$(1.11\pm0.59)\times 10^{-2}$ \\
41.75 --- 42.00 & $(1.20\pm0.59)\times 10^{-2}$\,\,\,\,\,$(1.67\pm1.07)\times 10^{-2}$ & $(6.07\pm2.70)\times 10^{-3}$\,\,\,\,\,$(6.87\pm3.85)\times 10^{-3}$ \\
42.00 --- 42.25 & $(6.95\pm3.42)\times 10^{-3}$\,\,\,\,\,$(1.05\pm0.75)\times 10^{-2}$ & $(3.94\pm1.26)\times 10^{-3}$\,\,\,\,\,$(4.25\pm2.31)\times 10^{-3}$ \\
42.25 --- 42.50 & $(4.07\pm1.62)\times 10^{-3}$\,\,\,\,\,$(6.48\pm4.28)\times 10^{-3}$ & $(2.44\pm0.82)\times 10^{-3}$\,\,\,\,\,$(2.70\pm1.22)\times 10^{-3}$ \\
42.50 --- 42.75 & $(2.31\pm0.63)\times 10^{-3}$\,\,\,\,\,$(3.93\pm2.16)\times 10^{-3}$ & $(1.55\pm0.33)\times 10^{-3}$\,\,\,\,\,$(1.63\pm0.49)\times 10^{-3}$ \\
42.75 --- 43.00 & $(1.23\pm0.20)\times 10^{-3}$\,\,\,\,\,$(2.35\pm0.99)\times 10^{-3}$ & $(7.63\pm2.21)\times 10^{-4}$\,\,\,\,\,$(8.60\pm1.94)\times 10^{-4}$ \\
43.00 --- 43.25 & $(4.80\pm0.66)\times 10^{-4}$\,\,\,\,\,$(1.27\pm0.40)\times 10^{-3}$ & $(3.30\pm0.93)\times 10^{-4}$\,\,\,\,\,$(3.91\pm0.90)\times 10^{-4}$ \\
43.25 --- 43.50 & $(1.63\pm0.58)\times 10^{-4}$\,\,\,\,\,$(6.22\pm1.57)\times 10^{-4}$ & $(1.38\pm0.91)\times 10^{-4}$\,\,\,\,\,$(1.38\pm0.43)\times 10^{-4}$ \\
43.50 --- 43.75 & $(4.74\pm1.71)\times 10^{-5}$\,\,\,\,\,$(2.61\pm0.88)\times 10^{-4}$ & $(3.44\pm2.23)\times 10^{-5}$\,\,\,\,\,$(5.13\pm2.41)\times 10^{-5}$ \\
43.75 --- 44.00 & $(1.34\pm0.58)\times 10^{-5}$\,\,\,\,\,$(9.53\pm3.03)\times 10^{-5}$ & $(2.11\pm1.89)\times 10^{-5}$\,\,\,\,\,$(1.24\pm1.01)\times 10^{-5}$ \\
\enddata
\tablenotetext{a}{Errors include the systematic uncertainty in the $\ebmv$ distribution for $\rs>25.5$ galaxies, as described in the text.}
\label{tab:half}
\end{deluxetable*}

\subsection{Predicted H$\alpha$ LFs at $z\sim 3$}

The correlation between UV and H$\alpha$ SFRs has only been tested
directly at redshifts $2.0\la z\la 2.6$, where the H$\alpha$ line
falls within the $K$-band, making it accessible to near-IR
spectrographs, such as Keck~II/NIRSPEC \citep{mclean98}.  It is
useful, nonetheless, to predict the form of the H$\alpha$ LF at $z\sim
3$ assuming that the correlation between UV and H$\alpha$ SFRs holds
at these higher redshifts.  This is a reasonable assumption to make
given that, to $\rs=25.5$, the $z\sim 2$ and $z\sim 3$ samples host
galaxies with a virtually identical range of $\ebmv$
(Figure~\ref{fig:ebmvcomp}), and galaxies in these respective samples
have similar average dust attenuation factors, represented as the
ratio of the dust-corrected UV and unobscured UV luminosities, of
$\sim 4-5$ based on X-ray and radio stacking analyses \citep{nandra02,
reddy04}.  Our predictions for the uncorrected and dust-corrected
H$\alpha$ LFs at $z\sim 3$, computed using the steps above and using
the combined ground-based and HDF samples to generate the UV LF
realizations, are shown in Figure~\ref{fig:half} and listed in
Table~\ref{tab:half}.  We briefly present a comparison of our
H$\alpha$ LFs and luminosity densities with others from the literature
in \S~\ref{sec:halfevol}.

\section{DISCUSSION}
\label{sec:discussion}

We have presented the most robust estimates of the rest-frame UV LFs
and moderate luminosity regime of the IR LFs of star forming galaxies
at redshifts $1.9\le z<3.4$ (\S~\ref{sec:uvlf}, \ref{sec:half},
\ref{sec:irlf}).  We have demonstrated how photometric redshifts over
this redshift range can introduce non-trivial biases in the LF,
underscoring the need for spectroscopy where it is feasible
(\S~\ref{sec:photoz}).  Further, our extensive spectroscopy allows us
to examine other systematic effects, including Ly$\alpha$ line
perturbations to the intrinsic rest-frame UV colors of galaxies
(\S~\ref{sec:lyas},\ref{sec:lyadist}).  In the next section, we will
examine more closely the trend between Ly$\alpha$ emission and
redshift and its dependence on the physical properties of galaxies.
In \S~\ref{sec:vvds}, we discuss recent results that have indicated an
excess of bright galaxies at $z\sim 3$ over that observed in the
initial LBG studies of \citet{steidel99} and \citet{dickinson98}.
Comparison of our UV, H$\alpha$, and IR LFs with those of previous
studies, and the evolution in the LF and luminosity density, are
discussed in \S~\ref{sec:uvhalfevol} and \ref{sec:irevol}.  Integral
constraints on the star formation rate density are presented in
\S~\ref{sec:sfrd}.

\subsection{$W_{\rm Ly\alpha}$ Distribution as a Function of Redshift}

The analysis of \S~\ref{sec:lyadist} demonstrated that the intrinsic
$W_{\rm Ly\alpha}$ distribution of galaxies at $1.9\le z<3.4$ was not
significantly modulated by the BX and LBG color criteria and,
furthermore, that the fraction of galaxies with $W_{\rm Ly\alpha}\ge
20$~\AA\, ($f20$) was larger at higher redshifts.  This trend was
recognized by comparing $f20$ between (a) BX galaxies and LBGs and (b)
BX galaxies at $z\le 2.48$ and $z>2.48$, with results summarized in
Table~\ref{tab:lyaew}.  In the latter case, the fact that we see a
trend in $f20$ with redshift even for galaxies selected using a single
set of color criteria (BX) further strengthens our conclusions that
$f20$ increases with redshift, irrespective of selection biases (see
Figure~\ref{fig:lya}).

To interpret this trend in a physical context, we assembled the
stellar population parameters for galaxies with measured $W_{\rm
Ly\alpha}$ where SED modeling was available from \citet{shapley05} and
\citet{reddy06b}.  The resulting sample includes 139 galaxies, 14 with
$W_{\rm Ly\alpha}\ge 20$~\AA.  We used KS-tests to determine whether
the SED parameters (star formation histories, ages, stellar masses,
and star formation rates) for galaxies with $W_{\rm
Ly\alpha}<20$~\AA\, are drawn from the same parent population as those
with $W_{\rm Ly\alpha}\ge 20$~\AA.  Doing this, we found no
significant differences in the star formation histories, ages, stellar
masses, and star formation rates of galaxies between these two
samples.  This result is not surprising given (1) the small sample
size analyzed here, (2) the significant systematic degeneracies
between SED parameters (e.g., \citealt{shapley05,erb06b,papovich01}),
and (3) the large uncertainty in the measured $W_{\rm Ly\alpha}$ for individual
galaxies.  Galaxies with $W_{\rm Ly\alpha}\ge 50$~\AA\, have an
average age of $\sim 300\pm 300$~Myr whereas those below this limit
have an average age of $\sim 460\pm 600$~Myr.  The difference in
average age is not significant given the large dispersion in ages
measured for the two samples.

Nonetheless, several previous studies at $z\ga 4$ suggest that
Ly$\alpha$ emitting galaxies are young ($\la 50$~Myr), low-metallicity
systems with small stellar masses (e.g., \citealt{stanway07,
pentericci07, dow07, pirzkal06, finkelstein07, lehnert03}),
particularly in relation to galaxies without Ly$\alpha$ in emission at
the same redshifts ($z\sim 6$; \citealt{stanway07,dow07}).  Results at
$z\sim 2$ also indicate that low stellar mass and low metallicity
galaxies have significantly stronger Ly$\alpha$ emission than high
stellar mass and high metallicity galaxies \citep{erb06a}.  Among
UV-continuum selected samples, both \citet{dow07} and
\citet{stanway07} find a fraction of $z\sim 6$ galaxies with $W_{\rm
Ly\alpha}\la 25$~\AA\, similar to the fraction found at $z\sim 3$
\citep{shapley03}, but the former find an excess of high equivalent
width galaxies ($W_{\rm Ly\alpha}>100$~\AA) compared to the $z\sim 3$
sample.  Further, the connection between Ly$\alpha$ profiles and the
physical properties of galaxies is well-known to be quite complicated,
with a sensitivity to ionizing flux, dust obscuration, and velocity of
outflowing material (e.g., \citealt{tapken07, reddy06a, hansen06,
shapley03, adel03}).  Despite these complications, the advantage of
our large spectroscopic analysis is that we can very accurately
quantify the trends between Ly$\alpha$ emission and redshift
(\S~\ref{sec:lyadist}).  The presently small samples at $z\sim 6$
prevent us from determining whether the trends observed at $1.9\le
z<3.4$ extend to higher redshift, but the general expectation is that
if younger and less dusty galaxies preferentially show Ly$\alpha$ in
emission, then the frequency of such Ly$\alpha$ emitting galaxies
should increase with increasing redshift as the average galaxy age
(and average dust-to-gas ratios; see \citealt{reddy06a}), decreases.

\subsection{VVDS-Inferred Excess of Bright $M(1700\AA)\la -22.0$ Galaxies
at Redshifts $2.7\la z\la3.4$}
\label{sec:vvds}

\begin{figure}[hbt]
\plotone{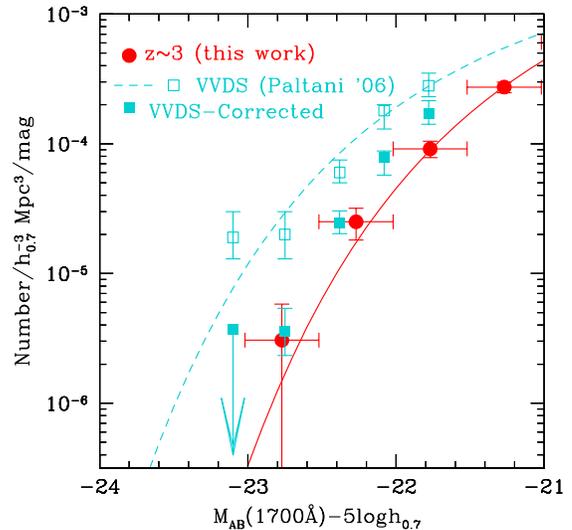}
\caption{Comparison of our $z\sim 3$ UV LF determination (circles for
data and solid line for best-fit Schechter function) and that of the
VVDS (open squares and dashed line).  Correcting the VVDS points for
(a) objects outside the redshift range $2.7\le z<3.4$ and (b) AGN/QSOs
at redshifts $2.7\le z<3.4$, as determined from our spectroscopic
sample, results in much better agreement between the two LFs, as
indicated by the filled squares.
\label{fig:vvds}}
\end{figure}

Recently, \citet{lefevre05} reported results from the $I$-band
magnitude limited VIMOS VLT Deep Survey (VVDS) that indicated a
significant excess of bright ($M(1700\AA)\la -22.0$) galaxies with
redshifts $2.7\la z\la 3.4$ compared with the earlier results of
\citet{steidel99} and those inferred here from our likelihood
analysis.  \citet{paltani07} further quantified this excess by casting
it into the form of a luminosity function, which we reproduce in
Figure~\ref{fig:vvds}.\footnote{Although the VVDS UV LF of
\citet{paltani07} is computed in a slightly different redshift range,
$3\la z\la 4$, from that considered for our $z\sim 3$ LF ($2.7\le
z<3.4$), the comparison between the two is valid since we find little
evolution in the number density of $M(1700\AA)\la -22.0$ galaxies
between $z\sim 3$ and $z\sim 4$.}

We suggest two reasons for the discrepancy between our LF and that of
the VVDS.  First, the frequency of objects with redshifts {\it
outside} the redshift range $2.7\le z<3.4$ (i.e, the contamination
fraction; $f_{\rm c}$) for $M(1700\AA)\la -22.0$ is significantly
larger than the value inferred by \citet{paltani07, lefevre05}, based
on the statistics of our much larger spectroscopic sample.  To
illustrate this, we must first consider how \citet{paltani07} weight
their galaxies in their computation of the LF to account for $f_{\rm
c}$.  The VVDS redshifts used to compute the LF fall within 4
categories.  Flag-1 and Flag-2 objects are considered to have the
least secure redshifts, whereas flag-3 and flag-4 objects are more
secure \citep{paltani07}.  \citet{paltani07} then determine a
contamination fraction of $f_{\rm c}\sim 0.54$ for the 254 flag-1/2
sources and assume a value of $f_{\rm c}=0$ for the 12 flag-3/4
sources that are used in their LF computation.  Weighting the
fractions according to the number of sources then yields a net
contamination rate of $[254\times0.54 + 12\times 0]/[254+12] \approx
0.52$.\footnote{This gives the same result as the ``photometric
rejection'' method, where $137$ of $266$ objects are likely to lie
outside the redshift range $2.7\le z<3.4$, as discussed in
\citet{paltani07}.}

The actual contamination rate among VVDS objects must be larger than
$f_{\rm c}=0.52$ for several reasons.  First, the BX and LBG criteria
account for the $\ugr$ colors of $\sim 70-80\%$ of VVDS objects
claimed to lie at $2.7\le z<3.4$ (see Figure~5 of \citealt{paltani07}
and Figure~3 of \citealt{lefevre05}, where most of the objects lie in
the {\it same} region of color space encompassed by either the BX or
LBG criteria).  Yet, $77\%$ of spectroscopically-confirmed candidates
in the BX and LBG samples with $M(1700\AA)\la -22.0$ are low redshift
interlopers (Table~\ref{tab:interlopertab}).\footnote{For the BX
sample, most of the contamination at bright magnitudes arises from
foreground galaxies.  For the LBG sample, most of the contamination
arises from stars.}

Second, VVDS objects that do not satisfy the BX and LBG color criteria
because of their redder $\gmr$ colors lie in the same region of color
space as low redshift star-forming galaxies (e.g.,
\citealt{adelberger04, reddy05a}).  Without additional $\ks$-band data
to exclude these low-z interlopers (e.g., $\bzk$ selection of
\citealt{daddi04}), the contamination rate among these red $\gmr$
objects is likely to be at least as large as the rate among objects
that are targeted by criteria specifically {\it designed} to selected
galaxies at $2.7\le z<3.4$ (e.g., the LBG color criteria).  To
quantify this further, we turned to the magnitude limited Team Keck
Redshift Survey database in the GOODS-N field (TKRS;
\citealt{wirth04,cowie04}).  There are 2471 TKRS sources with $\rs\le
24$ (roughly corresponding to the VVDS magnitude limit) with matching
$\ugr$ photometry in our catalog.  Of these matches, there are 755
sources that satisfy the following conditions: (a) spectroscopically
observed in the TKRS, (b) do not satisfy either the BX or LBG
criteria, and (c) have $U_{\rm n}-G > 0.4$ and $\gmr < 1.8$.  These
limits define the region in color space of the $\approx 20\%$ of VVDS
objects that do not satisfy the BX or LBG criteria.\footnote{Note that
although nominally the BX criteria are designed to select galaxies
with $z<2.7$, we include them in the discussion here since a
significant fraction of VVDS objects claimed to lie at $z\ga 3$ fall
in the same region of color space as BX candidates.}  Of these 755
bright sources that were observed in TKRS, 581, or $\approx 77\%$, are
spectroscopically-confirmed to lie at redshifts $z\la 1.4$.  This is a
strict lower limit to the contamination rate since there will be (a)
some $1.4<z<2.7$ galaxies that are missed simply because the lines
used for redshift identification are shifted out of the TKRS
wavelength coverage and (b) some galaxies with $z<1.4$ that are
unidentified because of poor weather, bad reduction, and other
reasons.

Finally, the instrumental setup used by the VVDS to obtain
spectroscopy, resulting in observed wavelengths of $5500$ and
$9500$~\AA\, is highly non-optimal for selecting LBGs, particularly
because of the lack of coverage around the Ly$\alpha$ line.  In
summary, our value of $f_{\rm c}=0.77$ is significantly larger than
the value of $f_{\rm c}=0.52$ claimed by \citet{paltani07}, yet is
based on our sample of 285 {\it secure} spectroscopic redshifts versus
only 12 secure redshifts and 254 unsecure redshifts of the VVDS
survey.  For the reasons discussed above, the real contamination rate
among VVDS objects is likely to be larger than $f_{\rm c}=0.77$.  Even
so, assuming this fraction would conservatively lower the brightest
VVDS LF points by a factor of at least two.  We already alluded to in
\S~\ref{sec:photoz} how catastrophic redshift errors can artificially
boost the bright-end of the LF.

The excess of bright galaxies in the VVDS LF is also likely to be due
in part to the presence of non star-forming galaxies at redshifts
$2.7\le z < 3.4$, in other words QSOs and other bright AGN.  We have
explicitly excluded AGN from our LF determination, as described in
\S~\ref{sec:interlopers}, but this was not done for the VVDS analysis.
Based on our spectroscopy, the AGN/QSO contamination rate is a strong
function of magnitude and is larger than $\sim 60\%$ in the brightest
luminosity bin of the VVDS analysis (Table~\ref{tab:specfrac}).  In
fact, of the $14$ spectroscopically-confirmed LBGs in our sample with
$M(1700\AA)\la -22.5$, $11$ (or roughly $80\%$) are QSOs or bright
AGN.  Applying the AGN/QSO contamination fractions (as determined from
our spectroscopy), will further reduce the brightest VVDS LF points by
a factor of $2.5$.

To conclude our comparison, the two primary causes for the excess of
bright galaxies inferred by the VVDS is (a) their underestimate of the
fraction of objects that lie outside the redshift range $2.7\le z<3.4$
and (b) the fraction of bright AGN and QSOs at redshifts $2.7\le
z<3.4$.  Our spectroscopy, which is the most extensive at the
redshifts in question and was obtained using the most optimal
instrumental setup to identify galaxies at $2.7\le z<3.4$, allow us to
very accurately quantify the magnitude of both sources of
contamination, all within a combined survey area that is roughly twice
as large as the VVDS field.  Figure~\ref{fig:vvds} demonstrates that
applying the contamination fractions determined from our spectroscopy
to the VVDS points (after factoring out the VVDS contamination
correction) results in a better agreement between the VVDS and our LF.
Taking all these results into consideration, we find no convincing
evidence for an excess of bright galaxies at $2.7\le z<3.4$.

Finally, the agreement between our corrected estimate of the VVDS LF
(derived from a magnitude-limited survey) with the one from our
likelihood analysis (derived from a color-selected survey), strongly
suggests that our LF must be reasonably complete for UV-bright
galaxies.

\subsection{Evolution of the UV and H$\alpha$ Luminosity Functions and Densities}
\label{sec:uvhalfevol}

\begin{figure*}[hbt]
\plottwo{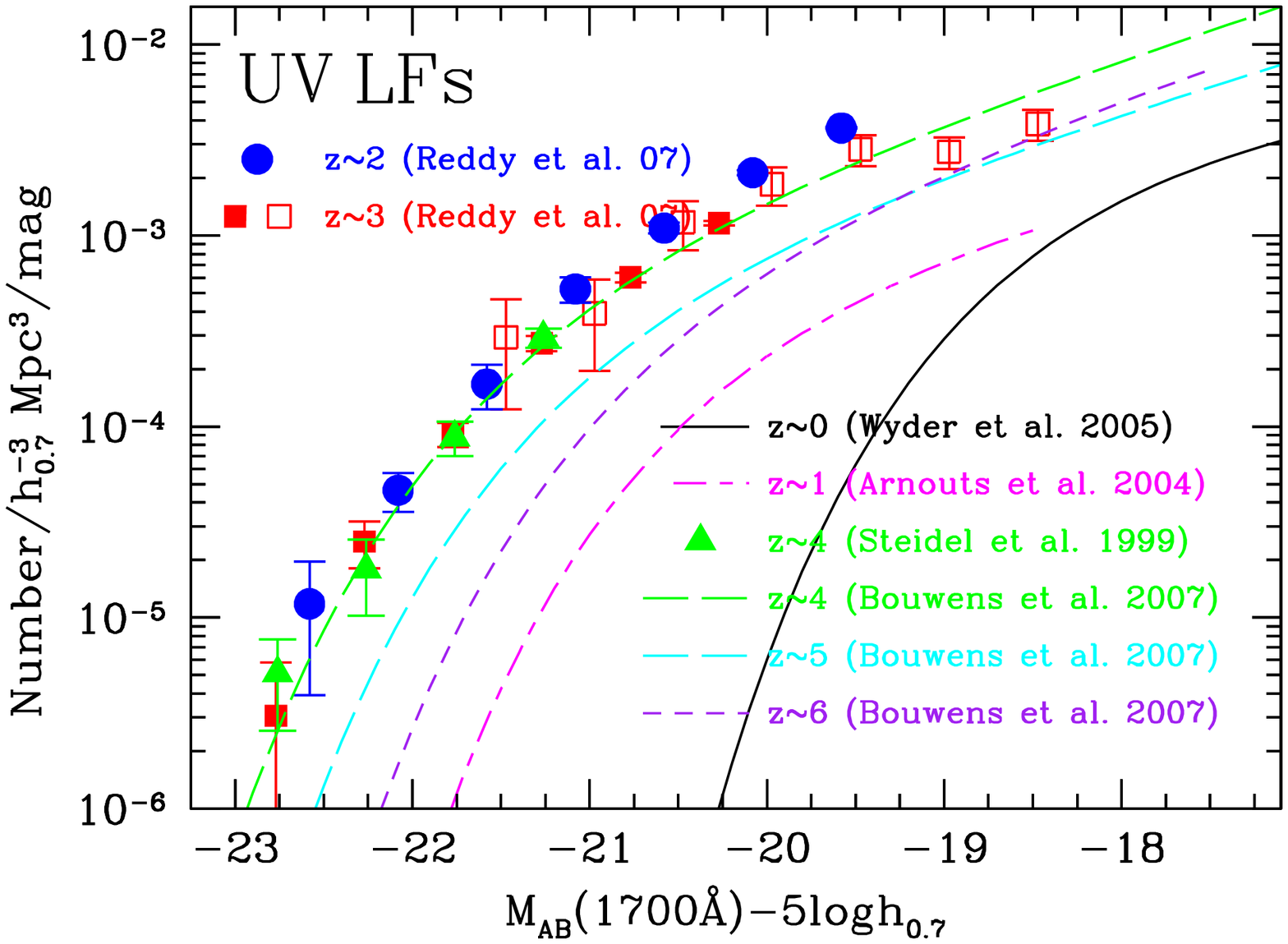}{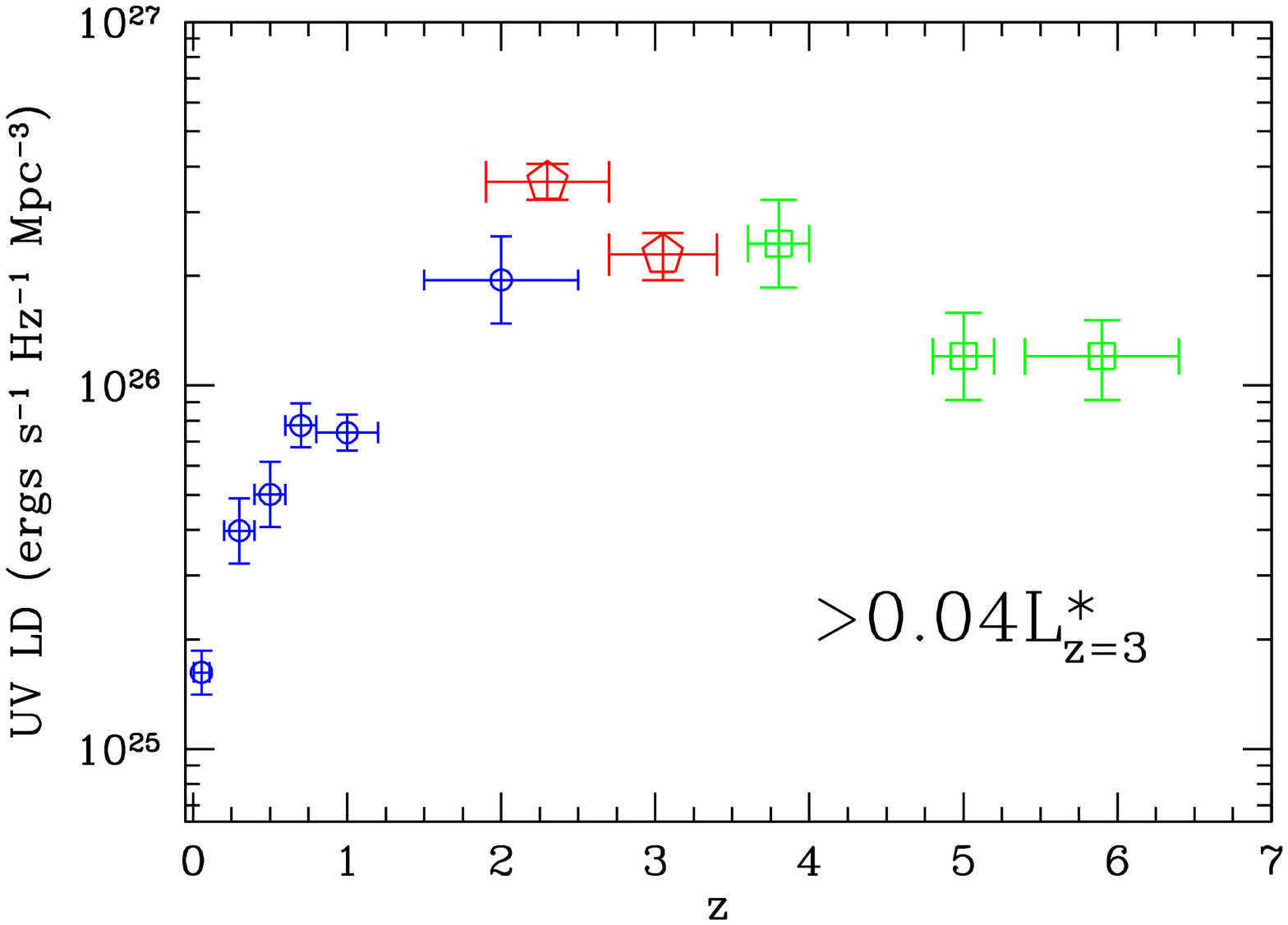}
\caption{({\it Left:}) Comparison of a {\it few} UV LFs from the
literature.  The local UV LF (derived from GALEX data) is shown
\citep{wyder05} along with the $z\sim 1$ UV LF from \citet{arnouts05},
as well as higher redshift determinations from the $B$, $V$, and $I$
dropout samples of Bouwens et~al. (2007; submitted). ({\it Right:})
Unobscured (uncorrected for extinction) UV luminosity density,
integrated to a fixed luminosity of $L_{\rm lim}=0.04L^{\ast}_{z=3}$,
from the following sources: \citet{wyder05} at $z\sim 0$,
\citet{schiminovich05} at $z\sim 0.3-1.0$, \citet{arnouts05} at $z\sim
2.0$ from a photometric redshift analysis of the HDF North and South
fields, and Bouwens et~al. (2007; submitted) at $z\sim 4-6$.  Our
determinations at $z\sim 2$ and $z\sim 3$ are shown by the large red
pentagons.
\label{fig:uvlfcomp}}
\end{figure*}

\begin{deluxetable*}{lccc}
\tabletypesize{\footnotesize}
\tablewidth{0pc}
\tablecaption{Summary of Total UV, H$\alpha$, and IR Luminosity Densities at $1.9\le z<3.4$}
\tablehead{
\colhead{} &
\colhead{UV LD\tablenotemark{a}} &
\colhead{H$\alpha$ LD} &
\colhead{IR LD\tablenotemark{b}} \\
\colhead{Redshift Range} &
\colhead{(ergs~s$^{-1}$~Hz$^{-1}$~Mpc$^{-3}$)} &
\colhead{(ergs~s$^{-1}$~Mpc$^{-3}$)} &
\colhead{(L$_{\odot}$~Mpc$^{-3}$)}}
\startdata
$1.9\le z<2.7$ & $(3.63\pm 0.40)\times 10^{26}$ & $(4.41\pm1.10)\times 10^{40}$ & $(1.23\pm0.21)\times 10^{9}$ \\ 
$2.7\le z<3.4$ & $(2.29\pm 0.34)\times 10^{26}$ & $(2.71\pm0.57)\times 10^{40}$ & $(6.61\pm1.24)\times 10^{8}$ \\
\enddata
%\tablenotetext{a}{All values are computed by integrating the LF to a luminosity corresponding
%to $0.5$~M$_{\odot}$~yr$^{-1}$.}
\tablenotetext{a}{Uncorrected for extinction and integrated to $0.04L^{\ast}_{z=3}$.}
\tablenotetext{b}{Values assume the \citet{caputi07} calibration between $\nu L_{\nu}(8\mu m)$
and $L_{\rm IR}$.  The IR LD at $z\sim 2$ includes the contribution from ULIRGs \citep{caputi07}.}
\label{tab:lumdenstab}
\end{deluxetable*}

\subsubsection{UV LFs}
\label{sec:uvlfevol}

Figure~\ref{fig:rlfg} summarizes our determinations of the rest-frame
UV LFs at redshifts $1.9\le z<3.4$, compared with the
\citet{steidel99} LF at $z\sim 4$.  As our method of constraining the
reddening and luminosity distributions takes into account a number of
systematic effects (e.g., contamination fraction particularly at the
bright end of the LF, photometric bias and errors, Ly$\alpha$ line
perturbations to the observed colors\footnote{Our determinations of
the LFs are insensitive to small changes in the assumed $W_{\rm
Ly\alpha}$ distributions, such as those caused by trends in $W_{\rm
Ly\alpha}$ with apparent magnitude and color.}) that have not been
considered in previous analyses (e.g., \citealt{gabasch04, lefevre05})
or were only partially considered \citep{steidel99, adel00,
sawicki06a}, we regard our LFs as the most robust determinations at
$z\sim 2$ and $z\sim 3$ to $\rs=25.5$.

Our analysis indicates that the rest-frame UV LF shows little
evolution for galaxies brighter than $M^{\ast}$ (at $z\sim 2$) between
redshifts $1.9\le z <4.5$: the number density of galaxies brighter
than $M^{\ast}=-20.97$ appears to be constant over the $\sim 1.3$~Gyr
timespan between $z\sim 4$ and $z\sim 2.3$.  This lack of evolution in
the bright-end of the UV LFs does not specifically address how a
galaxy of a particular luminosity will evolve.  For example, the lack
of evolution at the bright end ($M_{\rm AB}(1700\AA)\la -21$) of the
LF does not imply that there is a population of UV-bright
galaxies that is unevolving.  Rather, if galaxies follow an
exponentially-declining star formation history, then UV-bright
galaxies at $z\sim 3$ will become fainter in the UV by $z\sim 2$, but
will not necessarily be absent from the $z\sim 2$ sample.  A
precipitously declining star formation history may imply that some
UV-bright galaxies at $z\sim 3$ will be too faint to be included in
UV-selected samples at $z\sim 2$.  In any case, the lack of evolution
at the bright-end of the UV LF implies that whatever UV-bright
galaxies at $z\sim 3-4$ fall out of UV-selected samples by $z\sim 2$
must be made up in number by younger galaxies, those that are merging
and just ``turning on'', and/or those that are caught in an active
phase of star formation at $z\sim 2$.  The net effect is that the
number density of galaxies with ($M_{\rm AB}(1700\AA)\la -21$) remains
essentially constant.

For galaxies fainter than $M^{\ast}$, we do find evidence for a small
evolution between $z\sim 3$ and $z\sim 2$: the number density of
galaxies fainter than $M^{\ast}=-20.97$ is systematically larger at
$z\sim 2$ than at $z\sim 3$ to a lower luminosity limit of $\rs=25.5$
or $M_{\rm AB}(1700\AA)\sim -19$, although the result is not of great
significance given the generally overlapping error bars on the points
between $z\sim 2$ and $z\sim 3$.  What is clear is that the number
density of $-21\la M_{\rm AB}(1700\AA)\la -19$ galaxies at $z\sim 2$
is {\it at least} as large as the the corresponding number density at
$z\sim 3$.  To put these results in context, Figure~\ref{fig:uvlfcomp}
summarizes our results at $z\sim 2$ and $z\sim 3$ with a few results
at higher and lower redshifts.\footnote{The figure is not meant to be
comprehensive with respect to all determinations of the UV LF,
particularly at $z\ga 4$ where there are some differences between
studies (e.g., \citealt{sawicki06a, ouchi04, iwata07, beckwith06}).}

\begin{figure*}[hbt]
\plottwo{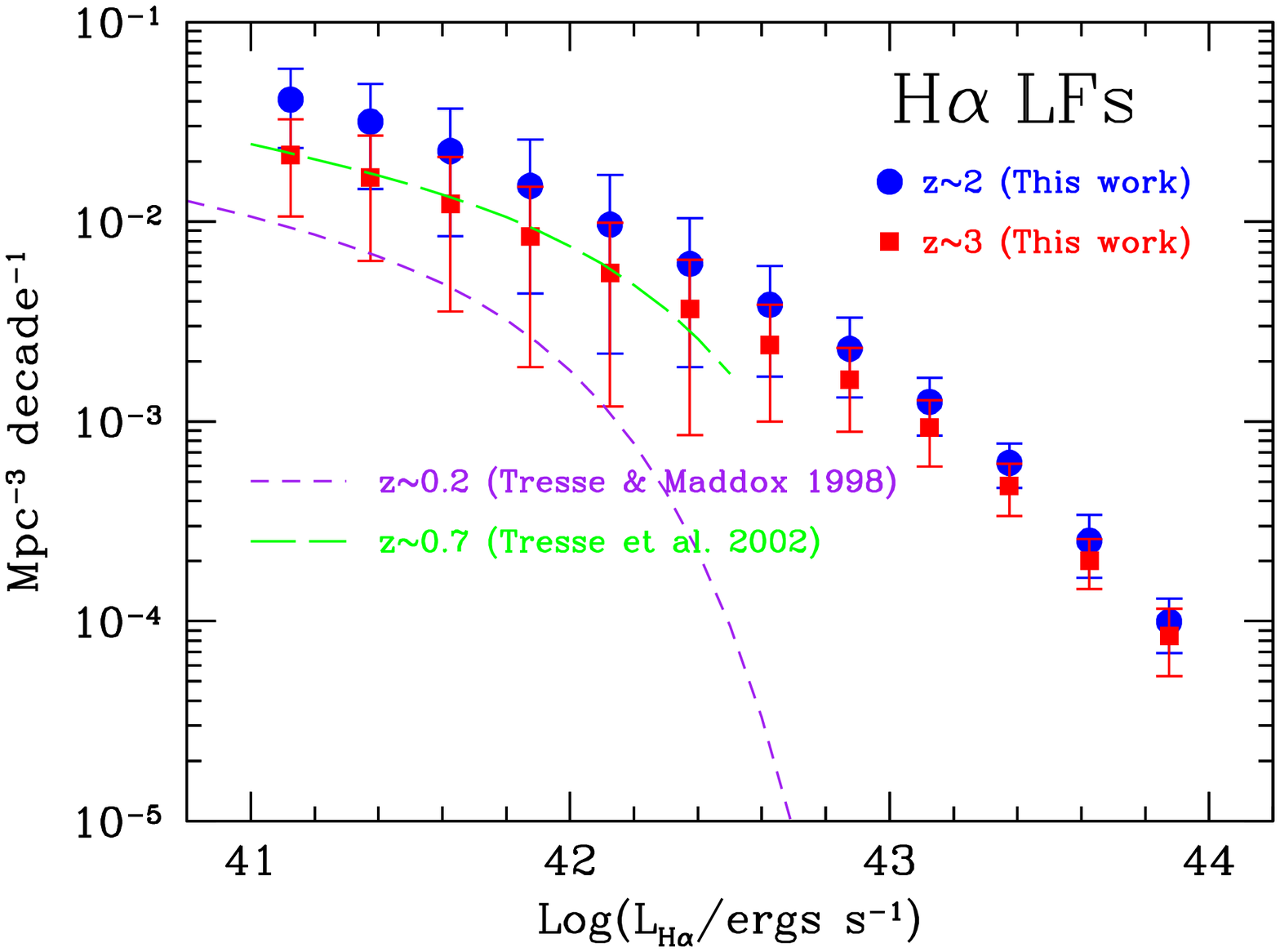}{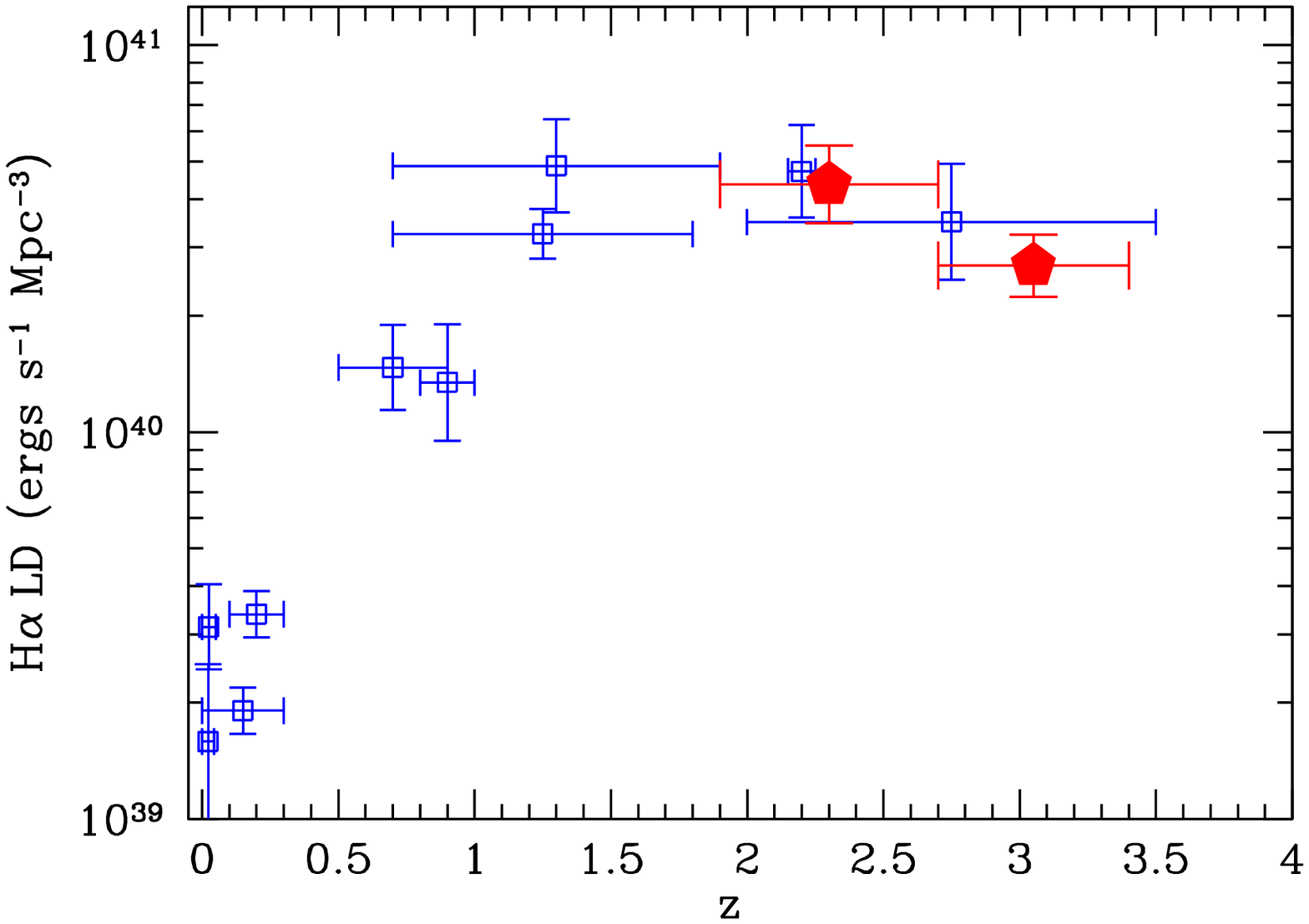}
\caption{({\it Left:}) Comparison of our inferred dust-corrected
H$\alpha$ luminosity function at $z\sim 2$ and predicted one at $z\sim
3$, with the direct H$\alpha$ LF determinations at lower redshift from
\citet{tresse02, tresse98}. ({\it Right:}) Extinction-corrected
H$\alpha$ luminosity density: open squares are from \citet{hopkins04}
and include determinations from \citet{gallego95, perez03, sullivan00,
tresse98, tresse02, glazebrook99, hopkins00, yan99, moorwood00}.  The
point at $z=2.75$ is the H$\beta$ determination from \citet{pettini98}.
The large red pentagons denote values from this work.
\label{fig:halfevol}}
\end{figure*}

Figure~\ref{fig:uvlfcomp} demonstrates the evolution of the UV LF.
The number density of bright galaxies increases from $z\sim 6$ to
$z\sim 4$ causing a brightening of $M^{\ast}$ by $\sim 0.6-0.7$~mag.  In
contrast, the number of faint galaxies appears to undergo less
evolution between $z\sim 6$ and $z\sim 4$.  To quantify this further,
we have calculated the rest-frame $1700$~\AA\, unobscured UV
luminosity density (LD).  The luminosity density and error are
estimated by simulating many realizations of the UV LF consistent
within the normally-distributed LF errors and evaluating the mean
luminosity-weighted integral of the LFs from each of these
realizations.  The mean value and dispersion of these integrated
values give the mean luminosity density and error, and values are
listed in Table~\ref{tab:lumdenstab}.  We have assumed a faint-end
slope of $\alpha=-1.6\pm0.11$ (i.e., as in \S~\ref{sec:bolmeas}) at
$z\sim 2$.  For consistency, all the LFs are integrated to a
luminosity limit of $L_{\rm lim} = 0.04L^{\ast}_{z=3}$.  This
corresponds to a luminosity of $L_{\rm lim} = 3.9\times
10^{27}$~ergs~s$^{-1}$~Hz$^{-1}$ at $1700$~\AA, is equivalent to
$\approx 0.05L^{\ast}_{z=2}$, and corresponds to an unobscured SFR of
$\sim 0.5$~M$_{\odot}$~yr$^{-1}$ assuming the \citet{kennicutt98}
relation.  The right panel of Figure~\ref{fig:uvlfcomp} summarizes the
integrated (unobscured) UV LD as a function of redshift including
several published values, showing the significant evolution between
$z\sim 2$ and $z=0$.

While the observed evolution in the UV LF is hardly surprising, we
have placed this evolution during the most active epoch of star
formation ($z\sim 2-4$) on a secure footing with our extensive
spectroscopic analysis and detailed completeness corrections.  Our
analysis covers a larger number of uncorrelated fields than what has
typically been considered in previous studies, thus enabling us to
mitigate the effects of sample variance.  We note that part of the
evolution of the unobscured UV LF may be a result of extinction, as we
will discuss shortly.

\subsubsection{H$\alpha$ LFs}
\label{sec:halfevol}

Figure~\ref{fig:halfevol} compares our inference of the H$\alpha$ LF
at $z\sim 2$ and our prediction at $z\sim 3$, with the direct
determinations at lower redshifts by \citet{tresse98} and
\citet{tresse02}.  The evolution of the dust-corrected H$\alpha$ LF
qualitatively mimics the evolution observed in the UV LF
(Figure~\ref{fig:uvlfcomp}) for redshifts $z\la 2-3$.  We see a factor
of two decline in the number density of moderately luminous galaxies
with $10^{41}<L(H\alpha)<10^{42.5}$~ergs~s$^{-1}$ from $z\sim 2$ to
$z\sim 0.7$.  The decline of moderately luminous galaxies is at least
a factor of $4-10$ between $z\sim 2$ and $z\sim 0.2$.

The systematic excess of $z\sim 2$ galaxies with respect to $z\sim 3$
for luminosities $L(H\alpha)\la 10^{43}$~ergs~s$^{-1}$ is primarily a
result of the fact that galaxies on the faint-end of the UV LF (where
we observed the same small systematic excess; see
Figure~\ref{fig:rlfg}) are scattered to correspondingly more luminous
bins of H$\alpha$ luminosity after correcting for extinction.  The
significance of the systematic excess at $z\sim 2$ with respect to
$z\sim 3$ is hard to judge since, unlike the case with the UV LFs,
there are no direct determinations of the H$\alpha$ LFs at $z\ga 2$.
What is certain is that the number (and luminosity) density of
moderately H$\alpha$ luminous galaxies at $z\sim 2$ is {\it at least}
as large as the number at $z\sim 3$.  In contrast, the significance of
the increased frequency of moderately luminous galaxies at $z\sim 2$
with respect to $z\sim 0.7$ is supported by the fact that the
H$\alpha$ luminosity density shows a decline from $z\sim 2$ to $z=0$.
To illustrate this, we have compiled estimates of the H$\alpha$ LD
from \citet{hopkins04} in the right panel of
Figure~\ref{fig:halfevol}, including our new inferences at $z\sim 2$
and $z\sim 3$ (the latter are listed in Table~\ref{tab:lumdenstab}).
Values from \citet{hopkins04} have been reddening-corrected assuming a
luminosity dependent obscuration correction.  Our values at $z\sim 2$
and $z\sim 3$ are computed from the dust-corrected H$\alpha$ LFs.
Quantitatively, the H$\alpha$ LD per comoving Mpc decreases by a
factor $\sim 25$ between $z\sim 2$ and the local value.  While
H$\beta$ observations have been performed for a handful of objects at
$z\sim 3$ \citep{pettini98}, larger samples are needed to directly
constrain the H$\alpha$ LF at $z\sim 3$.  Since our H$\alpha$ results
on the SFRD are degenerate with those estimated from the UV and IR
LFs, we will not discuss the H$\alpha$ LFs any further.

\begin{figure*}[hbt]
\plotone{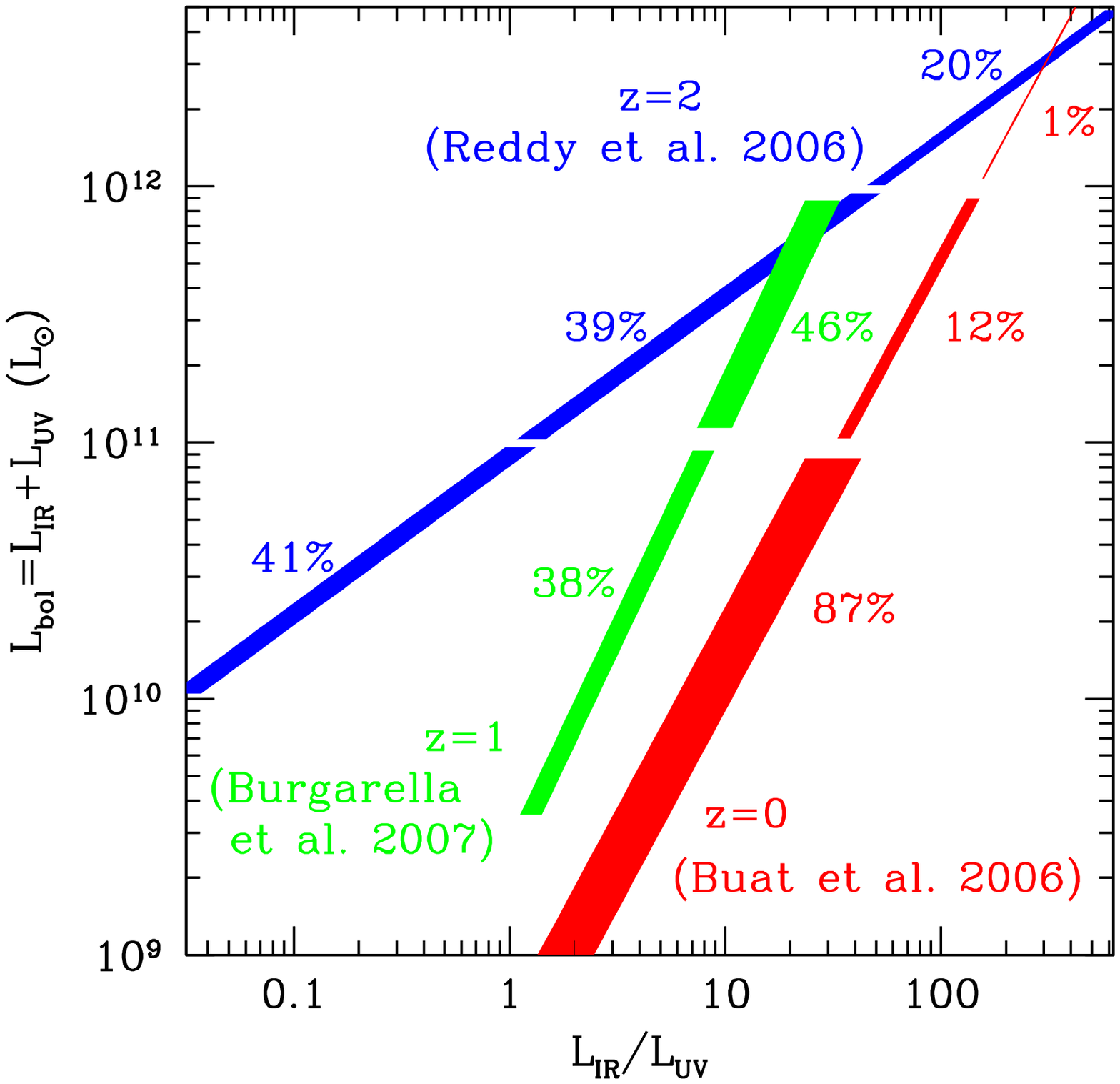}
\caption{Trend between bolometric luminosity and dust attenuation
based on the analyses of \citet{reddy06a} at $z\sim 2$,
\citet{burgarella07} at $z\sim 1$, and \citet{buat06} at $z\sim 0$.
The thickness of the lines show schematically the contribution of
galaxies in different ranges of bolometric luminosity to the
bolometric luminosity density, with the relative fractions listed.
Fractional contributions at $z\sim 2$ were determined from our
estimate of the bolometric LF at this epoch (Figure~\ref{fig:bollf}).
The contributions at $z\sim 0$ and $z\sim 1$ were taken from
\citet{lefloch05}.
\label{fig:dustevol}}
\end{figure*}

\subsection{Average Extinction and IR and Bolometric LFs}
\label{sec:irevol}

\subsubsection{Evolution of Dust Obscuration with Redshift}

As mentioned in \S~\ref{sec:uvlfevol}, the evolution observed in the
unobscured rest-frame UV LF between $z\sim 2$ and $z=0$ may be partly
a result of systematic differences in extinction with redshift.  It is
already known that galaxies of a fixed bolometric luminosity have an
average attenuation factor at $z\sim 2$ that is $\sim 8-10$ times
smaller than the attenuation factors of local galaxies
\citep{reddy06a, burgarella07, buat07}.  Figure~\ref{fig:dustevol},
adapted from \citet{reddy06a} and the GALEX results of \citet{buat06}
and \citet{burgarella07}, illustrates the offset between the $z\sim
2$, $z\sim 1$, and $z=0$ trends between $L_{\rm bol}$ and attenuation.
\citet{reddy06a} interpreted this trend as a result of the increasing
extinction per unit SFR (or increasing dust-to-gas ratio) as galaxies
age.  The dependence of attenuation on bolometric luminosity at low
redshifts has been discussed by many authors (e.g., \citealt{buat05,
perez03, afonso03, sullivan01, hopkins01, wang96}).  The analyses of
\citet{reddy06a, adel00} and Figure~\ref{fig:dustevol} demonstrate
that this dependence continues unabated from $z=0$ to $z\sim 2-3$.

\begin{figure*}[hbt]
\plottwo{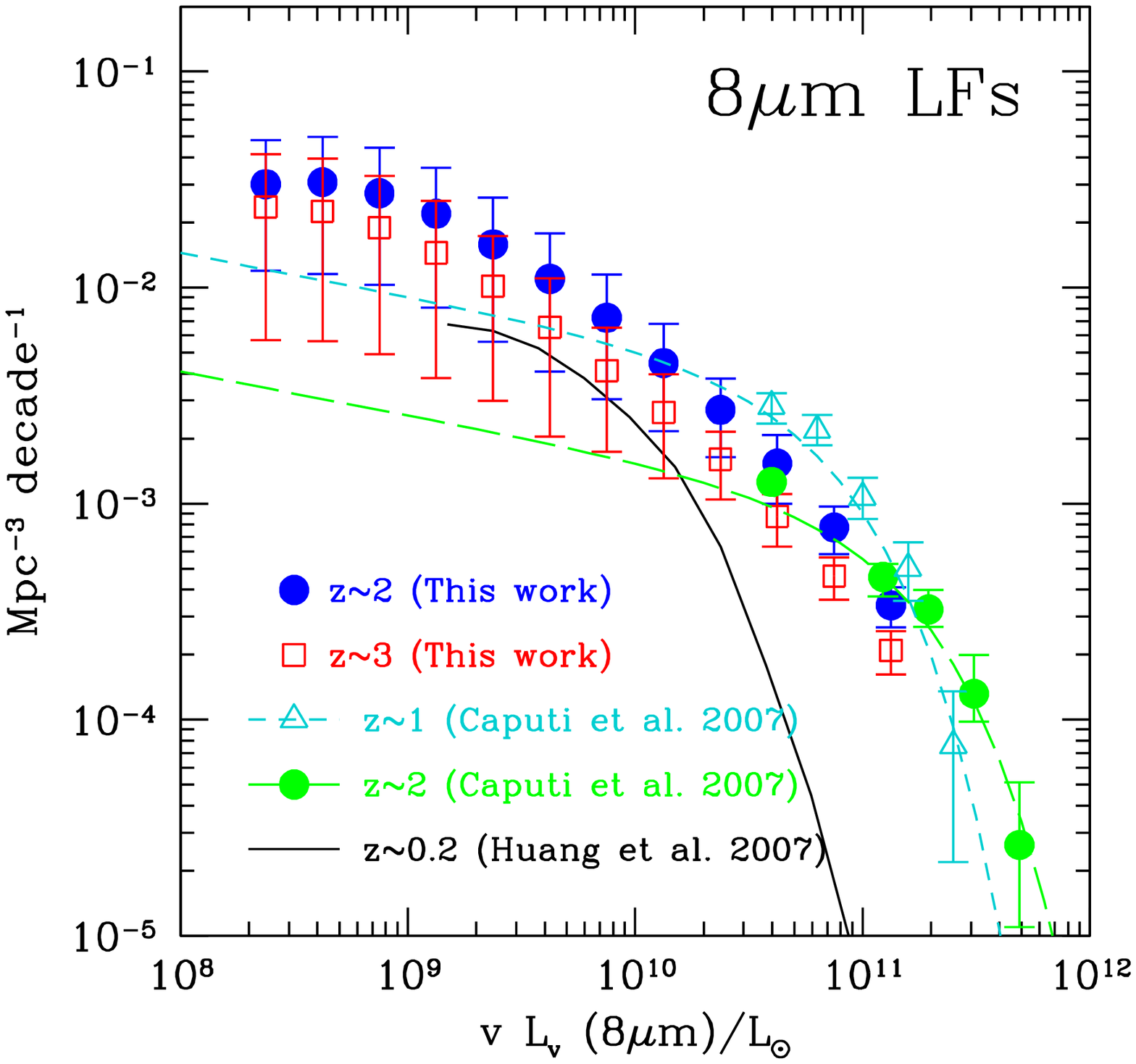}{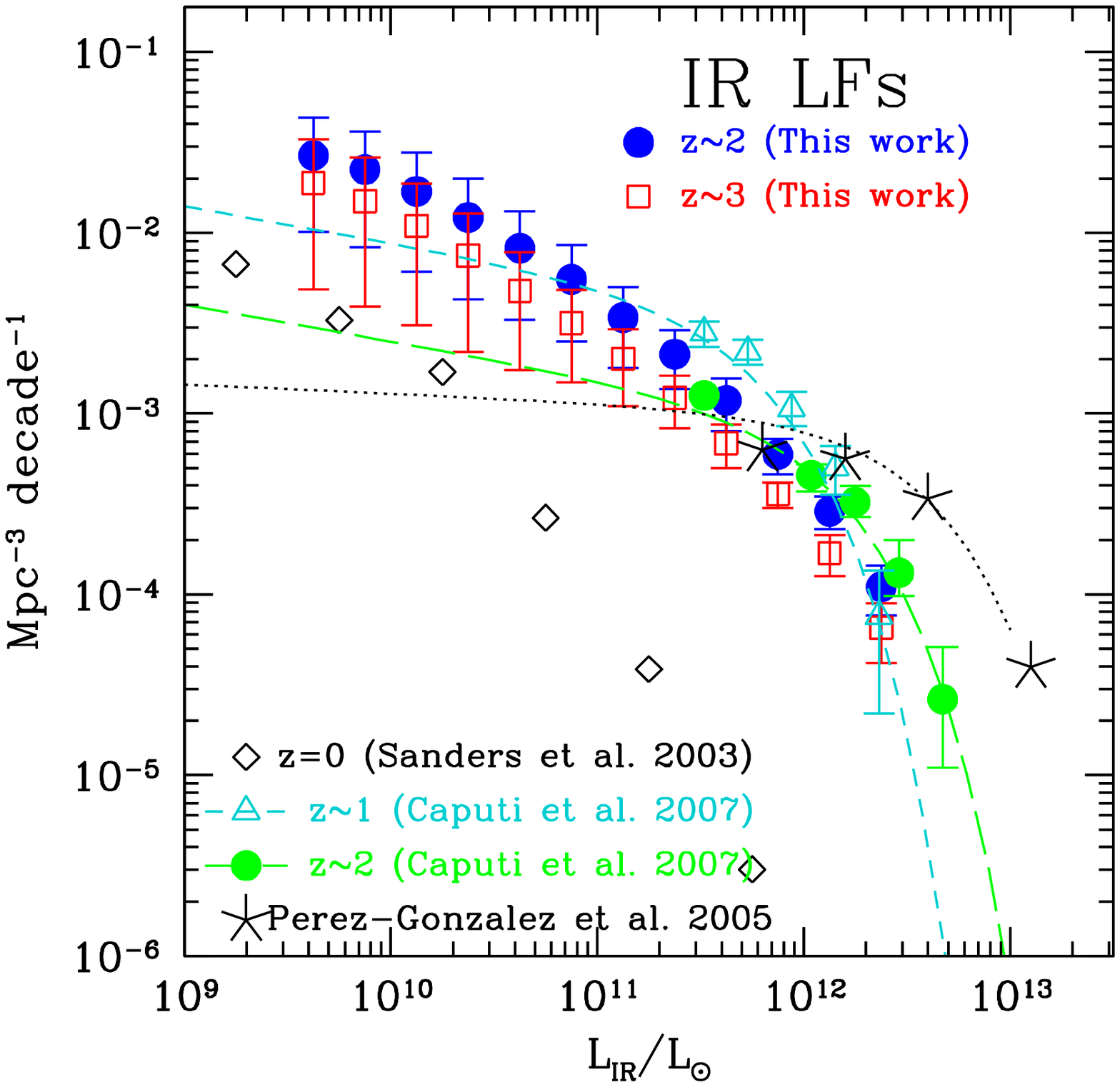}
\caption{Comparison of our inference of the $8$~$\mu$m (left) and IR
LFs (right) at $z\sim 2$ and our prediction at $z\sim 3$, with direct
measurements from the literature: $z\sim 1$ and $z\sim 2$ points from
\citet{caputi07} and \citet{perez05}, $8$~$\mu$m LF at $z\sim 0.2$
from Huang et al. (2007, submitted), and the local IR LF from the IRAS
BGS sample \citep{sanders03}.  The \citet{perez05} points at $z\sim 2$
are significantly larger than those of \citet{caputi07}, primarily
because the former exclude only the most extreme AGN from their
analysis and adopt a conversion between rest-frame mid-IR and IR
luminosity that has been shown to overpredict the bolometric
luminosities of $\ga 2\times 10^{12}$~L$_{\odot}$ galaxies by a factor
of $>2$ (\citealt{papovich07, daddi07a}; see text).
\label{fig:ircomp}}
\end{figure*}

\begin{figure*}[hbt]
\plotone{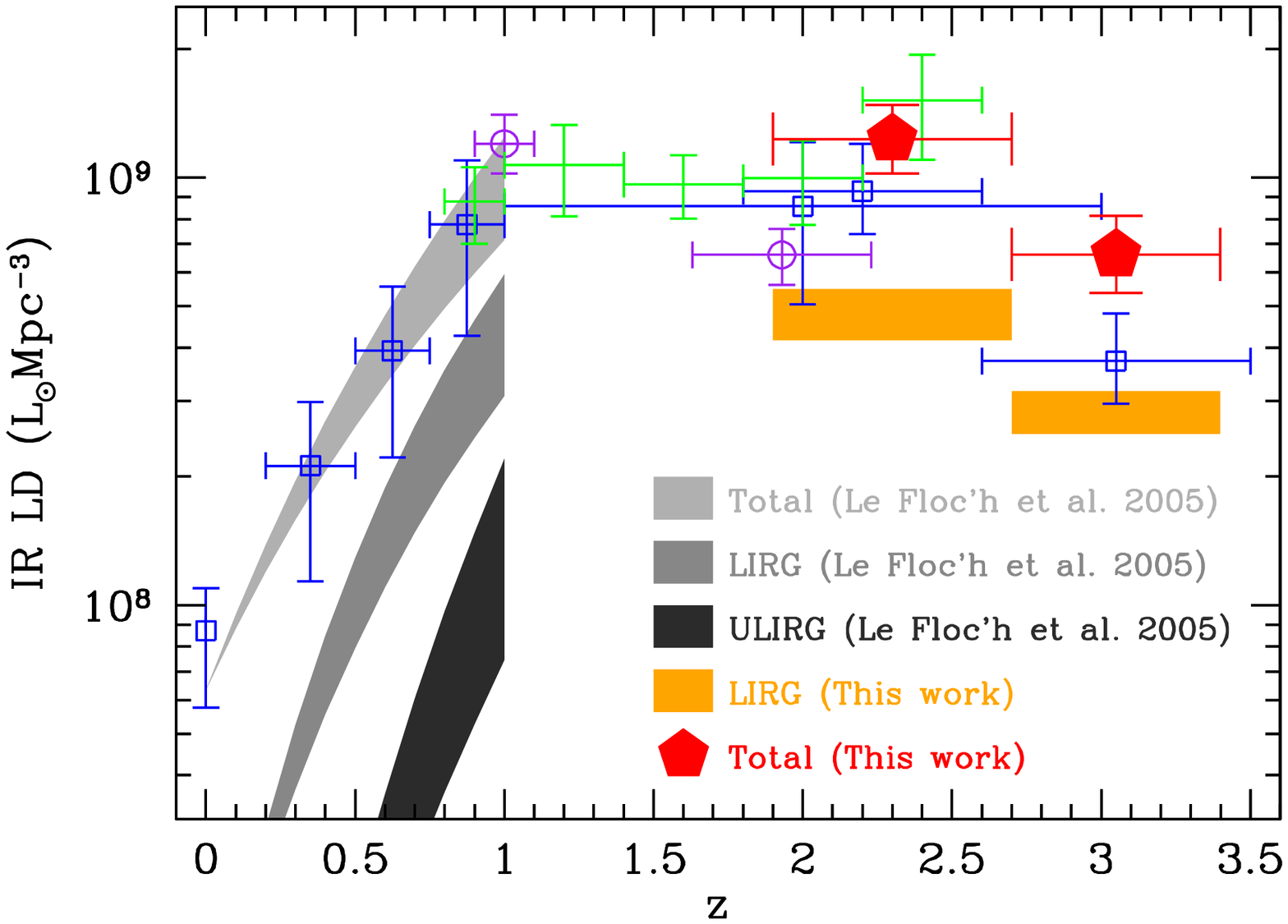}
\caption{IR luminosity density as a function of redshift, including
data from \citet{yun01} at $z=0$, \citet{flores99} at $0.2\la z \la
1.0$, \citet{barger00} at $z\sim 2$, and \citet{chapman05} at $z\ga
2$; all shown with open squares.  Results from \citet{caputi07} and
\citet{perez05} are shown by the open circles and green lines,
respectively.  Our points at $1.9\le z<2.7$ and $2.7\le z<3.4$ are
shown by the solid pentagons, assuming the \citet{caputi07}
calibration between $\nu L_{\nu} (8\mu m)$ and $L_{\rm IR}$.  The
\citet{reddy06a} calibration would raise the points by $\approx 30\%$.
The ULIRG, LIRG, and total contributions to the IR LD inferred by
\citet{lefloch05} are indicated by the light, medium, and dark gray
shaded regions, respectively, up to $z=1$.  The inferred contribution
of LIRGs from our analysis at $z\sim 2-3$ is denoted by the shaded
rectangles.
\label{fig:irlumdens}}
\end{figure*}

If the LF was unevolving between $z\sim 2$ and $z=0$, then the offset
shown in Figure~\ref{fig:dustevol} implies that, when integrating the
UV (or H$\alpha$) LF to a fixed luminosity, the extinction correction
will be larger at lower redshifts.  However, as
Figures~\ref{fig:uvlfcomp} and \ref{fig:halfevol} demonstrate, there
is a very strong evolution in the LF between $z\sim 2$ and $z\sim 0$.
We find a similar evolution in the IR LFs and IR LDs, shown in
Figures~\ref{fig:ircomp} and \ref{fig:irlumdens}, respectively, where
we summarize the comparison between the $z\sim 2$ and $z\sim 3$ IR LFs
and luminosity densities from our analysis with other published
measurements.  The important point is that the average extinction
correction needed to recover bolometric LDs from UV LDs will depend
both on the offset in the $L_{\rm bol}$ versus attenuation trends
between $z\sim 2$ and $z=0$, as well as the relative numbers of
galaxies in different luminosity ranges that contribute to the LD.
For example, LIRGs at $z\sim 2$ are on average $8-10$ times less dusty
than LIRGs at $z=0$, but there are many more LIRGs at $z\sim 2$ than
at $z\sim 0$.  The contribution of galaxies in different luminosity
ranges to the bolometric LD at $z\sim 2$, $z\sim 1$, and $z=0$ are
shown schematically in Figure~\ref{fig:dustevol}.

If the IR LF was unevolving between $z\sim 2$ and $z=0$, then the fact
that galaxies of a fixed $L_{\rm bol}$ are $8-10$ times less dusty at
$z\sim 2$ than at $z\sim 0$ would imply that the average extinction
correction needed to recover the bolometric LD from the UV LD would be
roughly $8-10$ times {\it larger} at $z\sim 0$ than at $z\sim 2$.
However, because of the evolution in number density, the average
correction at $z\sim 0$ is not $8-10$ times larger than the value at
$z \sim 2$.  \citet{schiminovich05} find an average attenuation factor
for UV-selected samples at $z\la 1$ of $\langle L_{\rm bol}/L_{\rm
UV}\rangle \sim 7$, which is only $1.6$ times larger than the value of
$4.5$ found for UV-selected galaxies at $z\sim 2-3$.  It is
interesting to note that even taking into account the evolution in
number density, the average correction at $z\sim 0$ is still larger
than the correction at $z\sim 2$, despite a greater fraction of the LD
in dustier galaxies at high redshift.  This result may be partly due
to the redshift evolution in $L_{\rm bol}$ versus attenuation trend
(Figure~\ref{fig:dustevol}), but the larger correction at $z\sim 0$
may also be needed to account for IR luminous galaxies that altogether
escape UV-selection at $z\sim 0$.

\subsubsection{Evolution of the Dust-Obscured and Bolometric Luminosity Densities}

There are several important conclusions to draw from our analysis of
the IR luminosity density.  First, the unobscured UV LD drops by a
factor $\sim 23$ between $z\sim 2$ and $z\sim 0$, whereas the IR LD
drops by a factor of $\sim 14$ between $z\sim 2$ and the present-day.
This difference can be partly accounted for by an evolution in the
average extinction correction, as discussed in the previous section.
The {\it independent} determinations of the average extinction
corrections at $z\sim 2$ ($\langle L_{\rm bol}/L_{\rm UV}\rangle \sim
4.5$; \citealt{reddy06a}) and at $z\sim 0$ ($\langle L_{\rm
bol}/L_{\rm UV}\rangle \sim 7$; \citealt{schiminovich05}) result in an
evolution of the dust-corrected UV LD that is in remarkable agreement
with the evolution of the IR LD between $z\sim 2$ and $z=0$ (see
\S~\ref{sec:sfrd}).

Second, the IR LD at $z\sim 2$ appears to be at least as large as the
value at $z\sim 3$, a result consistent with that of the UV LD
analysis.  This should not come as a surprise since the same
incompleteness-corrected rest-frame UV LF was used to infer the IR LF.
Nonetheless, our {\it spectroscopic} analysis puts the constraints on
the H$\alpha$ and IR LD on a more secure footing.  In particular, our
analysis provides the first spectroscopic constraints on the sub-ULIRG
regime of the IR LFs at $z\ga 2$.

Third, the constraints on the IR LD between $1.9\le z<3.4$ yield
critical information on the relative contribution of galaxies to the
IR LD as a function of luminosity.  Figure~\ref{fig:ircomp} hints that
previous studies of the IR LF at $z\sim 2$ underpredicted the number
(and luminosity) density of sub-ULIRGs with $L_{\rm
IR}<10^{12}$~L$_{\odot}$ (i.e., compare our IR LF points with the
extrapolation of \citet{perez05} and \citet{caputi07} for galaxies
with $L_{\rm IR}<10^{12}$~L$_{\odot}$ in Figure~\ref{fig:ircomp}).
When integrating the IR LF to account for {\it all} galaxies, we find
a total IR LD of $\sim 1.2\times 10^9$~L$_{\odot}$~Mpc$^{-1}$, about a
factor of 2 larger than the previous determination at $z\sim 2$.  Note
that the \citet{perez05} determination of the IR LD at $z\sim 2.4$ is
consistent with our measurement, primarily for two reasons.  First,
they only excluded the most extreme AGN, which by number made up $5\%$
of their sample, whereas other surveys indicate a larger contamination
rate of $10-20\%$ (see \S~3.2 of \citealt{perez05} and references
therein).  Second, \citet{perez05} use a conversion relation between
rest-frame $12$~$\mu$m and IR luminosity that has been shown to
overproduce by factors of $>2$ the IR luminosities of the most IR
luminous galaxies based on $70$ and $160$~$\mu$m stacking analysis
\citep{papovich07}.  This result is also supported by
\citet{daddi07a}, who also find a systematic excess of rest-frame
$8$~$\mu$m emission relative to UV and radio emission for a sample of
$z\sim 2$ star-forming ULIRGs.  Further, \citet{daddi07b} suggest that
the obscured AGN fraction among ULIRGs at $z\sim 2$ is significantly
larger than previously inferred.  A more conservative approach of
excluding IR-luminous AGN, e.g., such as that adopted by
\citet{caputi07}, and assuming their calibration between $8$~$\mu$m
and IR luminosity, results in an IR LD from ULIRGs that is roughly a
factor of $2-3$ lower than the \citet{perez05} determination.  The
critical point of this discussion is that while both the
\citet{perez05} and \citet{caputi07} studies were able to probe the
ULIRG regime of the IR LF with a high degree of completeness, the
shallowness of their data precluded strong constraints on galaxies
significantly fainter than $10^{12}$~L$_{\odot}$.  Our sample and
analysis yield the first constraints on the IR and bolometric LFs for
{\it moderately} luminous galaxies at redshifts $1.9\le z<3.4$, thus
allowing us to evaluate the unobscured SFRD over a larger range of
intrinsic luminosity and redshift than previously possible.  Our
results suggest that the luminosity density contributed by sub-ULIRGs
with $L_{\rm IR}<10^{12}$~L$_{\odot}$ at $z\sim 2$ is larger than
previously inferred.  Table~\ref{tab:contributions} indicates that
such galaxies would comprise roughly $70\%$ of the total IR LD at
$z\sim 2$.

It is important to take a step back at this point and re-examine the
systematic effects induced by the unknown attenuation distribution of
UV faint galaxies, since this is the dominant uncertainty in our
determination of the faint end of the IR LF.  The only way to
reconcile the relatively steep faint-end slope of the UV LF
($\alpha=-1.6$) with the shallow faint-end slope of the IR LF as
suggested by \citet{caputi07} and \citet{perez05} is if there is a
sudden change in the attenuation properties of UV galaxies such that
those with $\rs>25.5$ have negligible dust attenuation.  The
implication is that such UV-faint galaxies would be forming stars in
pristine metal-poor ISM with very low dust-to-gas ratios, akin to
``dwarf'' galaxies in the local universe.  While there are certainly
likely to be such dwarfs at high redshift, they would have to dominate
the number counts on the faint-end of the UV LF.  Future deep
spectroscopic observations should place both the determination of the
UV faint-end slope and the extinction properties of sub-L$^{\ast}$
galaxies on a more secure footing.  However, as we discuss in the next
section, comparison of our {\it total} IR LDs with those corresponding
to SFRD values estimated in previous studies (e.g.,
\citealt{reddy05a}) suggest the IR LD cannot be much lower than the
value derived here, effectively placing a constraint on the
attenuation of sub-ULIRGs.

Lastly, we note that UV emission comprises a non-negligible fraction
of the bolometric luminosity of LIRGs.  This is an effect that
becomes more pronounced at higher redshift as the average dust
attenuation of galaxies of a given bolometric luminosity decreases, as
already discussed.  The bolometric LFs derived in \S~\ref{sec:irlf}
allow us to assess the contribution of galaxies to the total LD,
taking into account both the obscured (IR) and unobscured (UV)
luminosity densities.  The total bolometric luminosity density can be
calculated by integrating our bolometric LF from $L\sim 0$ to
$L=10^{12}$~L$_{\odot}$, and adding the contribution of ULIRGs from
\citet{caputi07}.  Assuming that the fraction of bolometric luminosity
emergent in the UV in ULIRGs is negligible, then the total bolometric
LD at $z\sim 2$ is $\approx 1.5\times 10^{9}$~L$_{\odot}$~Mpc$^{-3}$.
Roughly $80\%$ of this bolometric LD arises from galaxies with $L_{\rm
bol}\la 10^{12}$~L$_{\odot}$.  This bolometric LD is more than $25\%$
larger than what we would have inferred from the IR LF alone because
the former includes the contribution of the LD emergent at UV
wavelengths.

To summarize, there are essentially four points worth keeping in mind
from our analysis of the IR and bolometric LD.  First, the evolution
of the UV LD shows a marked difference from the evolution of the
dust-corrected H$\alpha$ and IR LDs between $z\sim 2$ and $z\sim 0$.
The difference can be explained by an evolution in the average
extinction correction between $z\sim 0$ and $z\sim 2$.  Second, we
find that the IR LD at $z\sim 2$ is at least as large as the value at
$z\sim 3$, implying that the decline in SFRD to the local value must
have occurred after $z\sim 2$.  Third, while our analysis becomes
increasingly incomplete for the most luminous galaxies at $z\sim 2-3$,
it {\it does} provide the first spectroscopic constraints on the
moderate luminosity (e.g., LIRG) regime of the IR LF.  Even taking
into account the significant uncertainties associated with the dust
obscuration of UV-faint galaxies, these results suggest that previous
studies have significantly underestimated the contribution of galaxies
with $L_{\rm IR}\la 10^{12}$~L$_{\odot}$ to the IR luminosity density.
Finally, taking into account the emergent UV luminosity density of
galaxies, we find that sub-ULIRG galaxies comprise roughly $80\%$ of
the total {\it bolometric} LD at $z\sim 2$.  In the next section, we
will discuss these results in the context of the global SFRD.

\subsection{Constraints on the Global Star Formation Rate Density}
\label{sec:sfrd}

\begin{figure*}[hbt]
\plotone{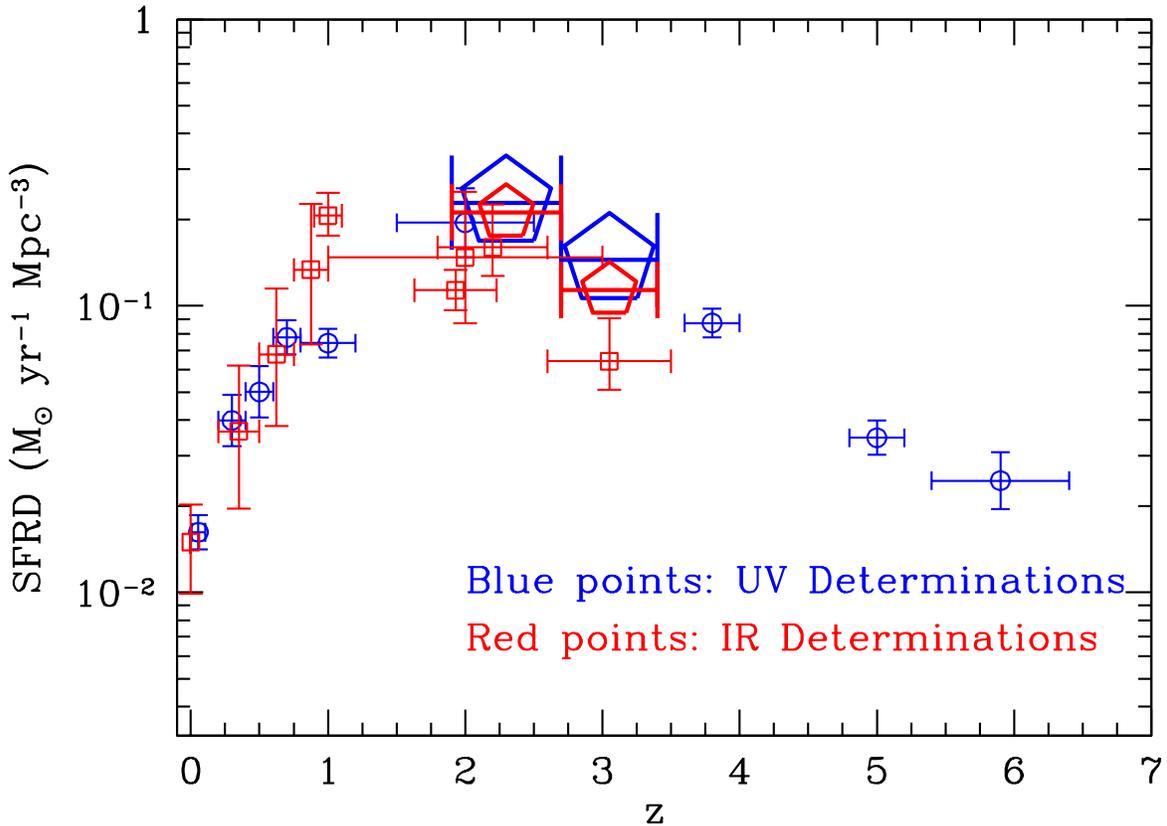}
\caption{Summary of UV (blue open circles) and IR (red open squares)
estimates of the global star formation rate density as a function of
redshift.  The UV points have been corrected for extinction in a
differential manner (i.e., as a function of redshift): determinations
at $z\la 1$ are corrected by a factor of $7$; our values at $z\sim 2$
and $z\sim 3$ are corrected by an average factor of $4.5$; the high
redshift ($z\ga 4$) determinations are corrected by the factors given
in Bouwens et al. (2007, submitted): the dust corrections at $z\sim
4-5$ are determined from a linear interpolation of the $z\sim 3$ and
$z\sim 6$ values.  Our dust-corrected UV determinations are indicated
by the large blue pentagons.  Our IR estimates, computed using a
differential extinction recipe with UV magnitude (see
\S~\ref{sec:attfaint}) and which includes the contribution of ULIRGs
from \citet{caputi07} at $z\sim 2$, are indicated by the red open
pentagons.  The errors on our UV and IR determinations at $z\sim 2$
and $z\sim 3$ are $\sim 20\%$ (derived from the errors on the
luminosity density) and are smaller than the size of the large
pentagons.
\label{fig:sfrd}}
\end{figure*}

We have converted the results of Figure~\ref{fig:uvlfcomp} and
\ref{fig:irlumdens} to star formation rates (SFRs) using the
\citet{kennicutt98} relations and assuming a \citet{salpeter55} IMF
from $0.1$ to $100$~M$_{\odot}$, with the results summarized in
Figure~\ref{fig:sfrd}.\footnote{Assuming the more realistic
\citet{chabrier03} IMF will reduce the SFRD by a factor of $\sim
1.8$.}  The UV points at low redshift ($z\la 1$) are corrected
assuming a factor of $7$ attenuation \citep{schiminovich05}.  Our UV
points are corrected by an average factor of $4.5$ (e.g.,
\citealt{reddy06a, reddy04, nandra02, steidel99}).  Note the
difference in extinction correction between the low redshift points
and our $z\sim 2-3$ determinations, again reflecting the dependence of
extinction on redshift.  The higher redshift points of Bouwens
et~al. (2007, submitted) are corrected according to the extinction
factors published therein.  While there are a number of systematics
that can affect our interpretation of SFRD plots such as
Figure~\ref{fig:sfrd} (e.g., changes in the IMF with redshift), our
analysis has allowed us to carefully and quantitatively assess two of
these systematics: accounting for sample incompleteness and
extinction.  Our measurements indicate an SFRD at $z\sim 2$ that is
{\it at least} as large as the value at $z\sim 3$.  Applying a factor
of $4.5$, as suggested by stacking analyses (e.g.,\citealt{reddy04,
reddy06a, nandra02}), to correct our UV estimates for extinction
yields values that are in general accord with the IR estimates.  On
the one hand, we might not have expected such good agreement given
that the factor of $4.5$ has only been measured for LIRGs, and the
average factor applied to the {\it total} UV LD may be lower depending
on the extinction of UV-faint galaxies.  However, we note that the
IR-estimated SFRD includes the contribution from ULIRGs (whereas the
UV-estimated SFRD does not explicitly take them into account), so the
difference in the two estimates is less than we might have expected.
The important point is that despite the significant uncertainties
regarding the attenuation of UV-faint galaxies, applying a factor of
$4.5$ to total UV LDs as advocated in many studies is not too far off
from the value obtained from IR estimates.

As alluded to previously, independent estimates of the SFRD based on
``census'' studies of high redshift galaxies \citep{reddy05a} can be
compared with our completeness corrected estimate.  Such census
studies estimate the global SFRD by adding up the contribution from
galaxies directly targeted by various selection criteria.  For
example, \citet{reddy05a} estimated the SFRD at $z\sim 2$ by taking
into account the overlap between galaxies selected using the BX,
$\bzk$ \citep{daddi04}, Distant Red Galaxy (DRG; \citealt{franx03})
criteria, and submillimeter selection (e.g., \citealt{chapman05}).
Taking into account the overlap between galaxies selected using these
methods, \citet{reddy05a} compute an SFRD at $z\sim 2$ of
$0.15\pm0.03$~M$_{\odot}$~yr$^{-1}$~Mpc$^{-3}$ for galaxies with
$\rs<25.5$, $\ks(AB)<23.8$, including the contribution from
submillimeter-bright ($S_{\rm 850\mu}\ga 5$~mJy) galaxies.  Since the
total SFRD cannot be smaller than that derived from census studies of
directly detected (optical, near-IR, and submm bright) galaxies, it
suggests that our SFRD measurement cannot be significantly lower than
the value derived here, $\sim 0.2$~M$_{\odot}$~yr$^{-1}$~Mpc$^{-3}$.
This in turn implies that the attenuation of sub-ULIRGs cannot be so
low as to bring down our total SFRD estimate to the point where it is
in violation of the census-computed SFRD.  As an example, integrating
the \citet{caputi07} IR LF at $z\sim 2$ yields an IR LD of $\sim
6.6^{+1.2}_{-1.0}\times 10^{8}$~L$_{\odot}$~Mpc$^{-3}$, corresponding
to an SFRD of $0.11^{+0.2}_{-0.1}$~M$_{\odot}$~yr$^{-1}$~Mpc$^{-3}$.
This ``total'' value is already comparable to, if not smaller, than
the census value of $0.15\pm0.03$~M$_{\odot}$~yr$^{-1}$~Mpc$^{-3}$
\citep{reddy05a}; the latter can be treated as a lower limit to the
SFRD.  The implications are that the faint-end of the IR LF is
unlikely to be as shallow as that predicted from previous IR surveys,
that such UV-faint galaxies are likely to have non-negligible dust
attenuation, and that the contribution of sub-ULIRGs to the total LD
and SFRD must be larger than previously inferred.  Finally, we note
that there are no observational constraints on the presence of dusty
LIRGs at $z\sim 2-3$ that are both faint in the UV {\it and} fall
below the detection threshold of the {\it Spitzer} surveys.  However,
large numbers of such UV-faint dusty LIRGs would serve to increase the
average attenuation of UV-faint galaxies and would strengthen our
conclusions regarding the increased contribution of sub-ULIRGs to the
total LD and SFRD at $z\sim 2-3$.

One of the principle conclusions from this work is that galaxies with
$L_{\rm IR} \approx L_{\rm bol} \la 10^{12}$~L$_{\odot}$ account for
$\approx 70\%$ of the SFRD at $1.9\le z<2.7$.  The fraction rises to
$\sim 80\%$ if we take into account the emergent UV luminosity density
of sub-ULIRGs.  Our analysis suggests that much of the star formation
activity at $z\sim 2$ may take place in faint and moderately-luminous
galaxies.  As a consequence, such faint to moderate luminosity
galaxies must have accounted for a significant fraction of the stellar
mass density formed between $z\sim 3$ and $z\sim 2$.  Assuming a
constant SFRD of $\sim 0.16$~M$_{\odot}$~yr$^{-1}$~Mpc$^{-3}$ (i.e.,
the average of our IR estimates of the SFRD at $z\sim 2$ and $z\sim
3$) between $z=3.4$ and $z=1.9$ (the limiting redshifts of our
analysis) implies that $\sim 45\%$ of the total stellar mass density
of the Universe (e.g., \citealt{cole01}) formed between these
redshifts (e.g., see also \citealt{dickinson03}).  If the fraction of
the luminosity density contributed by galaxies with $L_{\rm bol}\la
10^{12}$~L$_{\odot}$ is roughly constant ($\approx 80\%$) between
$z=3.05$ and $z=2.3$, as suggested by our analysis, then it implies
that one-third of the present-day stellar mass density was formed in
such galaxies.\footnote{We assume a Salpeter IMF in these
calculations.  A Chabrier IMF will reduce the stellar mass density
estimates by a factor of $\sim 1.8$, but the relative contribution of
sub-ULIRGs between $z=3.05$ and $z=2.3$ to the present-day stellar
mass density will be the same, assuming the IMF does not evolve
between $z\sim 2-3$ and the present-day.}

Finally, including the $z\ga 4$ rest-frame UV determinations (no such
corresponding H$\alpha$ or IR data exist at corresponding redshifts),
suggests that the SFRD falls off at these early epochs (e.g.,
\citealt{bouwens06} and references therein).  Assuming a constant IMF
implies that the SFRD at $z\sim 2$ is a factor of $\sim 9$ larger than
that observed at $z\sim 6$ (corrected for extinction).  This trend may
be suggesting a hierarchical buildup of galaxies at early times.  The
important result of our analysis is that this ``growth'' appears to
halt and stay constant for roughly $1.2$~Gyr, between $z\sim 4$ and
$z\sim 2$ (Figure~\ref{fig:uvlfcomp}).  This result is not far from
expectations if the rate of galaxies that are commencing rapid star
formation and populating the bright-end of the UV LF is offset by the
rate of galaxies that are fading because of gas exhaustion or some
other truncation of the star formation such as through AGN or
supernovae feedback (e.g., \citealt{kriek06b, reddy06a, reddy05a,
erb06b}).  In fact, this very epoch hosts the emergence of a
significant population of quiescent galaxies (e.g., \citealt{franx03,
rudnick03, vandokkum04, reddy05a, reddy06a, kriek06b}).  The balance
between rapidly star forming galaxies and those that are fading
appears to saturate around $z\sim 1-2$, at which point the latter
becomes the dominant effect, leading to a decrease in the number
density of bright galaxies between $z\sim 2$ and $z=0$
(Figure~\ref{fig:uvlfcomp}).  As discussed above, this reversal in the
evolution of the UV LF is likely due to gas exhaustion resulting from
any number of processes (e.g., SN-driven outflows, AGN feedback, see
also \citealt{bell05b}).  Once gas exhaustion has occurred and star
formation proceeds quiescently, at least a couple of mechanisms have
been proposed to explain why gas is unable to cool onto galaxies with
the largest stellar masses, including AGN feedback \citep{croton06,
scannapieco05, granato04} and dilution of infalling gas due to virial
shock \citep{deckel06}.  It is interesting to note that it is around
this epoch, $z\sim 2$, that AGN activity appears to peak (e.g.,
\citealt{hopkins07, fan01, shaver96}).
 
\section{Conclusions}
\label{sec:conclusions}

We have used the largest existing sample of spectroscopic redshifts in
the range $1.9\le z<3.4$ to evaluate the luminosity functions (LFs) of
star-forming galaxies at rest-frame UV, H$\alpha$, and infrared (IR)
wavelengths.  The sample of rest-frame UV selected galaxies includes
$\sim 15000$ photometric candidates in 29 independent fields (for a
total area of $\sim 0.9$~square degrees) of which $\sim 2000$ have
spectroscopic redshifts $1.9\le z<3.4$.  The large spectroscopic
database yields critical constraints on the contamination fraction of
our sample from objects at lower redshifts ($z<1.4$) and AGN/QSOs;
statistics that are vital to accurately estimate the bright-end of the
LFs.  We use our extensive sample to correct for incompleteness and
recover the intrinsic rest-frame UV LF at $z\sim 2$ and $z\sim 3$.
Combining this result with H$\alpha$ and {\it Spitzer} MIPS data in
several of our fields enables us to infer the H$\alpha$ and IR LFs of
star-forming galaxies.  The principle conclusions of this work are as
follows:

1. The fraction of star-forming galaxies with rest-frame Ly$\alpha$
  equivalent widths $W_{\rm Ly\alpha}>20$~\AA\, in emission ($f20$)
  increases with redshift in the range $1.9\le z<3.4$, independent of
  selection bias, from a value of $8\%$ at $1.90\le z<2.17$ to $23\%$
  for Lyman-break galaxies at $2.70\le z<3.40$.  If the general
  expectation is that young and less dusty galaxies show Ly$\alpha$ in
  emission, then the trend of increasing $f20$ with redshift reflects
  the decrease in average galaxy age and metallicity with increasing
  redshift.

2. Based on integrating our maximum-likelihood LFs, the fraction of
  star-forming galaxies with redshifts $1.9\le z<2.7$ and $M_{\rm
  AB}(1700\AA) < -19.33$ that have colors that satisfy BX criteria is
  $\approx 58\%$.  Similarly, the fraction of star-forming galaxies
  with redshifts $2.7\le z<3.4$ and $M_{\rm AB}(1700\AA) < -20.02$
  that have colors that satisfy the LBG criteria is $\approx 47\%$.
  The total fraction of $1.9\le z<3.4$ galaxies with $\rs<25.5$ that
  satisfy either the BX or LBG criteria is $55\%$.  We find little
  evolution in the rest-frame UV LF between $z\sim 3$ and $z\sim 2$.
  Correcting for extinction implies the dust-corrected UV luminosity
  density at $z\sim 2$ is at least as large as the value at $z\sim 3$,
  and roughly $9$ times larger than the value at $z\sim 6$.  This
  evolution reverses at redshifts $z\la 2$, where gas exhaustion is
  likely to dominate the evolution of UV-bright galaxies.

3. The incompleteness-corrected estimates of the $\ebmv$ distribution
  indicate very little evolution in the average dust extinction of
  galaxies between $z\sim 3$ and $z\sim 2$, and that such a
  distribution is approximately constant to our spectroscopic limit of
  $\rs=25.5$.  These results are in agreement with stacked X-ray
  analyses of $z\sim 2-3$ galaxies \citep{reddy04, nandra02}.
  However, examining the attenuation distribution of galaxies over a
  larger dynamic range in lookback time indicates an increasing
  extinction per unit star formation rate with decreasing redshift
  \citep{reddy06a, buat07}.  The implication of this trend in average
  extinction is that the evolution in the dust-corrected UV LD is less
  and more pronounced, respectively, at low ($z\la 2$) and high ($z\ga
  2$) redshift than what one would have predicted from the observed
  UV LD (e.g., see also \citealt{bouwens06}).

4. Factoring in the contamination rate of our sample from galaxies at
   lower redshifts and AGN/QSOs with redshifts $2.7\le z<3.4$, we find
   no evidence for an excess of UV-bright galaxies over what was
   inferred in the initial LBG studies of \citet{steidel99} and
   \citet{dickinson98}, as has been recently claimed.

5. The incompleteness-corrected rest-frame UV selected sample and deep
  {\it Spitzer} MIPS data in multiple fields are combined to yield the
  first spectroscopic constraints on the faint and moderate luminosity
  sub-ULIRG ($L_{\rm IR}\la 10^{12}$~L$_{\odot}$) regime of the total
  infrared luminosity functions at $z\sim 2$ and $z\sim 3$.  We use
  this information to show that the number density of $L_{\rm IR}\la
  10^{12}$~L$_{\odot}$ galaxies has been significantly under-estimated
  by previous studies that have relied on shallower IR data.  After
  accounting for the emergent UV luminosity, and assuming a realistic
  range of attenuation for UV faint galaxies, we find that $\approx
  80\%$ of the bolometric (IR+UV) luminosity density at $z\sim 2$
  comes from galaxies with $L_{\rm bol}<10^{12}$~L$_{\odot}$.
  Assuming a constant SFRD of $0.16$~M$_{\odot}$~yr$^{-1}$~Mpc$^{-3}$
  between $z=3.4$ and $z=1.9$ (the limiting redshifts of our $z\sim
  2-3$ analysis), suggests that $L_{\rm bol}<10^{12}$~L$_{\odot}$
  galaxies at these redshifts were responsible for approximately
  one-third of the buildup of the present-day stellar mass density.

6. Our estimate of the total SFRD at $z\sim 2$ is consistent with the
  lower limit on the SFRD provided by census studies of optical,
  near-IR, and submillimeter bright galaxies at similar redshifts
  \citep{reddy05a}.  Our analysis takes into account the systematics
  associated with the attenuation of UV-faint ($\rs>25.5$) galaxies.
  Assuming more extreme changes in the dust attenuation of UV-faint
  galaxies than considered here would be required to reconcile the
  steep and shallow faint-end slopes of the UV and IR LFs,
  respectively, implying that the vast majority of UV-faint galaxies
  would be forming stars from chemically pristine gas.  However, such
  an extreme scenario would result in a {\it total} SFRD that is
  comparable to, if not smaller than, the lower limit from census
  studies \citep{reddy05a}.  The implications are that some
  significant fraction of sub-ULIRGs must have non-negligible dust
  extinction, the faint-end of the IR LF must be steeper than what
  previous studies have suggested, and the contribution of sub-ULIRGs
  to the total SFRD must be significantly larger than previously
  inferred.

\acknowledgements 

We acknowledge useful conversations with Rychard Bouwens, Emeric Le
Floc'h, Casey Papovich, Marcin Sawicki, and Lin Yan.  We thank Emeric
Le Floc'h for providing data from \citet{lefloch05} in electronic
format, Rychard Bouwens for a careful reading of the manuscript, and
the referee for helpful suggestions to improve the clarity of the
paper.  This work would not have been possible without the support of
the staff of the Keck Observatory.  The work presented here has been
supported by grants AST 03-07263 and AST 06-06912 from the National
Science Foundation and by the David and Lucile Packard Foundation.

\bibliographystyle{apj}
%\bibliography{apj-jour,myrefs}

\begin{thebibliography}{134}
\expandafter\ifx\csname natexlab\endcsname\relax\def\natexlab#1{#1}\fi

\bibitem[{{Abraham} {et~al.}(2004){Abraham}, {Glazebrook}, {McCarthy},
  {Crampton}, {Murowinski}, {J{\o}rgensen}, {Roth}, {Hook}, {Savaglio}, {Chen},
  {Marzke}, \& {Carlberg}}]{abraham04}
{Abraham}, R.~G., et~al. 2004, \aj, 127, 2455

\bibitem[{{Adelberger}(2002)}]{adel02t}
{Adelberger}, K.~L. 2002, PhD thesis, California Institute of Technology

\bibitem[{{Adelberger} {et~al.}(2005{\natexlab{a}}){Adelberger}, {Erb},
  {Steidel}, {Reddy}, {Pettini}, \& {Shapley}}]{adelberger05a}
{Adelberger}, K.~L., {Erb}, D.~K., {Steidel}, C.~C., {Reddy}, N.~A., {Pettini},
  M., \& {Shapley}, A.~E. 2005{\natexlab{a}}, \apjl, 620, L75

\bibitem[{{Adelberger} {et~al.}(2005{\natexlab{b}}){Adelberger}, {Shapley},
  {Steidel}, {Pettini}, {Erb}, \& {Reddy}}]{adelberger05b}
{Adelberger}, K.~L., {Shapley}, A.~E., {Steidel}, C.~C., {Pettini}, M., {Erb},
  D.~K., \& {Reddy}, N.~A. 2005{\natexlab{b}}, \apj, 629, 636

\bibitem[{{Adelberger} \& {Steidel}(2000)}]{adel00}
{Adelberger}, K.~L. \& {Steidel}, C.~C. 2000, \apj, 544, 218

\bibitem[{{Adelberger} {et~al.}(2004){Adelberger}, {Steidel}, {Shapley},
  {Hunt}, {Erb}, {Reddy}, \& {Pettini}}]{adelberger04}
{Adelberger}, K.~L., {Steidel}, C.~C., {Shapley}, A.~E., {Hunt}, M.~P., {Erb},
  D.~K., {Reddy}, N.~A., \& {Pettini}, M. 2004, \apj, 607, 226

\bibitem[{{Adelberger} {et~al.}(2003){Adelberger}, {Steidel}, {Shapley}, \&
  {Pettini}}]{adel03}
{Adelberger}, K.~L., {Steidel}, C.~C., {Shapley}, A.~E., \& {Pettini}, M. 2003,
  \apj, 584, 45

\bibitem[{{Afonso} {et~al.}(2003){Afonso}, {Hopkins}, {Mobasher}, \&
  {Almeida}}]{afonso03}
{Afonso}, J., {Hopkins}, A., {Mobasher}, B., \& {Almeida}, C. 2003, \apj, 597,
  269

\bibitem[{{Alexander} {et~al.}(2003){Alexander}, {Bauer}, {Brandt},
  {Schneider}, {Hornschemeier}, {Vignali}, {Barger}, {Broos}, {Cowie},
  {Garmire}, {Townsley}, {Bautz}, {Chartas}, \& {Sargent}}]{alexander03}
{Alexander}, D.~M., et~al. 2003, \aj, 126, 539

\bibitem[{{Arnouts} {et~al.}(2005){Arnouts}, {Schiminovich}, {Ilbert},
  {Tresse}, {Milliard}, {Treyer}, {Bardelli}, {Budavari}, {Wyder}, {Zucca}, {Le
  F{\`e}vre}, {Martin}, {Vettolani}, {Adami}, {Arnaboldi}, {Barlow}, {Bianchi},
  {Bolzonella}, {Bottini}, {Byun}, {Cappi}, {Charlot}, {Contini}, {Donas},
  {Forster}, {Foucaud}, {Franzetti}, {Friedman}, {Garilli}, {Gavignaud},
  {Guzzo}, {Heckman}, {Hoopes}, {Iovino}, {Jelinsky}, {Le Brun}, {Lee},
  {Maccagni}, {Madore}, {Malina}, {Marano}, {Marinoni}, {McCracken}, {Mazure},
  {Meneux}, {Merighi}, {Morrissey}, {Neff}, {Paltani}, {Pell{\`o}}, {Picat},
  {Pollo}, {Pozzetti}, {Radovich}, {Rich}, {Scaramella}, {Scodeggio},
  {Seibert}, {Siegmund}, {Small}, {Szalay}, {Welsh}, {Xu}, {Zamorani}, \&
  {Zanichelli}}]{arnouts05}
{Arnouts}, S., et~al. 2005, \apjl, 619, L43

\bibitem[{{Barger} {et~al.}(2000){Barger}, {Cowie}, \& {Richards}}]{barger00}
{Barger}, A.~J., {Cowie}, L.~L., \& {Richards}, E.~A. 2000, \aj, 119, 2092

\bibitem[{{Barger} {et~al.}(1998){Barger}, {Cowie}, {Sanders}, {Fulton},
  {Taniguchi}, {Sato}, {Kawara}, \& {Okuda}}]{barger98}
{Barger}, A.~J., {Cowie}, L.~L., {Sanders}, D.~B., {Fulton}, E., {Taniguchi},
  Y., {Sato}, Y., {Kawara}, K., \& {Okuda}, H. 1998, \nat, 394, 248

\bibitem[{{Beckwith} {et~al.}(2006){Beckwith}, {Stiavelli}, {Koekemoer},
  {Caldwell}, {Ferguson}, {Hook}, {Lucas}, {Bergeron}, {Corbin}, {Jogee},
  {Panagia}, {Robberto}, {Royle}, {Somerville}, \& {Sosey}}]{beckwith06}
{Beckwith}, S.~V.~W., et~al. 2006, \aj, 132, 1729

\bibitem[{{Bell} {et~al.}(2005){Bell}, {Papovich}, {Wolf}, {Le Floc'h},
  {Caldwell}, {Barden}, {Egami}, {McIntosh}, {Meisenheimer},
  {P{\'e}rez-Gonz{\'a}lez}, {Rieke}, {Rieke}, {Rigby}, \& {Rix}}]{bell05b}
{Bell}, E.~F., et~al. 2005, \apj, 625, 23

\bibitem[{{Blain} {et~al.}(2002){Blain}, {Smail}, {Ivison}, {Kneib}, 
  \& {Frayer}}]{blain02} {Blain}, A.~W., {Smail}, I., {Ivison}, R.~J.,
  {Kneib}, J.-P., \& {Frayer}, D.~T. 2002, \physrep, 369, 111

\bibitem[{{Bolzonella} {et~al.}(2000){Bolzonella}, {Miralles}, \& {Pell{\'
  o}}}]{bolzonella00}
{Bolzonella}, M., {Miralles}, J.-M., \& {Pell{\' o}}, R. 2000, \aap, 363, 476

\bibitem[{{Bouwens} {et~al.}(2006){Bouwens}, {Illingworth}, {Blakeslee}, \&
  {Franx}}]{bouwens06}
{Bouwens}, R.~J., {Illingworth}, G.~D., {Blakeslee}, J.~P., \& {Franx}, M.
  2006, \apj, 653, 53

\bibitem[{{Bouwens} {et~al.}(2005){Bouwens}, {Illingworth}, {Thompson}, \&
  {Franx}}]{bouwens05}
{Bouwens}, R.~J., {Illingworth}, G.~D., {Thompson}, R.~I., \& {Franx}, M. 2005,
  \apjl, 624, L5

\bibitem[{{Bouwens} {et~al.}(2004){Bouwens}, {Thompson}, {Illingworth},
  {Franx}, {van Dokkum}, {Fan}, {Dickinson}, {Eisenstein}, \&
  {Rieke}}]{bouwens04}
{Bouwens}, R.~J., et~al. 2004, \apjl, 616, L79

\bibitem[{{Bruzual} \& {Charlot}(1996)}]{bruzual96}
{Bruzual}, G. \& {Charlot}, S. 1996, private communication (BC96 manual)

\bibitem[{{Buat} {et~al.}(2005){Buat}, {Iglesias-P{\'a}ramo}, {Seibert},
  {Burgarella}, {Charlot}, {Martin}, {Xu}, {Heckman}, {Boissier}, {Boselli},
  {Barlow}, {Bianchi}, {Byun}, {Donas}, {Forster}, {Friedman}, {Jelinski},
  {Lee}, {Madore}, {Malina}, {Milliard}, {Morissey}, {Neff}, {Rich},
  {Schiminovitch}, {Siegmund}, {Small}, {Szalay}, {Welsh}, \& {Wyder}}]{buat05}
{Buat}, V., et~al. 2005, \apjl, 619, L51

\bibitem[{{Buat} {et~al.}(2007){Buat}, {Marcillac}, {Burgarella}, {Le Floc'h},
  {Rieke}, {Takeuchi}, {Iglesias-Paramo}, \& {Xu}}]{buat07}
{Buat}, V., {Marcillac}, D., {Burgarella}, D., {Le Floc'h}, E., {Rieke}, G.,
  {Takeuchi}, T.~T., {Iglesias-Paramo}, J., \& {Xu}, C.~K. 2007, \aap, 469, 19

\bibitem[{{Buat} {et~al.}(2006){Buat}, {Takeuchi}, {Iglesias-Paramo}, {Xu},
  {Burgarella}, {Boselli}, {Barlow}, {Bianchi}, {Donas}, {Forster}, {Friedman},
  {Heckman}, {Lee}, {Madore}, {Martin}, {Milliard}, {Morissey}, {Neff}, {Rich},
  {Schiminovich}, {Seibert}, {Small}, {Szalay}, {Welsh}, {Wyder}, \&
  {Yi}}]{buat06}
{Buat}, V., et~al. 2006, ArXiv Astrophysics e-prints, 0609738

\bibitem[{{Bunker} {et~al.}(2006){Bunker}, {Stanway}, {Ellis}, {McMahon},
  {Eyles}, \& {Lacy}}]{bunker06}
{Bunker}, A., {Stanway}, E., {Ellis}, R., {McMahon}, R., {Eyles}, L., \&
  {Lacy}, M. 2006, New Astronomy Review, 50, 94

\bibitem[{{Bunker} {et~al.}(2004){Bunker}, {Stanway}, {Ellis}, \&
  {McMahon}}]{bunker04}
{Bunker}, A.~J., {Stanway}, E.~R., {Ellis}, R.~S., \& {McMahon}, R.~G. 2004,
  \mnras, 355, 374

\bibitem[{{Burgarella} {et~al.}(2007){Burgarella}, {Le Floc'h}, {Takeuchi},
  {Huang}, {Buat}, {Rieke}, \& {Tyler}}]{burgarella07}
{Burgarella}, D., {Le Floc'h}, E., {Takeuchi}, T.~T., {Huang}, J.~S., {Buat},
  V., {Rieke}, G.~H., \& {Tyler}, K.~D. 2007, ArXiv e-prints, 0706.0810

\bibitem[{{Calzetti} {et~al.}(2000){Calzetti}, {Armus}, {Bohlin}, {Kinney},
  {Koornneef}, \& {Storchi-Bergmann}}]{calzetti00}
{Calzetti}, D., {Armus}, L., {Bohlin}, R.~C., {Kinney}, . A.~L., {Koornneef},
  J., \& {Storchi-Bergmann}, T. 2000, \apj, 533, 682

\bibitem[{{Caputi} {et~al.}(2007){Caputi}, {Lagache}, {Yan}, {Dole},
  {Bavouzet}, {Le Floc'h}, {Choi}, {Helou}, \& {Reddy}}]{caputi07}
{Caputi}, K.~I., et~al. 2007, \apj, 660, 97

\bibitem[{{Chabrier}(2003)}]{chabrier03}
{Chabrier}, G. 2003, \pasp, 115, 763

\bibitem[{{Chapman} {et~al.}(2003){Chapman}, {Blain}, {Ivison}, \&
  {Smail}}]{chapman03}
{Chapman}, S.~C., {Blain}, A.~W., {Ivison}, R.~J., \& {Smail}, I.~R. 2003,
  \nat, 422, 695

\bibitem[{{Chapman} {et~al.}(2005){Chapman}, {Blain}, {Smail}, \&
  {Ivison}}]{chapman05}
{Chapman}, S.~C., {Blain}, A.~W., {Smail}, I., \& {Ivison}, R.~J. 2005, \apj,
  622, 772

\bibitem[{{Cole} {et~al.}(2001){Cole}, {Norberg}, {Baugh}, {Frenk},
  {Bland-Hawthorn}, {Bridges}, {Cannon}, {Colless}, {Collins}, {Couch},
  {Cross}, {Dalton}, {De Propris}, {Driver}, {Efstathiou}, {Ellis},
  {Glazebrook}, {Jackson}, {Lahav}, {Lewis}, {Lumsden}, {Maddox}, {Madgwick},
  {Peacock}, {Peterson}, {Sutherland}, \& {Taylor}}]{cole01}
{Cole}, S., et~al. 2001, \mnras, 326, 255

\bibitem[{{Conselice} {et~al.}(2004){Conselice}, {Grogin}, {Jogee}, {Lucas},
  {Dahlen}, {de Mello}, {Gardner}, {Mobasher}, \& {Ravindranath}}]{conselice04}
{Conselice}, C.~J., et~al. 2004,
  \apjl, 600, L139

\bibitem[{{Cowie} {et~al.}(2004){Cowie}, {Barger}, {Hu}, {Capak}, \&
  {Songaila}}]{cowie04}
{Cowie}, L.~L., {Barger}, A.~J., {Hu}, E.~M., {Capak}, P., \& {Songaila}, A.
  2004, \aj, 127, 3137

\bibitem[{{Croton} {et~al.}(2006){Croton}, {Springel}, {White}, {De Lucia},
  {Frenk}, {Gao}, {Jenkins}, {Kauffmann}, {Navarro}, \& {Yoshida}}]{croton06}
{Croton}, D.~J., et~al. 2006, \mnras, 365, 11

\bibitem[{{Daddi} {et~al.}(2007{\natexlab{a}}){Daddi}, {Alexander},
  {Dickinson}, {Gilli}, {Renzini}, {Elbaz}, {Cimatti}, {Chary}, {Frayer},
  {Bauer}, {Brandt}, {Giavalisco}, {Grogin}, {Huynh}, {Kurk}, {Mignoli},
  {Morrison}, {Pope}, \& {Ravindranath}}]{daddi07b}
{Daddi}, E.,et~al. 2007{\natexlab{a}}, ArXiv e-prints, 0705.2832

\bibitem[{{Daddi} {et~al.}(2004{\natexlab{a}}){Daddi}, {Cimatti}, {Renzini},
  {Fontana}, {Mignoli}, {Pozzetti}, {Tozzi}, \& {Zamorani}}]{daddi04b}
{Daddi}, E., {Cimatti}, A., {Renzini}, A., {Fontana}, A., {Mignoli}, M.,
  {Pozzetti}, L., {Tozzi}, P., \& {Zamorani}, G. 2004{\natexlab{a}}, \apj, 617,
  746

\bibitem[{{Daddi} {et~al.}(2004{\natexlab{b}}){Daddi}, {Cimatti}, {Renzini},
  {Vernet}, {Conselice}, {Pozzetti}, {Mignoli}, {Tozzi}, {di Serego Alighieri},
  {Fontana}, {Nonino}, {Rosati}, \& {Zamorani}}]{daddi04}
{Daddi}, E., et~al. 2004{\natexlab{b}}, \apjl, 600, L127

\bibitem[{{Daddi} {et~al.}(2007{\natexlab{b}}){Daddi}, {Dickinson}, {Morrison},
  {Chary}, {Cimatti}, {Elbaz}, {Frayer}, {Renzini}, {Pope}, {Alexander},
  {Bauer}, {Giavalisco}, {Huynh}, {Kurk}, \& {Mignoli}}]{daddi07a}
{Daddi}, E., et~al. 2007{\natexlab{b}}, ArXiv e-prints, 0705.2831

\bibitem[{{Dekel} \& {Birnboim}(2006)}]{deckel06}
{Dekel}, A. \& {Birnboim}, Y. 2006, \mnras, 368, 2

\bibitem[{{Di Matteo} {et~al.}(2003){Di Matteo}, {Croft}, {Springel}, \&
  {Hernquist}}]{dimatteo03}
{Di Matteo}, T., {Croft}, R.~A.~C., {Springel}, V., \& {Hernquist}, L. 2003,
  \apj, 593, 56

\bibitem[{{Dickinson}(1998)}]{dickinson98}
{Dickinson}, M. 1998, in The Hubble Deep Field, ed. M.~{Livio}, S.~M. {Fall},
  \& P.~{Madau}, 219--+

\bibitem[{{Dickinson} {et~al.}(2003){Dickinson}, {Papovich}, {Ferguson}, \&
  {Budav{\' a}ri}}]{dickinson03}
{Dickinson}, M., {Papovich}, C., {Ferguson}, H.~C., \& {Budav{\' a}ri}, T.
  2003, \apj, 587, 25

\bibitem[{{Dickinson} {et~al.}(2004){Dickinson}, {Stern}, {Giavalisco},
  {Ferguson}, {Tsvetanov}, {Chornock}, {Cristiani}, {Dawson}, {Dey},
  {Filippenko}, {Moustakas}, {Nonino}, {Papovich}, {Ravindranath}, {Riess},
  {Rosati}, {Spinrad}, \& {Vanzella}}]{dickinson04}
{Dickinson}, M., et~al. 2004, \apjl, 600,
  L99

\bibitem[{{Dow-Hygelund} {et~al.}(2007){Dow-Hygelund}, {Holden}, {Bouwens},
  {Illingworth}, {van der Wel}, {Franx}, {van Dokkum}, {Ford}, {Rosati},
  {Magee}, \& {Zirm}}]{dow07}
{Dow-Hygelund}, C.~C., et~al. 2007, \apj, 660, 47

\bibitem[{{Erb} {et~al.}(2006{\natexlab{a}}){Erb}, {Shapley}, {Pettini},
  {Steidel}, {Reddy}, \& {Adelberger}}]{erb06a}
{Erb}, D.~K., {Shapley}, A.~E., {Pettini}, M., {Steidel}, C.~C., {Reddy},
  N.~A., \& {Adelberger}, K.~L. 2006{\natexlab{a}}, \apj, 644, 813

\bibitem[{{Erb} {et~al.}(2006{\natexlab{b}}){Erb}, {Steidel}, {Shapley},
  {Pettini}, {Reddy}, \& {Adelberger}}]{erb06c}
{Erb}, D.~K., {Steidel}, C.~C., {Shapley}, A.~E., {Pettini}, M., {Reddy},
  N.~A., \& {Adelberger}, K.~L. 2006{\natexlab{b}}, \apj, 647, 128

\bibitem[{{Erb} {et~al.}(2006{\natexlab{c}}){Erb}, {Steidel}, {Shapley},
  {Pettini}, {Reddy}, \& {Adelberger}}]{erb06b}
---. 2006{\natexlab{c}}, \apj, 646, 107

\bibitem[{{Fan} {et~al.}(2001){Fan}, {Strauss}, {Schneider}, {Gunn}, {Lupton},
  {Becker}, {Davis}, {Newman}, {Richards}, {White}, {Anderson}, {Annis},
  {Bahcall}, {Brunner}, {Csabai}, {Hennessy}, {Hindsley}, {Fukugita}, {Kunszt},
  {Ivezi{\' c}}, {Knapp}, {McKay}, {Munn}, {Pier}, {Szalay}, \& {York}}]{fan01}
{Fan}, X., et~al. 2001, \aj, 121, 54

\bibitem[{{Finkelstein} {et~al.}(2007){Finkelstein}, {Rhoads}, {Malhotra},
  {Pirzkal}, \& {Wang}}]{finkelstein07}
{Finkelstein}, S.~L., {Rhoads}, J.~E., {Malhotra}, S., {Pirzkal}, N., \&
  {Wang}, J. 2007, \apj, 660, 1023

\bibitem[{{Flores} {et~al.}(1999){Flores}, {Hammer}, {Thuan}, {C{\'e}sarsky},
  {Desert}, {Omont}, {Lilly}, {Eales}, {Crampton}, \& {Le
  F{\`e}vre}}]{flores99}
{Flores}, H., et~al. 1999, \apj, 517, 148

\bibitem[{{Franx} {et~al.}(2003){Franx}, {Labb{\' e}}, {Rudnick}, {van Dokkum\
  }, {Daddi}, {F{\" o}rster Schreiber}, {Moorwood}, {Rix}, {R{\" o}ttgering},
  {van de Wel}, {van der Werf}, \& {van Starkenburg}}]{franx03}
{Franx}, M., et~al. 2003, \apjl, 587, L79

\bibitem[{{Gabasch} {et~al.}(2004){Gabasch}, {Salvato}, {Saglia}, {Bender},
  {Hopp}, {Seitz}, {Feulner}, {Pannella}, {Drory}, {Schirmer}, \&
  {Erben}}]{gabasch04}
{Gabasch}, A.,et~al. 2004, \apjl, 616, L83

\bibitem[{{Gallego} {et~al.}(1995){Gallego}, {Zamorano}, {Aragon-Salamanca}, \&
  {Rego}}]{gallego95}
{Gallego}, J., {Zamorano}, J., {Aragon-Salamanca}, A., \& {Rego}, M. 1995,
  \apjl, 455, L1+

\bibitem[{{Giavalisco} {et~al.}(2004){Giavalisco}, {Ferguson}, {Koekemoer},
  {Dickinson}, {Alexander}, {Bauer}, {Bergeron}, {Biagetti}, {Brandt},
  {Casertano}, {Cesarsky}, {Chatzichristou}, {Conselice}, {Cristiani}, {Da
  Costa}, {Dahlen}, {de Mello}, {Eisenhardt}, {Erben}, {Fall}, {Fassnacht},
  {Fosbury}, {Fruchter}, {Gardner}, {Grogin}, {Hook}, {Hornschemeier}, {Idzi},
  {Jogee}, {Kretchmer}, {Laidler}, {Lee}, {Livio}, {Lucas}, {Madau},
  {Mobasher}, {Moustakas}, {Nonino}, {Padovani}, {Papovich}, {Park},
  {Ravindranath}, {Renzini}, {Richardson}, {Riess}, {Rosati}, {Schirmer},
  {Schreier}, {Somerville}, {Spinrad}, {Stern}, {Stiavelli}, {Strolger},
  {Urry}, {Vandame}, {Williams}, \& {Wolf}}]{giavalisco04}
{Giavalisco}, M., et~al. 2004, \apjl, 600, L93

\bibitem[{{Giavalisco} {et~al.}(1996){Giavalisco}, {Koratkar}, \&
  {Calzetti}}]{giavalisco96}
{Giavalisco}, M., {Koratkar}, A., \& {Calzetti}, D. 1996, \apj, 466, 831

\bibitem[{{Glazebrook} {et~al.}(1999){Glazebrook}, {Blake}, {Economou},
  {Lilly}, \& {Colless}}]{glazebrook99}
{Glazebrook}, K., {Blake}, C., {Economou}, F., {Lilly}, S., \& {Colless}, M.
  1999, \mnras, 306, 843

\bibitem[{{Granato} {et~al.}(2004){Granato}, {De Zotti}, {Silva}, {Bressan}, \&
  {Danese}}]{granato04}
{Granato}, G.~L., {De Zotti}, G., {Silva}, L., {Bressan}, A., \& {Danese}, L.
  2004, \apj, 600, 580

\bibitem[{{Hansen} \& {Oh}(2006)}]{hansen06}
{Hansen}, M. \& {Oh}, S.~P. 2006, \mnras, 367, 979

\bibitem[{{Hopkins}(2004)}]{hopkins04}
{Hopkins}, A.~M. 2004, \apj, 615, 209

\bibitem[{{Hopkins} {et~al.}(2001{\natexlab{a}}){Hopkins}, {Connolly},
  {Haarsma}, \& {Cram}}]{hopkins01}
{Hopkins}, A.~M., {Connolly}, A.~J., {Haarsma}, D.~B., \& {Cram}, L.~E.
  2001{\natexlab{a}}, \aj, 122, 288

\bibitem[{{Hopkins} {et~al.}(2001{\natexlab{b}}){Hopkins}, {Connolly},
  {Haarsma}, \& {Cram}}]{wang96}
---. 2001{\natexlab{b}}, \aj, 122, 288

\bibitem[{{Hopkins} {et~al.}(2000){Hopkins}, {Connolly}, \&
  {Szalay}}]{hopkins00}
{Hopkins}, A.~M., {Connolly}, A.~J., \& {Szalay}, A.~S. 2000, \aj, 120, 2843

\bibitem[{{Hopkins} {et~al.}(2007){Hopkins}, {Richards}, \&
  {Hernquist}}]{hopkins07}
{Hopkins}, P.~F., {Richards}, G.~T., \& {Hernquist}, L. 2007, \apj, 654, 731

\bibitem[{{Hughes} {et~al.}(1998){Hughes}, {Serjeant}, {Dunlop},
  {Rowan-Robinson}, {Blain}, {Mann}, {Ivison}, {Peacock}, {Efstathiou}, {Gear},
  {Oliver}, {Lawrence}, {Longair}, {Goldschmidt}, \& {Jenness}}]{hughes98}
{Hughes}, D.~H., et~al. 1998, \nat, 394, 241

\bibitem[{{Iwata} {et~al.}(2007){Iwata}, {Ohta}, {Tamura}, {Akiyama}, {Aoki},
  {Ando}, {Kiuchi}, \& {Sawicki}}]{iwata07}
{Iwata}, I., {Ohta}, K., {Tamura}, N., {Akiyama}, M., {Aoki}, K., {Ando}, M.,
  {Kiuchi}, G., \& {Sawicki}, M. 2007, \mnras, 376, 1557

\bibitem[{{Kennicutt}(1998)}]{kennicutt98}
{Kennicutt}, R.~C. 1998, \araa, 36, 189

\bibitem[{{Kriek} {et~al.}(2006){Kriek}, {van Dokkum}, {Franx}, {Illingworth},
  {Coppi}, {Forster Schreiber}, {Gawiser}, {Labbe}, {Lira}, {Marchesini},
  {Quadri}, {Rudnick}, {Taylor}, {Urry}, \& {van der Werf}}]{kriek06b}
{Kriek}, M., et~al. 2006, ArXiv Astrophysics e-prints, astro-ph/0611724

\bibitem[{{Law} {et~al.}(2007){Law}, {Steidel}, {Erb}, {Pettini}, {Reddy},
  {Shapley}, {Adelberger}, \& {Simenc}}]{law07}
{Law}, D.~R., {Steidel}, C.~C., {Erb}, D.~K., {Pettini}, M., {Reddy}, N.~A.,
  {Shapley}, A.~E., {Adelberger}, K.~L., \& {Simenc}, D.~J. 2007, \apj, 656, 1

\bibitem[{{Le F{\`e}vre} {et~al.}(2005){Le F{\`e}vre}, {Paltani}, {Arnouts},
  {Charlot}, {Foucaud}, {Ilbert}, {McCracken}, {Zamorani}, {Bottini},
  {Garilli}, {Le Brun}, {Maccagni}, {Picat}, {Scaramella}, {Scodeggio},
  {Tresse}, {Vettolani}, {Zanichelli}, {Adami}, {Bardelli}, {Bolzonella},
  {Cappi}, {Ciliegi}, {Contini}, {Franzetti}, {Gavignaud}, {Guzzo}, {Iovino},
  {Marano}, {Marinoni}, {Mazure}, {Meneux}, {Merighi}, {Pell{\`o}}, {Pollo},
  {Pozzetti}, {Radovich}, {Zucca}, {Arnaboldi}, {Bondi}, {Bongiorno},
  {Busarello}, {Gregorini}, {Lamareille}, {Mathez}, {Mellier}, {Merluzzi},
  {Ripepi}, \& {Rizzo}}]{lefevre05}
{Le F{\`e}vre}, O., et~al. 2005, \nat, 437, 519

\bibitem[{{Le Floc'h} {et~al.}(2005){Le Floc'h}, {Papovich}, {Dole}, {Bell},
  {Lagache}, {Rieke}, {Egami}, {P{\'e}rez-Gonz{\'a}lez}, {Alonso-Herrero},
  {Rieke}, {Blaylock}, {Engelbracht}, {Gordon}, {Hines}, {Misselt}, {Morrison},
  \& {Mould}}]{lefloch05}
{Le Floc'h}, E., et~al. 2005, \apj, 632, 169

\bibitem[{{Lehnert} \& {Bremer}(2003)}]{lehnert03}
{Lehnert}, M.~D. \& {Bremer}, M. 2003, \apj, 593, 630

\bibitem[{{Lilly} {et~al.}(1996){Lilly}, {Le Fevre}, {Hammer}, \&
  {Crampton}}]{lilly96}
{Lilly}, S.~J., {Le Fevre}, O., {Hammer}, F., \& {Crampton}, D. 1996, \apjl,
  460, L1+

\bibitem[{{Lilly} {et~al.}(1995){Lilly}, {Tresse}, {Hammer}, {Crampton}, \& {Le
  Fevre}}]{lilly95}
{Lilly}, S.~J., {Tresse}, L., {Hammer}, F., {Crampton}, D., \& {Le Fevre}, O.
  1995, \apj, 455, 108

\bibitem[{{Madau}(1995)}]{madau95}
{Madau}, P. 1995, \apj, 441, 18

\bibitem[{{Madau} {et~al.}(1996){Madau}, {Ferguson}, {Dickinson}, {Giavalisco},
  {Steidel}, \& {Fruchter}}]{madau96}
{Madau}, P., {Ferguson}, H.~C., {Dickinson}, M.~E., {Giavalisco}, M.,
  {Steidel}, C.~C., \& {Fruchter}, A. 1996, \mnras, 283, 1388

\bibitem[{{Maraston} {et~al.}(2006){Maraston}, {Daddi}, {Renzini}, {Cimatti},
  {Dickinson}, {Papovich}, {Pasquali}, \& {Pirzkal}}]{maraston06}
{Maraston}, C., {Daddi}, E., {Renzini}, A., {Cimatti}, A., {Dickinson}, M.,
  {Papovich}, C., {Pasquali}, A., \& {Pirzkal}, N. 2006, \apj, 652, 85

\bibitem[{{Marchesini} {et~al.}(2007){Marchesini}, {van Dokkum}, {Quadri},
  {Rudnick}, {Franx}, {Lira}, {Wuyts}, {Gawiser}, {Christlein}, \&
  {Toft}}]{marchesini07}
{Marchesini}, D., et~al. 2007, \apj, 656, 42

\bibitem[{{McLean} {et~al.}(1998){McLean}, {Becklin}, {Bendiksen}, {Brims},
  {Canfield}, {Figer}, {Graham}, {Hare}, {Lacayanga}, {Larkin}, {Larson},
  {Levenson}, {Magnone}, {Teplitz}, \& {Wong}}]{mclean98}
{McLean}, I.~S., et~al. 1998, in Proc. SPIE Vol. 3354, p. 566-578, Infrared Astronomical
  Instrumentation, Albert M. Fowler; Ed., ed. A.~M. {Fowler}, 566--578

\bibitem[{{Meurer} {et~al.}(1999){Meurer}, {Heckman}, \& {Calzetti}}]{meurer99}
{Meurer}, G.~R., {Heckman}, T.~M., \& {Calzetti}, D. 1999, \apj, 521, 64

\bibitem[{{Moorwood} {et~al.}(2000){Moorwood}, {van der Werf}, {Cuby}, \&
  {Oliva}}]{moorwood00}
{Moorwood}, A.~F.~M., {van der Werf}, P.~P., {Cuby}, J.~G., \& {Oliva}, E.
  2000, \aap, 362, 9

\bibitem[{{Nandra} {et~al.}(2002){Nandra}, {Mushotzky}, {Arnaud}, {Steidel},
  {Adelberger}, {Gardner}, {Teplitz}, \& {Windhorst}}]{nandra02}
{Nandra}, K., {Mushotzky}, R.~F., {Arnaud}, K., {Steidel}, C.~C., {Adelberger},
  K.~L., {Gardner}, J.~P., {Teplitz}, H.~I., \& {Windhorst}, R.~A. 2002, \apj,
  576, 625

\bibitem[{{Oke} \& {Gunn}(1983)}]{oke83}
{Oke}, J.~B. \& {Gunn}, J.~E. 1983, \apj, 266, 713

\bibitem[{{Ouchi} {et~al.}(2004){Ouchi}, {Shimasaku}, {Okamura}, {Furusawa},
  {Kashikawa}, {Ota}, {Doi}, {Hamabe}, {Kimura}, {Komiyama}, {Miyazaki},
  {Miyazaki}, {Nakata}, {Sekiguchi}, {Yagi}, \& {Yasuda}}]{ouchi04}
{Ouchi}, M., et~al. 2004, \apj, 611, 660

\bibitem[{{Paltani} {et~al.}(2007){Paltani}, {Le F{\`e}vre}, {Ilbert},
  {Arnouts}, {Bardelli}, {Tresse}, {Zamorani}, {Zucca}, {Bottini}, {Garilli},
  {Le Brun}, {Maccagni}, {Picat}, {Scaramella}, {Scodeggio}, {Vettolani},
  {Zanichelli}, {Adami}, {Bolzonella}, {Cappi}, {Charlot}, {Ciliegi},
  {Contini}, {Foucaud}, {Franzetti}, {Gavignaud}, {Guzzo}, {Iovino},
  {McCracken}, {Marano}, {Marinoni}, {Mazure}, {Meneux}, {Merighi},
  {Pell{\`o}}, {Pollo}, {Pozzetti}, {Radovich}, {Bondi}, {Bongiorno},
  {Brinchmann}, {Cucciati}, {de La Torre}, {Lamareille}, {Mellier}, {Merluzzi},
  {Temporin}, {Vergani}, \& {Walcher}}]{paltani07}
{Paltani}, S., et~al. 2007, \aap, 463, 873

\bibitem[{{Papovich} {et~al.}(2001){Papovich}, {Dickinson}, \&
  {Ferguson}}]{papovich01}
{Papovich}, C., {Dickinson}, M., \& {Ferguson}, H.~C. 2001, \apj, 559, 620

\bibitem[{{Papovich} {et~al.}(2003){Papovich}, {Giavalisco}, {Dickinson},
  {Conselice}, \& {Ferguson}}]{papovich03}
{Papovich}, C., {Giavalisco}, M., {Dickinson}, M., {Conselice}, C.~J., \&
  {Ferguson}, H.~C. 2003, \apj, 598, 827

\bibitem[{{Papovich} {et~al.}(2006){Papovich}, {Moustakas}, {Dickinson}, {Le
  Floc'h}, {Rieke}, {Daddi}, {Alexander}, {Bauer}, {Brandt}, {Dahlen}, {Egami},
  {Eisenhardt}, {Elbaz}, {Ferguson}, {Giavalisco}, {Lucas}, {Mobasher},
  {P{\'e}rez-Gonz{\'a}lez}, {Stutz}, {Rieke}, \& {Yan}}]{papovich06}
{Papovich}, C., et~al. 2006, \apj, 640, 92

\bibitem[{{Papovich} {et~al.}(2007){Papovich}, {Rudnick}, {Le Floc'h},
  {van Dokkum}, {Rieke}, {Taylor}, {Armus}, {Gawiser}, {Huang}, {Marcillac},
  \& {Franx}}]{papovich07}
{Papovich}, C., et~al. 2007, ArXiv e-prints, 0706.2164

\bibitem[{{Pentericci} {et~al.}(2007){Pentericci}, {Grazian}, {Fontana},
  {Salimbeni}, {Santini}, {De Santis}, {Gallozzi}, \&
  {Giallongo}}]{pentericci07}
{Pentericci}, L., {Grazian}, A., {Fontana}, A., {Salimbeni}, S., {Santini}, P.,
  {De Santis}, C., {Gallozzi}, S., \& {Giallongo}, E. 2007, ArXiv Astrophysics
  e-prints, astro-ph/0703013

\bibitem[{{P{\'e}rez-Gonz{\'a}lez} {et~al.}(2005){P{\'e}rez-Gonz{\'a}lez},
  {Rieke}, {Egami}, {Alonso-Herrero}, {Dole}, {Papovich}, {Blaylock}, {Jones},
  {Rieke}, {Rigby}, {Barmby}, {Fazio}, {Huang}, \& {Martin}}]{perez05}
{P{\'e}rez-Gonz{\'a}lez}, P.~G., et~al. 2005, \apj, 630, 82

\bibitem[{{P{\'e}rez-Gonz{\'a}lez} {et~al.}(2003){P{\'e}rez-Gonz{\'a}lez},
  {Zamorano}, {Gallego}, {Arag{\'o}n-Salamanca}, \& {Gil de Paz}}]{perez03}
{P{\'e}rez-Gonz{\'a}lez}, P.~G., {Zamorano}, J., {Gallego}, J.,
  {Arag{\'o}n-Salamanca}, A., \& {Gil de Paz}, A. 2003, \apj, 591, 827

\bibitem[{{Pettini} {et~al.}(1998){Pettini}, {Kellogg}, {Steidel}, {Dickins\
  on}, {Adelberger}, \& {Giavalisco}}]{pettini98}
{Pettini}, M., {Kellogg}, M., {Steidel}, C.~C., {Dickins\ on}, M.,
  {Adelberger}, K.~L., \& {Giavalisco}, M. 1998, \apj, 508, 539

\bibitem[{{Pirzkal} {et~al.}(2006){Pirzkal}, {Malhotra}, {Rhoads}, \&
  {Xu}}]{pirzkal06}
{Pirzkal}, N., {Malhotra}, S., {Rhoads}, J.~E., \& {Xu}, C. 2006, ArXiv
  Astrophysics e-prints, astro-ph/0612513

\bibitem[{{Reddy} {et~al.}(2005){Reddy}, {Erb}, {Steidel}, {Shapley},
  {Adelberger}, \& {Pettini}}]{reddy05a}
{Reddy}, N.~A., {Erb}, D.~K., {Steidel}, C.~C., {Shapley}, A.~E., {Adelberger},
  K.~L., \& {Pettini}, M. 2005, \apj, 633, 748

\bibitem[{{Reddy} \& {Steidel}(2004)}]{reddy04}
{Reddy}, N.~A. \& {Steidel}, C.~C. 2004, \apjl, 603, L13

\bibitem[{{Reddy} {et~al.}(2006{\natexlab{a}}){Reddy}, {Steidel}, {Erb},
  {Shapley}, \& {Pettini}}]{reddy06b}
{Reddy}, N.~A., {Steidel}, C.~C., {Erb}, D.~K., {Shapley}, A.~E., \& {Pettini},
  M. 2006{\natexlab{a}}, \apj, 653, 1004

\bibitem[{{Reddy} {et~al.}(2006{\natexlab{b}}){Reddy}, {Steidel}, {Fadda},
  {Yan}, {Pettini}, {Shapley}, {Erb}, \& {Adelberger}}]{reddy06a}
{Reddy}, N.~A., {Steidel}, C.~C., {Fadda}, D., {Yan}, L., {Pettini}, M.,
  {Shapley}, A.~E., {Erb}, D.~K., \& {Adelberger}, K.~L. 2006{\natexlab{b}},
  \apj, 644, 792

\bibitem[{{Rudnick} {et~al.}(2003){Rudnick}, {Rix}, {Franx}, {Labb{\' e}},
  {Blanton}, {Daddi}, {F{\" o}rster Schreiber}, {Moorwood}, {R{\" o}ttgering},
  {Trujillo}, {van de Wel}, {van der Werf}, {van Dokkum}, \& {van
  Starkenburg}}]{rudnick03}
{Rudnick}, G., et~al. 2003, \apj, 599, 847

\bibitem[{{Salpeter}(1955)}]{salpeter55}
{Salpeter}, E.~E. 1955, \apj, 121, 161

\bibitem[{{Sanders} {et~al.}(2003){Sanders}, {Mazzarella}, {Kim}, {Surace}, \&
  {Soifer}}]{sanders03}
{Sanders}, D.~B., {Mazzarella}, J.~M., {Kim}, D.-C., {Surace}, J.~A., \&
  {Soifer}, B.~T. 2003, \aj, 126, 1607

\bibitem[{{Sawicki} \& {Thompson}(2006)}]{sawicki06a}
{Sawicki}, M. \& {Thompson}, D. 2006, \apj, 642, 653

\bibitem[{{Scannapieco} {et~al.}(2005){Scannapieco}, {Silk}, \&
  {Bouwens}}]{scannapieco05}
{Scannapieco}, E., {Silk}, J., \& {Bouwens}, R. 2005, \apjl, 635, L13

\bibitem[{{Schechter}(1976)}]{schechter76}
{Schechter}, P. 1976, \apj, 203, 297

\bibitem[{{Schiminovich} {et~al.}(2005){Schiminovich}, {Ilbert}, {Arnouts},
  {Milliard}, {Tresse}, {Le F{\`e}vre}, {Treyer}, {Wyder}, {Budav{\'a}ri},
  {Zucca}, {Zamorani}, {Martin}, {Adami}, {Arnaboldi}, {Bardelli}, {Barlow},
  {Bianchi}, {Bolzonella}, {Bottini}, {Byun}, {Cappi}, {Contini}, {Charlot},
  {Donas}, {Forster}, {Foucaud}, {Franzetti}, {Friedman}, {Garilli},
  {Gavignaud}, {Guzzo}, {Heckman}, {Hoopes}, {Iovino}, {Jelinsky}, {Le Brun},
  {Lee}, {Maccagni}, {Madore}, {Malina}, {Marano}, {Marinoni}, {McCracken},
  {Mazure}, {Meneux}, {Morrissey}, {Neff}, {Paltani}, {Pell{\`o}}, {Picat},
  {Pollo}, {Pozzetti}, {Radovich}, {Rich}, {Scaramella}, {Scodeggio},
  {Seibert}, {Siegmund}, {Small}, {Szalay}, {Vettolani}, {Welsh}, {Xu}, \&
  {Zanichelli}}]{schiminovich05}
{Schiminovich}, D., et~al. 2005, \apjl, 619, L47

\bibitem[{{Shapley} {et~al.}(2001){Shapley}, {Steidel}, {Adelberger},
  {Dickinson}, {Giavalisco}, \& {Pettini}}]{shapley01}
{Shapley}, A.~E., {Steidel}, C.~C., {Adelberger}, K.~L., {Dickinson}, M.,
  {Giavalisco}, M., \& {Pettini}, M. 2001, \apj, 562, 95

\bibitem[{{Shapley} {et~al.}(2005){Shapley}, {Steidel}, {Erb}, {Reddy},
  {Adelberger}, {Pettini}, {Barmby}, \& {Huang}}]{shapley05}
{Shapley}, A.~E., {Steidel}, C.~C., {Erb}, D.~K., {Reddy}, N.~A., {Adelberger},
  K.~L., {Pettini}, M., {Barmby}, P., \& {Huang}, J. 2005, \apj, 626, 698

\bibitem[{{Shapley} {et~al.}(2003){Shapley}, {Steidel}, {Pettini}, \&
  {Adelberger}}]{shapley03}
{Shapley}, A.~E., {Steidel}, C.~C., {Pettini}, M., \& {Adelberger}, K.~L. 2003,
  \apj, 588, 65

\bibitem[{{Shaver} {et~al.}(1996){Shaver}, {Wall}, {Kellermann}, {Jackson}, \&
  {Hawkins}}]{shaver96}
{Shaver}, P.~A., {Wall}, J.~V., {Kellermann}, K.~I., {Jackson}, C.~A., \&
  {Hawkins}, M.~R.~S. 1996, \nat, 384, 439

\bibitem[{{Smail}(2003)}]{smail03}
{Smail}, I. 2003, in IAU Symposium

\bibitem[{{Smail} {et~al.}(1997){Smail}, {Ivison}, \& {Blain}}]{smail97}
{Smail}, I., {Ivison}, R.~J., \& {Blain}, A.~W. 1997, \apjl, 490, L5+

\bibitem[{{Stanway} {et~al.}(2007){Stanway}, {Bunker}, {Glazebrook}, {Abraham},
  {Rhoads}, {Malhotra}, {Crampton}, {Colless}, \& {Chiu}}]{stanway07}
{Stanway}, E.~R., et~al. 2007, \mnras, 91

\bibitem[{{Steidel} {et~al.}(1999){Steidel}, {Adelberger}, {Giavalisco},
  {Dickinson}, \& {Pettini}}]{steidel99}
{Steidel}, C.~C., {Adelberger}, K.~L., {Giavalisco}, M., {Dickinson}, M., \&
  {Pettini}, M. 1999, \apj, 519, 1

\bibitem[{{Steidel} {et~al.}(2005){Steidel}, {Adelberger}, {Shapley}, {Erb},
  {Reddy}, \& {Pettini}}]{steidel05}
{Steidel}, C.~C., {Adelberger}, K.~L., {Shapley}, A.~E., {Erb}, D.~K., {Reddy},
  N.~A., \& {Pettini}, M. 2005, \apj, 626, 44

\bibitem[{{Steidel} {et~al.}(2003){Steidel}, {Adelberger}, {Shapley},
  {Pettini}, {Dickinson}, \& {Giavalisco}}]{steidel03}
{Steidel}, C.~C., {Adelberger}, K.~L., {Shapley}, A.~E., {Pettini}, M.,
  {Dickinson}, M., \& {Giavalisco}, M. 2003, \apj, 592, 728

\bibitem[{{Steidel} {et~al.}(2002){Steidel}, {Hunt}, {Shapley}, {Adelberger},
  {Pettini}, {Dickinson}, \& {Giavalisco}}]{steidel02}
{Steidel}, C.~C., {Hunt}, M.~P., {Shapley}, A.~E., {Adelberger}, K.~L.,
  {Pettini}, M., {Dickinson}, M., \& {Giavalisco}, M. 2002, \apj, 576, 653

\bibitem[{{Steidel} {et~al.}(1995){Steidel}, {Pettini}, \&
  {Hamilton}}]{steidel95}
{Steidel}, C.~C., {Pettini}, M., \& {Hamilton}, D. 1995, \aj, 110, 2519

\bibitem[{{Steidel} {et~al.}(2004){Steidel}, {Shapley}, {Pettini},
  {Adelberger}, {Erb}, {Reddy}, \& {Hunt}}]{steidel04}
{Steidel}, C.~C., {Shapley}, A.~E., {Pettini}, M., {Adelberger}, K.~L., {Erb},
  D.~K., {Reddy}, N.~A., \& {Hunt}, M.~P. 2004, \apj, 604, 534

\bibitem[{{Sullivan} {et~al.}(2001){Sullivan}, {Mobasher}, {Chan}, {Cram},
  {Ellis}, {Treyer}, \& {Hopkins}}]{sullivan01}
{Sullivan}, M., {Mobasher}, B., {Chan}, B., {Cram}, L., {Ellis}, R., {Treyer},
  M., \& {Hopkins}, A. 2001, \apj, 558, 72

\bibitem[{{Sullivan} {et~al.}(2000){Sullivan}, {Treyer}, {Ellis}, {Bridges},
  {Milliard}, \& {Donas}}]{sullivan00}
{Sullivan}, M., {Treyer}, M.~A., {Ellis}, R.~S., {Bridges}, T.~J., {Milliard},
  B., \& {Donas}, J. 2000, \mnras, 312, 442

\bibitem[{{Takeuchi} {et~al.}(2005){Takeuchi}, {Buat}, \&
  {Burgarella}}]{takeuchi05}
{Takeuchi}, T.~T., {Buat}, V., \& {Burgarella}, D. 2005, \aap, 440, L17

\bibitem[{{Tapken} {et~al.}(2007){Tapken}, {Appenzeller}, {Noll}, {Richling},
  {Heidt}, {Meink{\"o}hn}, \& {Mehlert}}]{tapken07}
{Tapken}, C., {Appenzeller}, I., {Noll}, S., {Richling}, S., {Heidt}, J.,
  {Meink{\"o}hn}, E., \& {Mehlert}, D. 2007, \aap, 467, 63

\bibitem[{{Tresse} \& {Maddox}(1998)}]{tresse98}
{Tresse}, L. \& {Maddox}, S.~J. 1998, \apj, 495, 691

\bibitem[{{Tresse} {et~al.}(2002){Tresse}, {Maddox}, {Le F{\`e}vre}, \&
  {Cuby}}]{tresse02}
{Tresse}, L., {Maddox}, S.~J., {Le F{\`e}vre}, O., \& {Cuby}, J.-G. 2002,
  \mnras, 337, 369

\bibitem[{{Valdes}(1982)}]{valdes82}
{Valdes}, F. 1982, {FOCAS} User's Manual (NOAO, Tucson).

\bibitem[{{van Dokkum} {et~al.}(2003){van Dokkum}, {F{\" o}rster Schreiber},
  {Franx}, {Daddi}, {Illingworth}, {Labb{\' e}}, {Moorwood}, {Rix}, {R{\"
  o}ttgering}, {Rudnick}, {van der Wel}, {van der Werf}, \& {van
  Starkenburg}}]{vandokkum03}
{van Dokkum}, P.~G., et~al. 2003, \apjl, 587, L83

\bibitem[{{van Dokkum} {et~al.}(2004){van Dokkum}, {Franx}, {F{\" o}rster
  Schreiber}, {Illingworth}, {Daddi}, {Knudsen}, {Labb{\' e}}, {Moorwood},
  {Rix}, {R{\" o}ttgering}\, {Rudnick}, {Trujillo}, {van der Werf}, {van der
  Wel}, {van Starkenburg}, \& {Wuyts}}]{vandokkum04}
{van Dokkum}, P.~G., et~al. 2004, \apj, 611, 703

\bibitem[{{Williams} {et~al.}(2000){Williams}, {Baum}, {Bergeron}, {Bernstein},
  {Blacker}, {Boyle}, {Brown}, {Carollo}, {Casertano}, {Covarrubias}, {de
  Mello}, {Dickinson}, {Espey}, {Ferguson}, {Fruchter}, {Gardner}, {Gonnella},
  {Hayes}, {Hewett}, {Heyer}, {Hook}, {Irwin}, {Jones}, {Kaiser}, {Levay},
  {Lubenow}, {Lucas}, {Mack}, {MacKenty}, {Madau}, {Makidon}, {Martin},
  {Mazzuca}, {Mutchler}, {Norris}, {Perriello}, {Phillips}, {Postman}, {Royle},
  {Sahu}, {Savaglio}, {Sherwin}, {Smith}, {Stiavelli}, {Suntzeff}, {Teplitz},
  {van der Marel}, {Walker}, {Weymann}, {Wiggs}, {Williger}, {Wilson},
  {Zacharias}, \& {Zurek}}]{williams00}
{Williams}, R.~E., et~al. 2000, \aj, 120, 2735

\bibitem[{{Williams} {et~al.}(1996){Williams}, {Blacker}, {Dickinson}, {Dixon},
  {Ferguson}, {Fruchter}, {Giavalisco}, {Gilliland}, {Heyer}, {Katsanis},
  {Levay}, {Lucas}, {McElroy}, {Petro}, {Postman}, {Adorf}, \&
  {Hook}}]{williams96}
{Williams}, R.~E., et~al. 1996, \aj, 112, 1335

\bibitem[{{Wirth} {et~al.}(2004){Wirth}, {Willmer}, {Amico}, {Chaffee},
  {Goodrich}, {Kwok}, {Lyke}, {Mader}, {Tran}, {Barger}, {Cowie}, {Capak},
  {Coil}, {Cooper}, {Conrad}, {Davis}, {Faber}, {Hu}, {Koo}, {Le Mignant},
  {Newman}, \& {Songaila}}]{wirth04}
{Wirth}, G.~D., et~al. 2004, \aj, 127, 3121

\bibitem[{{Wyder} {et~al.}(2005){Wyder}, {Treyer}, {Milliard}, {Schiminovich},
  {Arnouts}, {Budav{\'a}ri}, {Barlow}, {Bianchi}, {Byun}, {Donas}, {Forster},
  {Friedman}, {Heckman}, {Jelinsky}, {Lee}, {Madore}, {Malina}, {Martin},
  {Morrissey}, {Neff}, {Rich}, {Siegmund}, {Small}, {Szalay}, \&
  {Welsh}}]{wyder05}
{Wyder}, T.~K., et~al. 2005, \apjl, 619, L15

\bibitem[{{Yan} {et~al.}(2003){Yan}, {Windhorst}, \& {Cohen}}]{yan03}
{Yan}, H., {Windhorst}, R.~A., \& {Cohen}, S.~H. 2003, \apjl, 585, L93

\bibitem[{{Yan} {et~al.}(1999){Yan}, {McCarthy}, {Freudling}, {Teplitz},
  {Malumuth}, {Weymann}, \& {Malkan}}]{yan99}
{Yan}, L., {McCarthy}, P.~J., {Freudling}, W., {Teplitz}, H.~I., {Malumuth},
  E.~M., {Weymann}, R.~J., \& {Malkan}, M.~A. 1999, \apjl, 519, L47

\bibitem[{{Yun} {et~al.}(2001){Yun}, {Reddy}, \& {Condon}}]{yun01}
{Yun}, M.~S., {Reddy}, N.~A., \& {Condon}, J.~J. 2001, \apj, 554, 803

\end{thebibliography}

\end{document}